\pdfoutput=1
\documentclass[aip, rsi, amsmath, twocolumn, longbibliography, nobibnotes, reprint]{revtex4-1}

\usepackage[caption=false]{subfig}
\usepackage{graphicx}
\usepackage{dcolumn}
\usepackage{pifont}
\usepackage{float}
\usepackage{colortbl}
\usepackage{multirow}
\usepackage{array}
\usepackage[colorlinks=true,citecolor=blue,urlcolor=blue,linkcolor=blue]{hyperref}

\newcommand{\aref}[1]{\hyperref[#1]{Appendix~\ref{#1}}}

\newcolumntype{d}[1]{D{.}{.}{#1}}

\begin{document}

\preprint{AIP/2017}

\title{FPGA Design Techniques for Stable Cryogenic Operation}

\author{Harald Homulle}
\email{h.a.r.homulle@tudelft.nl}
\author{Stefan Visser}
\author{Bishnu Patra}
\author{Edoardo Charbon}
\thanks{The following article has been submitted to Review of Scientific Instruments. If it is published, it will be available on \href{http://rsi.aip.org}{http://rsi.aip.org}.}
\affiliation{QuTech, Delft University of Technology, 2628CD Delft, The Netherlands}
 
\date{12 September 2017}
\revised{\today}

\begin{abstract}
In this paper we show how a deep-submicron FPGA can be modified to operate at extremely low temperatures through modifications in the supporting hardware and in the firmware programming it. Though FPGAs are not designed to operate at a few Kelvin, it is possible to do so on virtue of the extremely high doping levels found in deep-submicron CMOS technology nodes. First, any PCB component, that does not conform with this requirement, is removed. Both the majority of decoupling capacitor types and voltage regulators are not well behaved at cryogenic temperatures, asking for an ad-hoc solution to stabilize the FPGA supply voltage, especially for sensitive applications. Therefore, we have designed a firmware that enforces a constant power consumption, so as to stabilize the supply voltage in the interior of the FPGA chip. The FPGA is powered with a supply at several meters distance, causing significant IR drop and thus fluctuations on the local supply voltage. To achieve the stabilization, the variation in digital logic speed, which directly corresponds to changes in supply voltage, is constantly measured and corrected for through a tunable oscillator farm, implemented on the FPGA. The method is versatile and robust, enabling seamless porting to other FPGA families and configurations.
\end{abstract}

\maketitle

\section{Introduction}\label{sec:intro}

\noindent The cryogenic operation of electronic circuits in general and FPGAs in particular has been extensively studied over the past years \cite{Sheldon2011, Gong2012, Bakhshi2012, Hornibrook2015, ConwayLamb2016, Homulle2016Arxiv}. The majority of the FPGA building blocks, in Xilinx Artix~7 series for instance, has been shown to operate rather stably over the temperature range from 4~K to 300~K \cite{ConwayLamb2016, Homulle2016Arxiv}. For example, the delay change in both look-up tables (LUTs) and carry elements has been shown to change by less than 10\%.

Although the performance of FPGAs operating at deep-cryogenic temperatures is shown to be stable, this is only demonstrated in specific cases, e.g. when building blocks were tested individually. This tests ignored the influence of voltage drop due to long wires used to bring power supply and also the influence the blocks might have on one another. Furthermore, the behaviour of other components, ancillary to the FPGA, is not well known. This is of particular importance when, as it is the case in this work, a platform is to be implemented to control a quantum processor.

The most important components for such a platform are a power decoupling network and DC power supplies or voltage regulators. Capacitors in general tend to drop significantly in capacitance and they increase the effective series resistance by up to 1000$\times$ \cite{Patterson1998, Teverovsky2006, Niekerk2007, Teyssandier2010, Kirschman2014}. Previous studies \cite{Patterson2006, Basit2013} have shown some voltage regulators to operate at cryogenic temperatures, as low as 90~K, but no regulators were found to be working stably below that temperature due to either protection circuitries, bipolar transistors \cite{Ghanam2011, Song2016} or biasing. 

Power networks, that do not behave properly, can pose a burden on optimal FPGA performance, especially at cryogenic conditions where the power is supplied over long wires. While FPGAs switch significant amounts of current at high frequencies, fluctuating IR drop over these wires can be significant. 
While a static IR drop wouldn't cause significant problems, the dynamic IR drop alters the internal delays of the FPGA continuously, causing potential glitches and irregularities. Usually, these problems are mitigated using decoupling capacitors, but the performance of capacitors at cryogenic temperatures is not sufficient to compensate for these fluctuations. 

In this paper, we propose a novel methodology to implement hardware modifications in FPGAs via firmware design, so as to compensate for any potential fluctuations of power dissipation due to sub-optimal structures or biasing in the interior of the FPGA. The same technique can be applied to ASICs, where similar desired effects can be achieved at a wide variety of temperatures. The technique is based on real-time measurements of the variation of cell delay in the carry chain. This fluctuation is mainly caused by voltage drop of the logic supply voltage, but it is also caused by temperature fluctuations. A small farm of oscillators is used to flatten out the power consumption in real-time, so as to automatically stabilize the power supply as well. 

Similar techniques have been shown before, but mainly for the purpose of avoiding security attacks. In those systems, the power consumption or heat has to be flattened in order to prevent one from reading data through power or heat changes. For instance, \cite{He2015} proposes the use of a distributed oscillator farm to equalize the heat map from the FPGA seen from outside. Its control is based on the difference in frequency shift of two differently sized ring oscillators, which is proportional to temperature. 

In this paper, we first study the behaviour of (passive) decoupling capacitors and (active) voltage regulators in \autoref{sec:actpas}. In \autoref{sec:architecture} we propose an architecture, implemented inside the FPGA to stabilize power consumption. Results showing the effectiveness of the stabilization technique are presented in \autoref{sec:results}. \section{Passives and actives}\label{sec:actpas}

\begin{table*}[bt]
  \centering
  \scriptsize
  \caption{Summary of the performance change of several capacitors as used on the FPGA platform PCB and some potential better candidates. All values are measured at a test frequency of 100~Hz. }
    \begin{tabular}{d{3.2}|ll|d{3.2}d{3.2}|d{3.2}d{3.2}|d{3.2}d{3.2}}
	\hline
    \multicolumn{1}{c|}{Specified}      &       &       & \multicolumn{2}{c}{300~K} & \multicolumn{4}{|c}{4~K} \\
	\cline{4-9}
    \multicolumn{1}{l|}{capacitance [$\mu$F]} & Type  & \multicolumn{1}{l}{Part number} & \multicolumn{1}{|l}{Capacitance [$\mu$F]} & \multicolumn{1}{l}{ESR [$\Omega$]} & \multicolumn{1}{|l}{Capacitance [$\mu$F]} & \multicolumn{1}{l}{ESR [$\Omega$]} & \multicolumn{1}{|l}{$\Delta$C [\%]} & \multicolumn{1}{l}{$\Delta$ESR [\%]} \\
	\hline
    0.47  & NP0   & C2220C474J5GACTU      & 0.475 & 0     & 0.475 & 0.54  & 0     & \multicolumn{1}{c}{$\infty$} \\
    1     & PPS   & SMDIC04100TB00KQ00      & 1.01  & 0     & 1.02  & 0     & 1     & 0 \\
    4.7   & X8L & C1206C475J8NACTU      & 4.75  & 6     & 0.32  & 189   & -93   & 3050 \\
    47    & Tantalum & 16TQC47MW      & 44.5  & 0.8   & 38.2  & 0     & -14   & -100 \\
    100   & Tantalum & T495D107K010ATE050      & 102.8 & 0.2   & 91.9  & 1.8   & -11   & 800 \\
    330   & Tantalum & T495X337K010ATE035      & 328.2 & 0.1   & 265.4 & 0     & -19   & -100 \\
	\hline
	\hline
	 0.47  & PPS   & SMDIC03470TB00JQ00      & 0.475 & 0     & 0.478 & 0     & 1     & 0 \\
    1     & Silicon & 935133427710-T3N      & 1.03  & 1.4   & 1.01  & 0     & -2    & -100 \\
    \hline
    \end{tabular}
  \label{tab:caps}
\end{table*}

\subsection{Capacitors}
\noindent Passive components form an important part of any PCB design, whether the PCB houses an FPGA or ASIC as the main embodiment. Especially capacitors, which are not only used for the decoupling networks of power supplies, but also for analog filters, are important. 
At cryogenic conditions, the material properties can differ significantly, altering the dielectric values and thus the resulting junction capacitance and/or resistance. As this is such an important part of any system, a study was conducted to find the optimal capacitor materials that are commercially available. For some applications, it might be worth investigating the performance of cryogenic specialized materials. 

Several studies \cite{Patterson1998, Teverovsky2006, Niekerk2007, Teyssandier2010, Kirschman2014} revealed that significant changes are to be expected, with a large dependency on material dielectric properties. For example high k-dielectric constant materials, such as X5R or X7R, drop almost to zero capacitance, while low k materials, such as NP0, are very stable. 

FPGA systems in particular require a large decoupling network, with capacitance values ranging from 1 to 330~$\mu$F. With this large diversity, it is impossible to use only one dielectric (due to size and availability), and a solution for the higher values has to be found. Tantalum capacitors seem a good trade-off between performance and available values \cite{Teverovsky2006, Teyssandier2010}. There are also special tantalum capacitors available for cryogenic applications, but their performance is only optimal down to 77~K. 

In \autoref{tab:caps} an overview of the tested capacitors is given together with their type and values. Listed are the capacitors currently used on the FPGA PCB and some potential better candidates, with the measured capacitance and effective series resistance at 4~K and 300~K. Tests were executed using a KeySight LCR meter and a test frequency of 100~Hz.  The variance of both capacitance and effective series resistance (ESR) are mainly dictated by the material type. Clearly NP0/COG and PPS capacitor materials are ideal with both limited change in capacitance and no significant increase in ESR as a function of temperature. The worst capacitors are based on ceramic materials with high dielectric constants. In these devices, the capacitance drops over 90\% and ESR increases over 3000\%.

	\begin{figure}[b]
		\centering
		\null\hfill
		\subfloat[]{\label{fig:capacitors_a}
			\includegraphics[width=.225\textwidth]{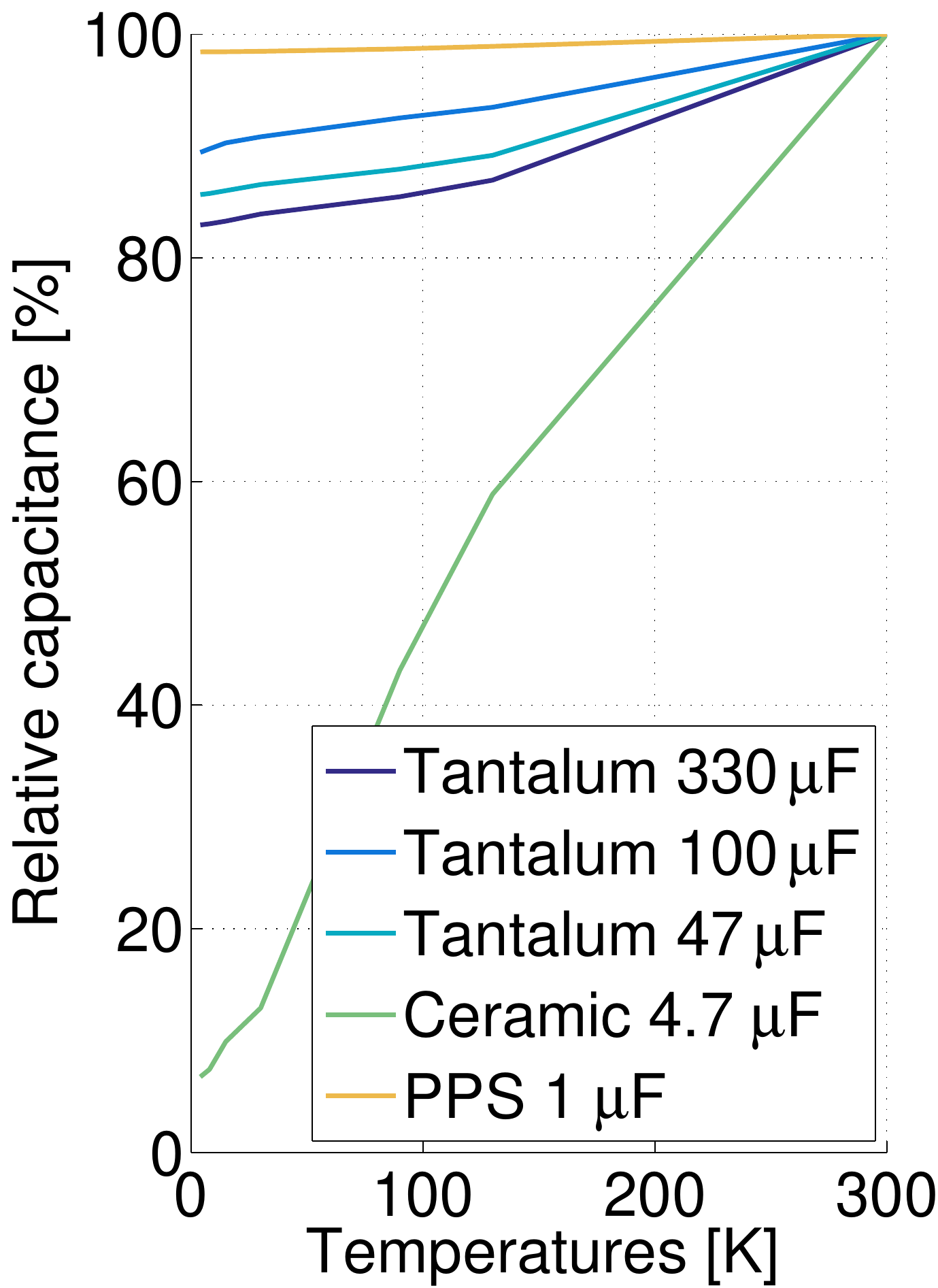}
		}
		\hfill
		\subfloat[]{\label{fig:capacitors_b}
			\includegraphics[width=.225\textwidth]{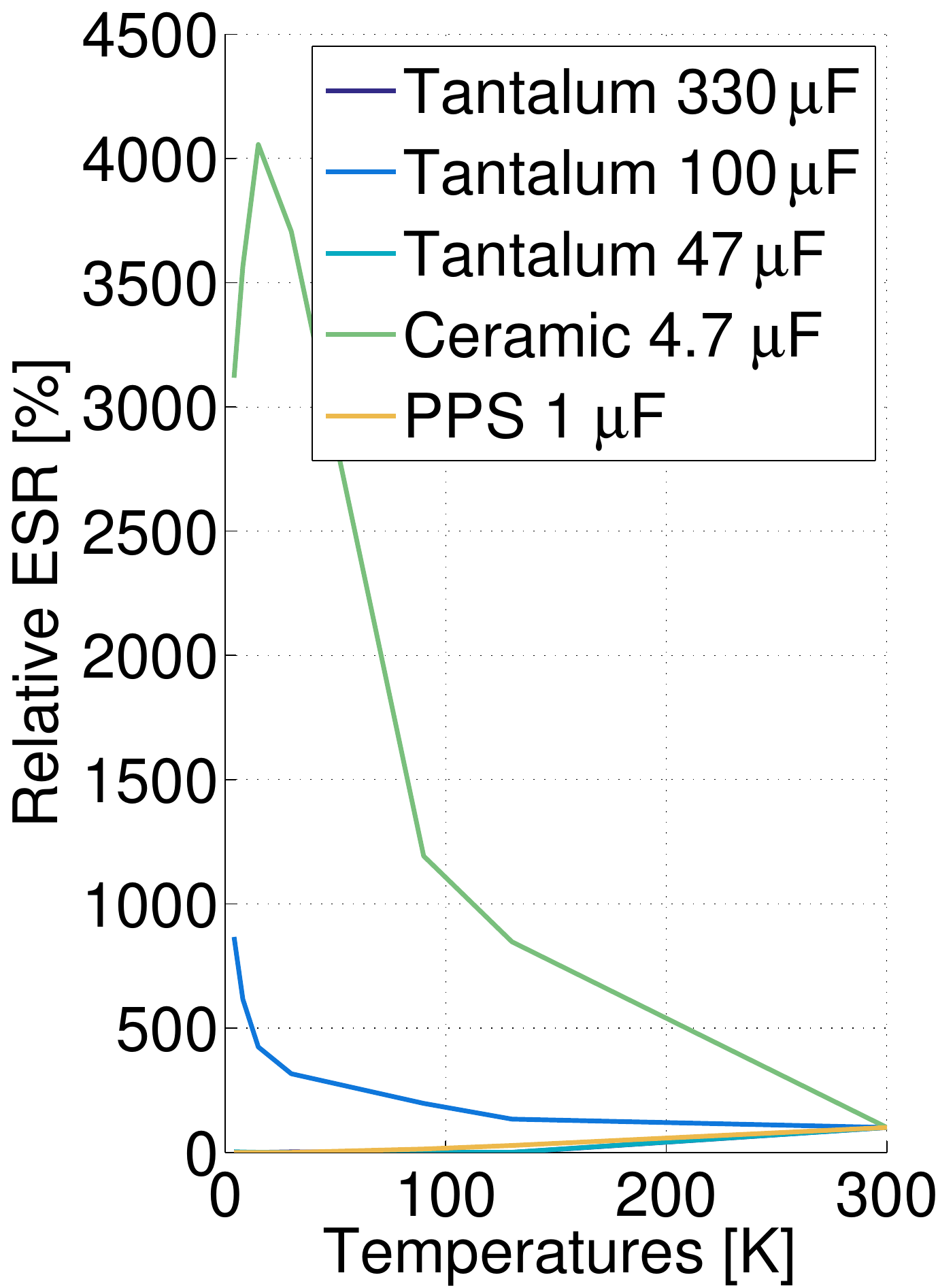}
		}
		\hfill\null
		\caption{Relative capacitance and ESR for some of the capacitors from \autoref{tab:caps} over temperature.}
		\label{fig:capacitors}
	\end{figure}

The relative capacitance and ESR over temperature are plotted in \autoref{fig:capacitors}\subref{fig:capacitors_a} for some of the capacitors from \autoref{tab:caps}. For the ceramic capacitor, the capacitance drops almost linearly with respect to temperature. The ESR on the other hand increases quasi exponentially while lowering temperature as is shown in \autoref{fig:capacitors}\subref{fig:capacitors_b}. More extensive characterization results are reported in \aref{sec:cap_details}.  \\

\noindent For optimal cryogenic performance of the decoupling network, selecting the appropriate capacitor type is important. Special cryogenic capacitors, based on tantalum EPPL2 \cite{Zednicek2013}, were tested and found to be generally better down to 77~K. At 4~K however, those capacitors significantly loose capacitance and increase ESR. Another type of capacitor is based on silicon, which is only available for small values under 1~$\mu$F. Those capacitors exhibit a very similar performance to NP0/COG types, with as main advantage, a significantly smaller footprint. As space is limited in dilution fridges, significantly smaller components are preferable. 

\subsection{Resistors}
\begin{figure}[t]
		\centering
		\includegraphics[width=0.45\textwidth]{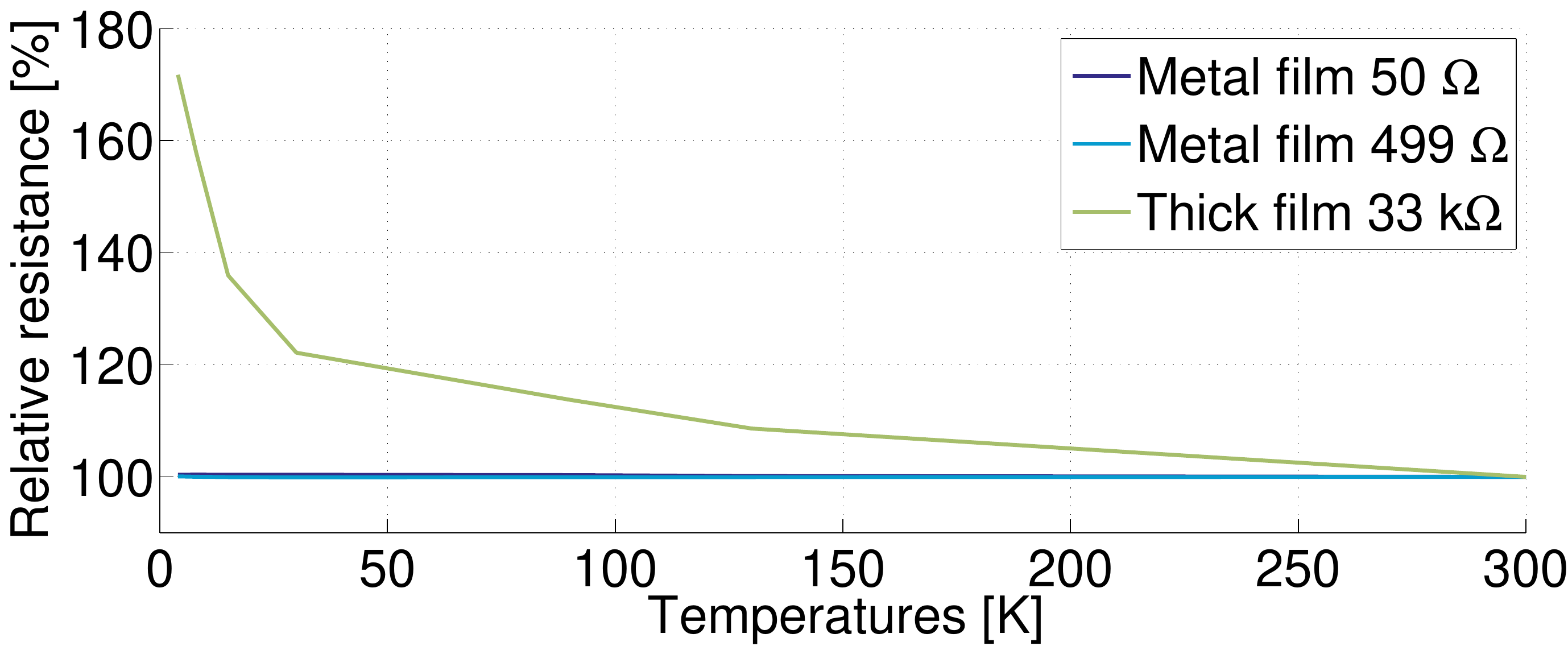}
		\caption{Relative resistance for different materials (metal and thick film) over temperature. More detailed plots are shown in \aref{sec:cap_details}.}
		\label{fig:resistors}
	\end{figure}

\noindent Although not as important as capacitors, resistors are commonly used for filters, protection circuits, termination etc. Therefore, the behaviour of resistances at cryogenic temperatures has to be well understood. As for capacitors, several resistive materials have been studied in the past \cite{Teyssandier2010}. Metal film resistors are found to be most stable over temperature, with fairly limited change in resistance ($<$1\%). The resistance over temperature for three SMD resistors, both metal and thick film, is shown in \autoref{fig:resistors}, confirming the extremely stable resistance of metal films over temperature. 

\subsection{Voltage regulators}

	\begin{figure*}[tb]
		\centering
		\null\hfill
		\subfloat[]{\label{fig:regulators_a}
			\includegraphics[width=.405\textwidth]{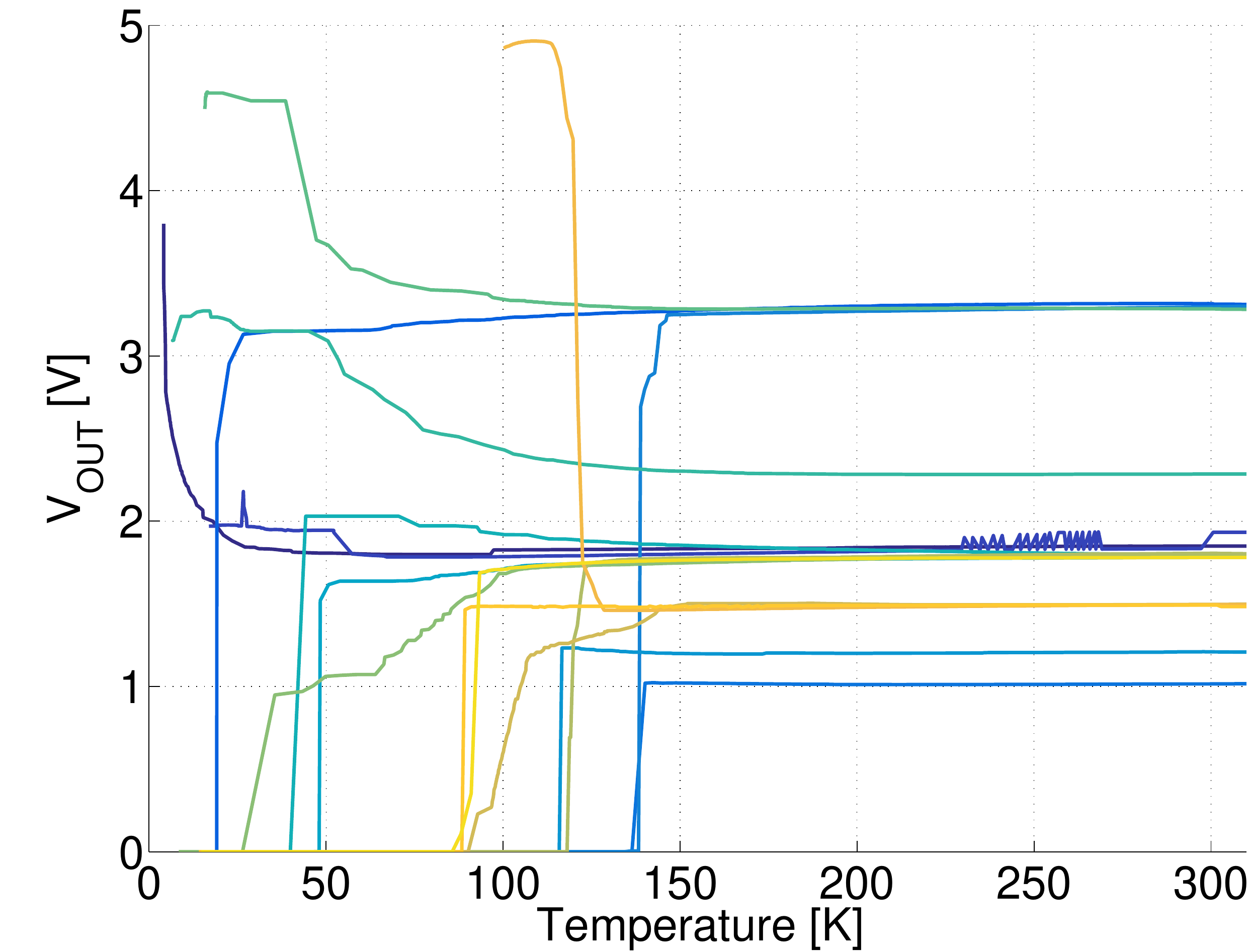}
		}
		\hfill
		\subfloat[]{\label{fig:regulators_b}
			\includegraphics[width=.405\textwidth]{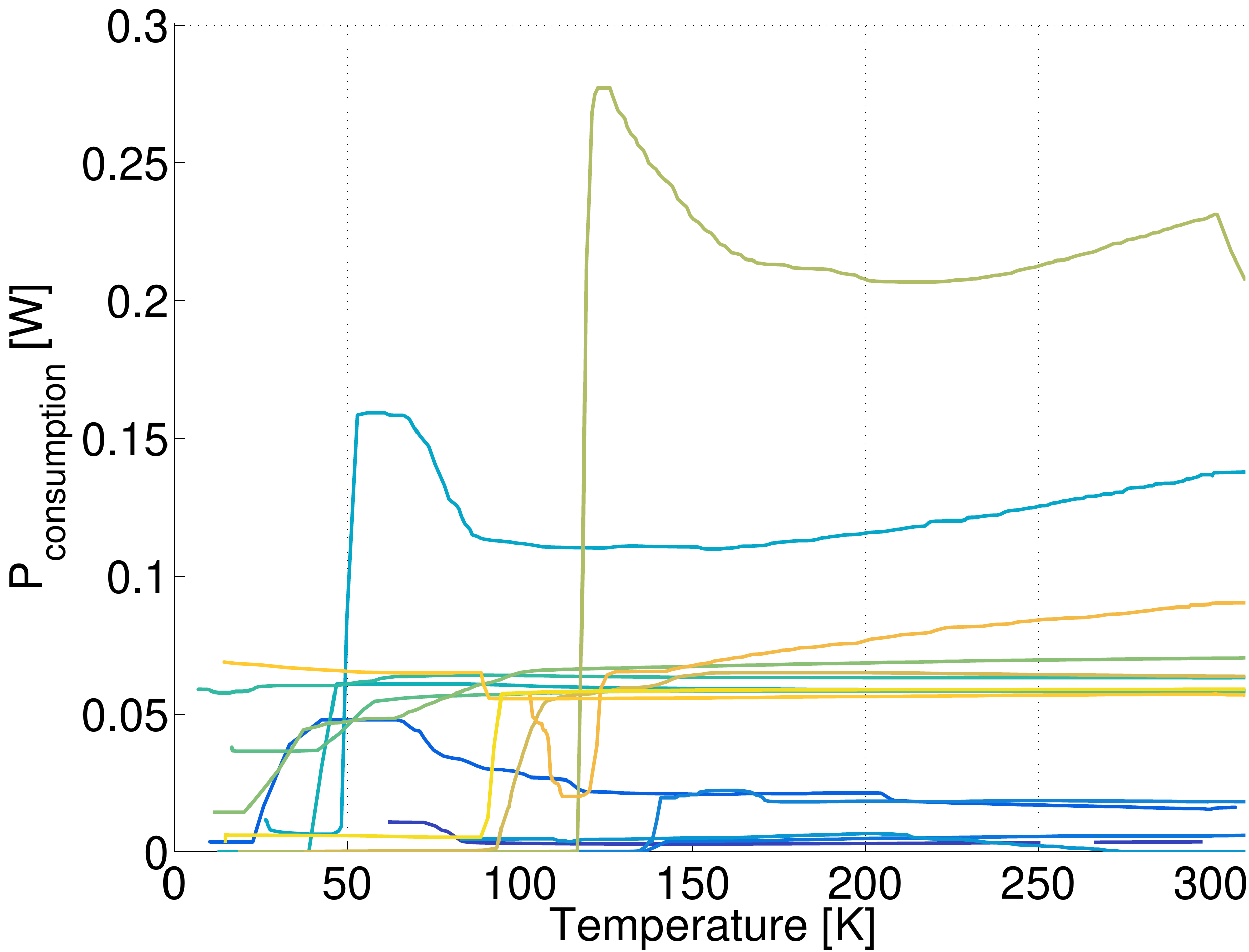}
		}
		\hspace{1pt}
		\includegraphics[width=.15\textwidth]{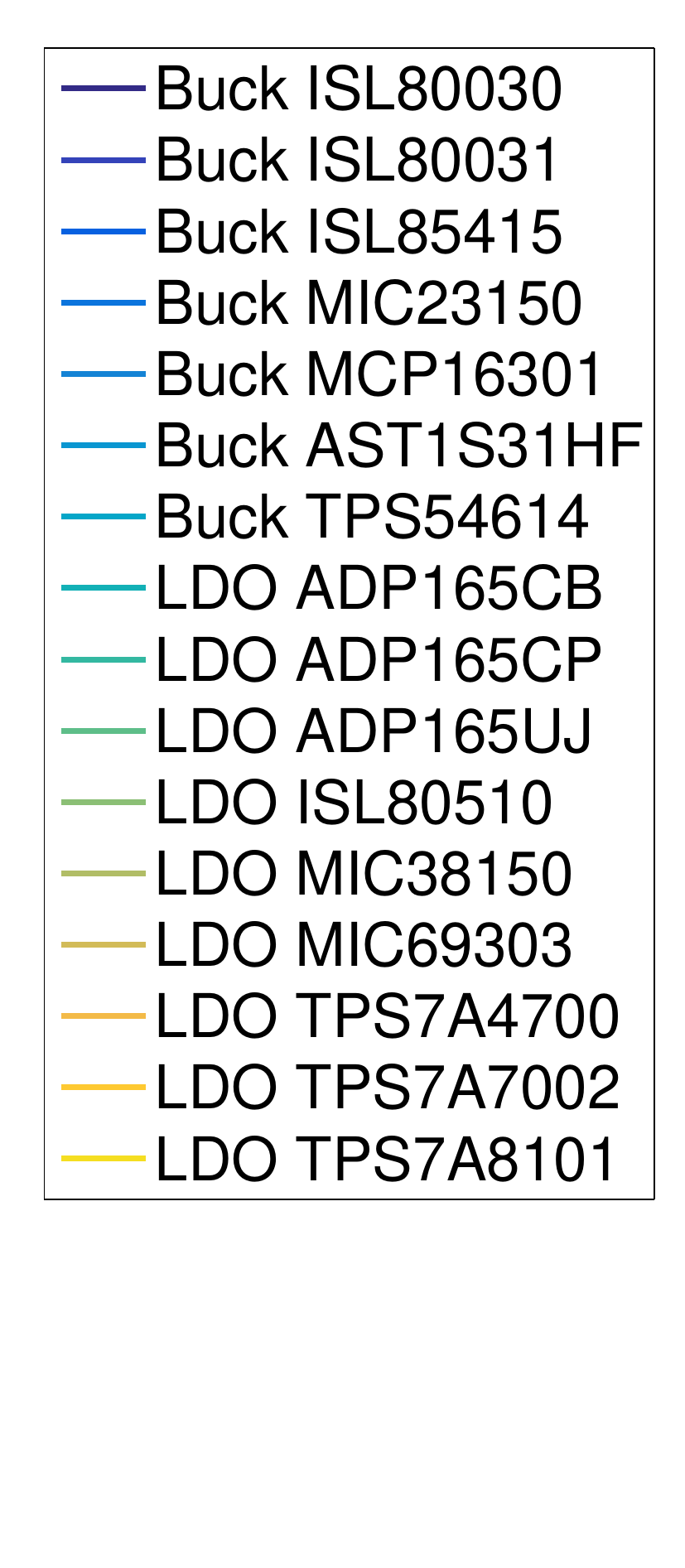}
		\hfill\null
		\caption{Output voltage and consumed power of several DC-DC converters / voltage regulators over temperature. All tested regulators fail operation at temperatures roughly below 90~K when assuming a 2.5\% margin on the voltage at 300~K. More complete characterization can be found in \aref{sec:reg_details}.}
		\label{fig:regulators}
	\end{figure*}

\noindent Voltage regulators or DC-DC converters and also voltage references form the core of any power distribution network. Especially for FPGA systems with significant switching currents, the voltage regulation has to be done as close as possible to the FPGA in order to reduce latency and thus increase stability and reduce IR drop. As the wire length into any cryogenic system is generally several meters long, voltage regulation at cryogenic temperatures can significantly improve system stability and performance. 

Previous studies have shown operating voltage regulators, however the temperature range was generally limited to 77~K. Simple shunt regulators based on Zener diodes have been shown to operate down to 100K with relatively stable performance \cite{Basit2013}. Also low drop-out (LDO) voltage regulators have been shown to operate down to 77~K \cite{Patterson2006}. 

Several voltage regulators have been tested in the temperature range from 4~K to 300~K and the output voltage is plotted versus temperature in \autoref{fig:regulators}\subref{fig:regulators_a}. Measurements were done at the specified V$_\text{in}$ and the regulators were loaded with 100~$\Omega$. As can be seen, none of the regulators are stable in voltage over the complete temperature range. 
To concretize functionality, a 2.5\% margin was allowed on the output voltage with respect to room temperature. The corresponding output voltage plots for all regulators are presented in \aref{sec:reg_details}. The best performing regulator was found to be an LDO from Texas Instruments (TPS7A7002, \autoref{fig:detailedregulators_LDO_voltage}\subref{fig:regulators_voltage_o}), at least in terms of output voltage. The device is extremely stable down to 90~K, at which temperature it completely turns off. 

Besides a stable output, the power consumption is extremely important at cryogenic temperatures, due to limited power budgets in any cryogenic system. The power consumed in the same set of regulators is shown in \autoref{fig:regulators}\subref{fig:regulators_b}. It is well known that switching voltage supplies outperform other regulator types in terms of power consumption, as is also shown here. 

Of course the main trade-off is regulator stability versus noise (output ripple etc.) versus power consumption. In general, switching supplies introduce more noise, while consuming less power. LDOs introduce very limited amounts of noise, while due to the drop-out behaviour, the power consumption is higher.

Extensive characterization results of the various voltage regulators are reported in \aref{sec:reg_details}.

In this study, no regulators were found to operate reliably below 60~K, however when operating the FPGA system, it is preferable to operate the regulators at 4~K to be able to integrate them as close as possible to the FPGA. This can be achieved either in the form of a discrete regulator, built from components operating at cryogenic temperatures, or by an FPGA based solution. 

In the next section, we propose a technique for stabilizing the FPGA supply voltage from within the FPGA, (partially) eliminating the need for an external regulator. 

 \section{Internal regulation architecture}\label{sec:architecture}

\subsection{Measurement set-up}
\begin{figure}[bt]
		\centering
		\includegraphics[width=0.3\textwidth]{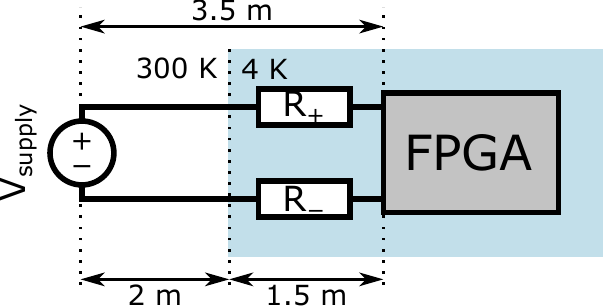}
		\caption{Schematic drawing of the FPGA operating at cryogenic temperatures while supplying its power through long wires coming from room temperature.  }
		\label{fig:FPGA_setup}
	\end{figure}

\noindent The problematic IR drop is mainly caused by powering an FPGA over extremely long wires (compared to regulating the voltage on the PCB itself). The wire length is needed, because the power has to be supplied from outside the cryogenic environment, as the power supplies don't work at these temperatures. The measurement set-up is schematically drawn in \autoref{fig:FPGA_setup}. As can be seen, there are roughly 2 meters of wires from supply towards the cryogenic set-up and roughly 1.5~m within the cryogenic set-up. While different cable types have to be used, low ohmic wires do not represent a problem at room temperature, but cause significant heat injection into the cryogenic environment. The main resistive path is therefore the cryogenic cable, which is estimated to be roughly 0.2~$\Omega$ for R$_+$ and 0.1~$\Omega$ for R$_-$. Besides the problems arising from these resistances, as will be discussed in the next section, latency is the second issue. While the power supplies will have finite reaction time, longer wires will inherently make the reaction time longer and thus the effect of IR drop worse, especially at the time of switching from low to high power consumption. 

\subsection{Operating principle}
\noindent To regulate the voltage from within the FPGA, the power consumed by the FPGA has to be made constant, so as to keep internal delays also constant. To achieve constant power consumption, an architecture was developed to monitor the FPGA's internal delays and act upon detected variations. 

	\begin{figure}[b]
		\centering
		\includegraphics[width=0.45\textwidth]{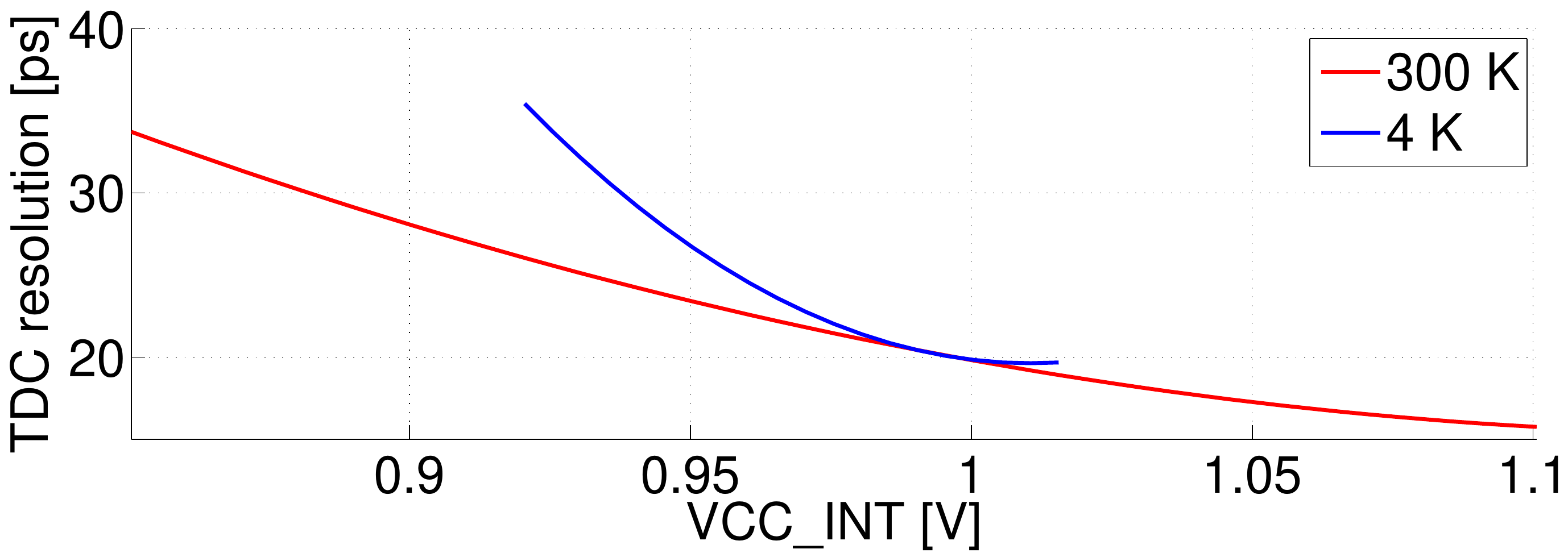}
		\caption{TDC resolution or carry delay versus the internal FPGA supply power (VCC\_INT) at both 300~K and 4~K.}
		\label{fig:TDC_resolution}
	\end{figure}

\begin{figure}[tb]
		\centering
		\null\hfill
		\subfloat[]{\label{fig:stab_principle_a}
			\includegraphics[width=.225\textwidth]{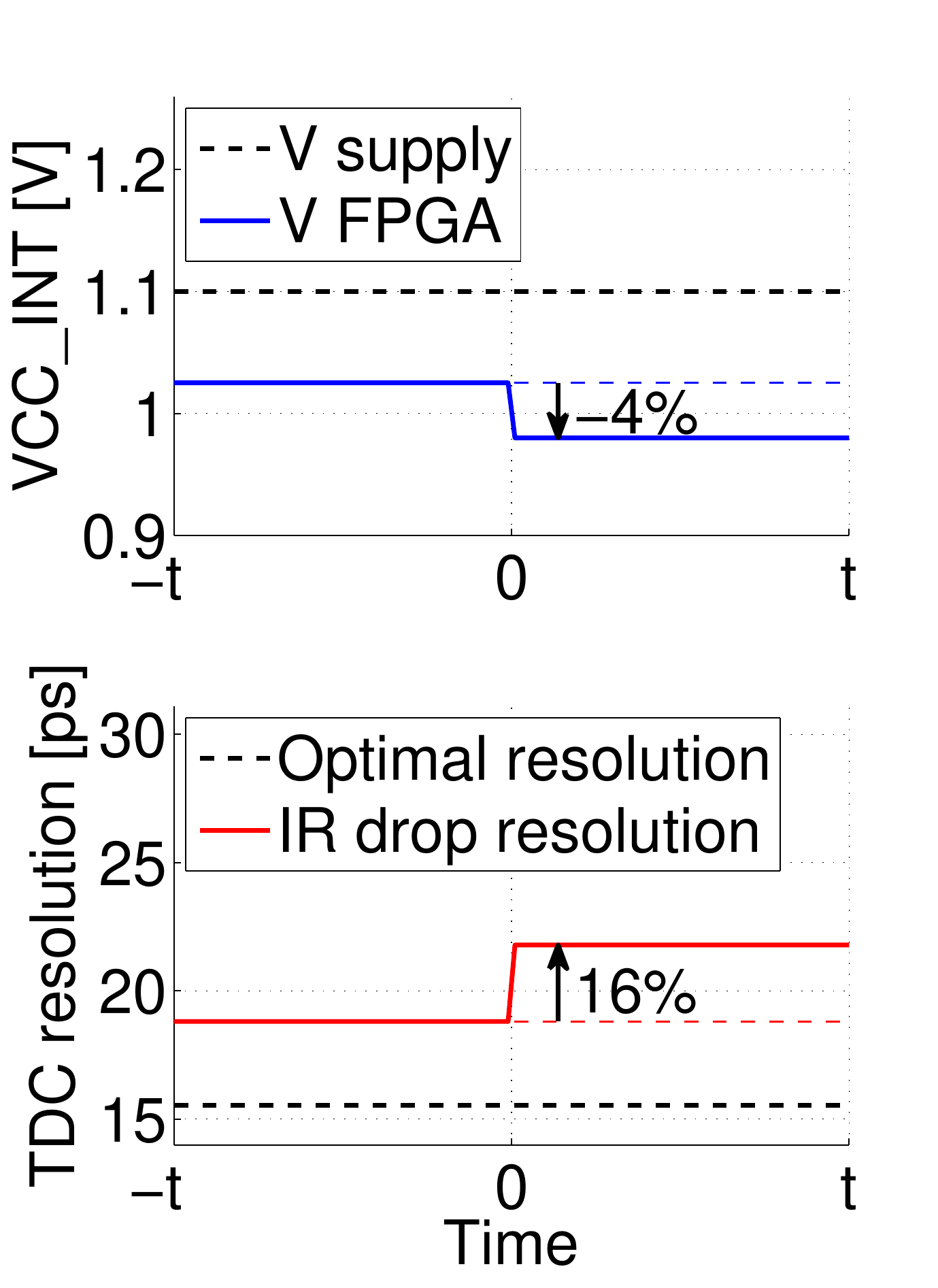}
		}
		\hfill
		\subfloat[]{\label{fig:stab_principle_b}
			\includegraphics[width=.225\textwidth]{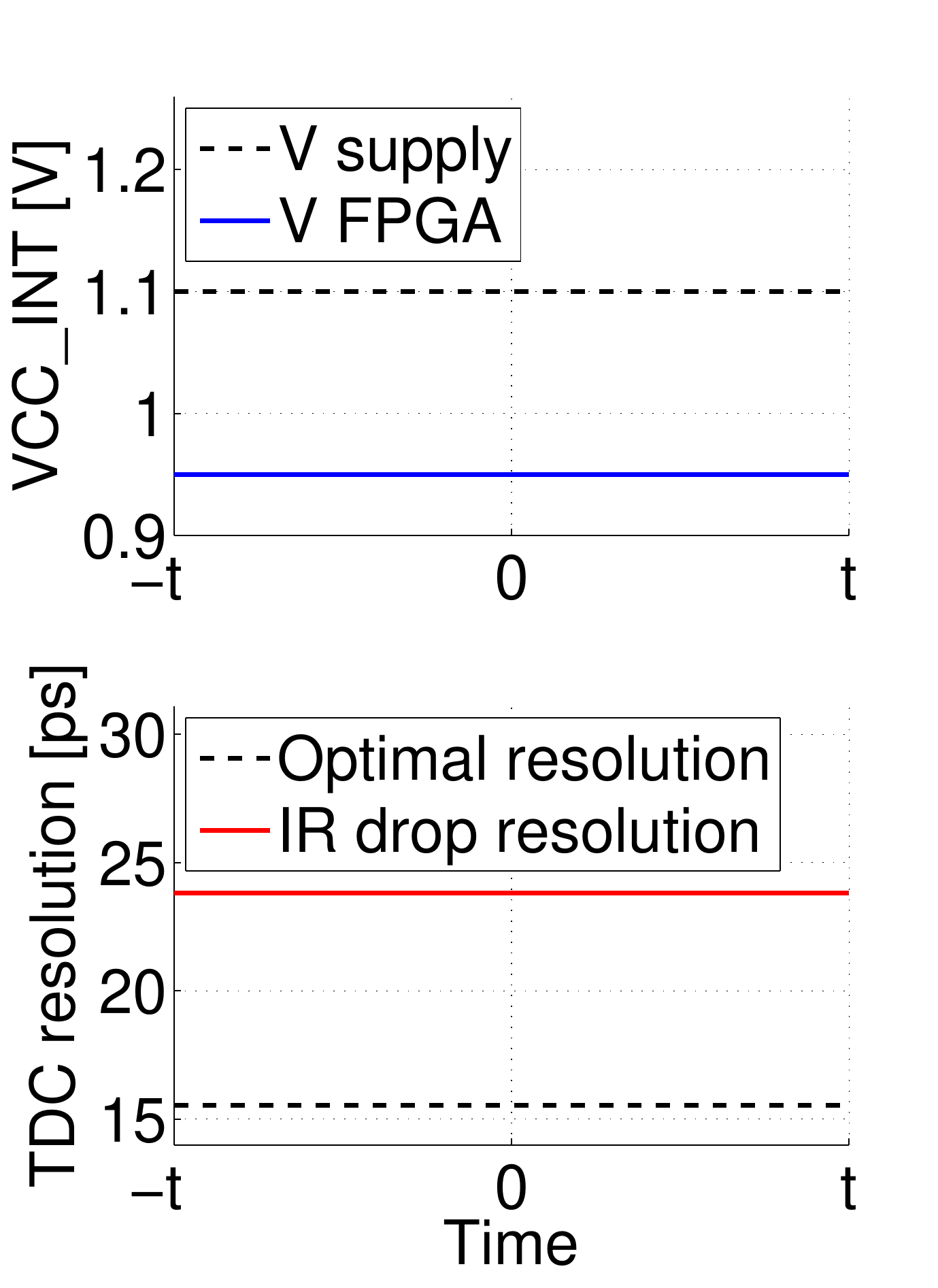}
		}
		\hfill\null
		\caption{Principal effects of the power stabilizer on the FPGA voltage and the resulting delay of the carry elements, i.e. the resolution of the TDC. The resistive loss in the cables is approximately 0.3~$\Omega$, while the power dissipated by the system is 0.25, 0.4 and 0.5~W, respectively in steady-state, after the trigger and with the stabilizer active. }
		\label{fig:stab_principle}
	\end{figure}

The operation and its results are schematically drawn in \autoref{fig:stab_principle}. In \subref{fig:stab_principle_a} the operation without regulation is shown. In this example, the system is triggered at a certain time to start with an operation. At that time, the power consumption is suddenly increased, in this example from 0.25 to 0.4~W. 
The resulting voltage on the FPGA is dropped, with 4\% when assuming a resistive drop in the cable of 0.3~$\Omega$ from 1.03 to 0.98~V. As a result, the propagation delay of the carry chain is increased by approximately 16\% or 4~ps. This change is significantly more than the 4\% drop in voltage thanks to their semi-exponential relationship as shown in \autoref{fig:TDC_resolution}. 

With the regulation turned on, this effect is effectively smoothed, as shown in \autoref{fig:stab_principle}\subref{fig:stab_principle_b}. The power consumption before the system is triggered is slightly higher than that in the non-regulated system, around 0.5~W, after the trigger, the power consumption remains stable. While this technique can achieve a more stable system, the main drawback is the increased power consumption. This is especially important for designs operating at cryogenic temperatures, for which the power budgets are limited. \\

\noindent These problems can be partially addressed with a supply capable of forcing and sensing the voltage through different terminals. In this way, the voltage is measured closer to the FPGA through high-ohmic contacts and the supply can therefore adjust its internal voltage to compensate for the IR drop in the cables. In practice, this technique is limited by the bandwidth of the feedback and the supply is not always capable of acting quickly and accurately enough, especially with a system switching current frequently and at a high rate.

\subsection{Implementation}
\begin{figure}[tb]
		\centering
		\subfloat[]{\label{fig:arch_a}
			\includegraphics[width=.49\textwidth]{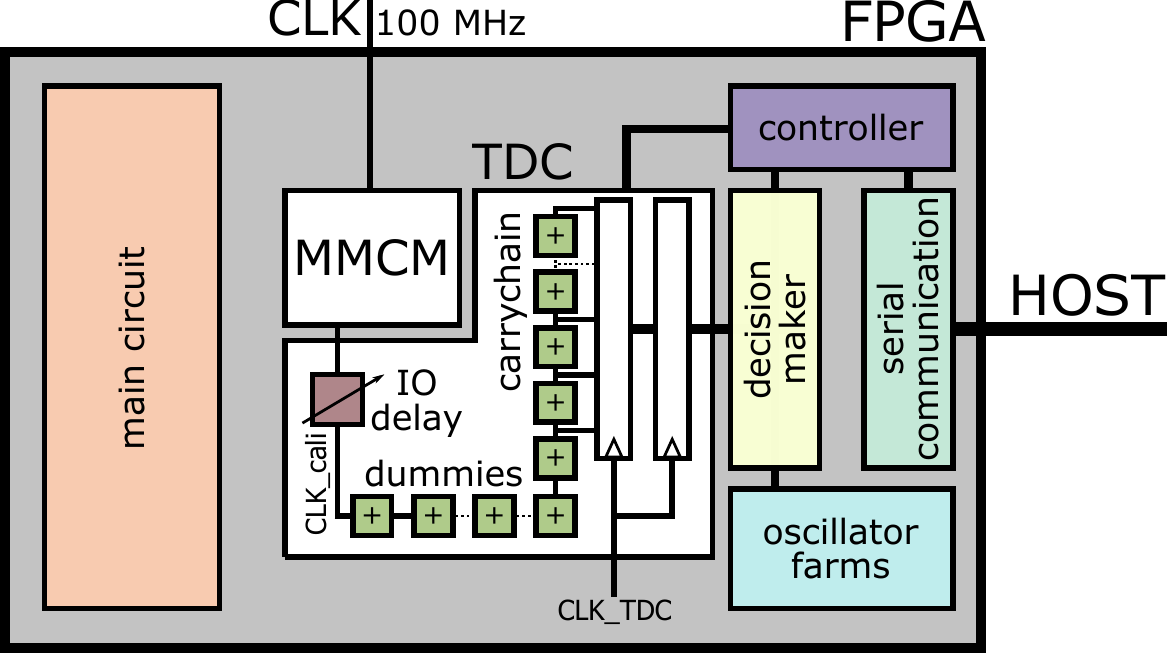}
		}
		\\
		\subfloat[]{\label{fig:arch_b}
			\includegraphics[width=.49\textwidth]{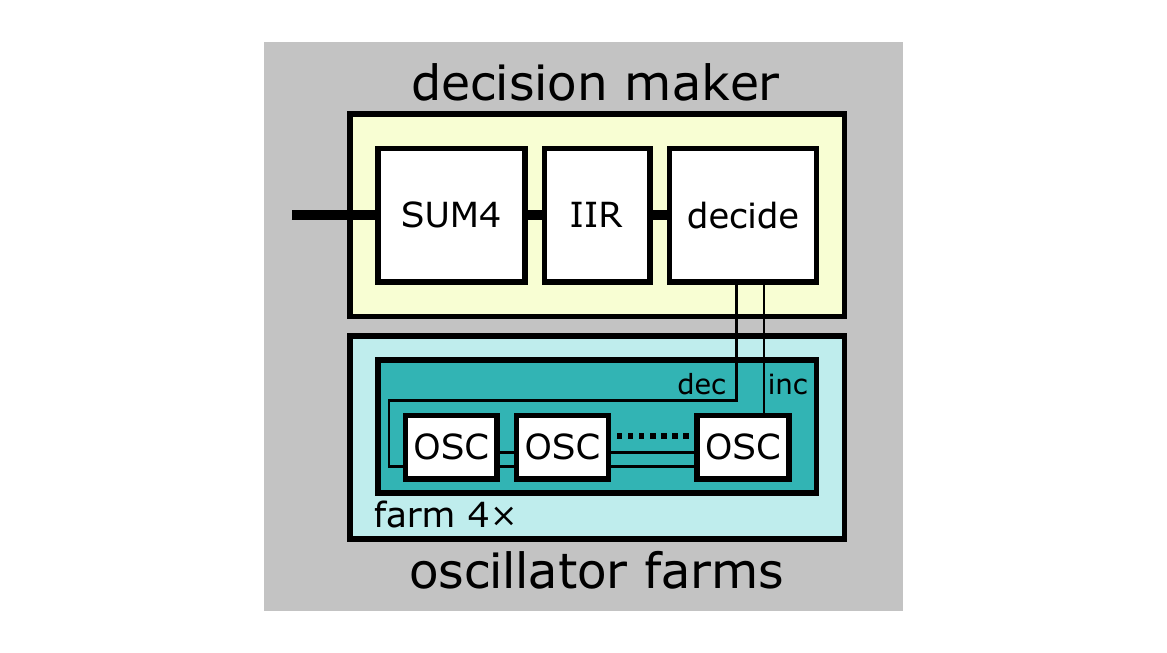}
		}
		\caption{FPGA internal stabilizer architecture. \protect\subref{fig:arch_a} The oscillator stabilizer control system is divided in the following parts: a TDC with IO delay and a dummy carry chain in order to calibrate the clock delay into the delay line and to generate extra impact of voltage change on the TDC outcome. The outcome is routed towards a decision maker which detects if the voltage is increased or decreased. Control signals are finally sent to the oscillator farms. \protect\subref{fig:arch_b} Detail of decision maker and farm. The decision maker first integrates and filters the TDC data, leading to increment (inc) or decrement (dec) signals that are sent to four oscillator farms, each consisting of 128~oscillators.  }
		\label{fig:arch}
	\end{figure}

\noindent To achieve the stabilization of the supply voltage, a small oscillator farm is employed together with a regulator consisting of a time-to-digital converter, an IO delay block and some control logic. The complete circuit is shown in \autoref{fig:arch}. 

The internal clock signal, coming from an MMCM (mixed-mode clock manager) digital clock manager, is used as stable reference. It is routed through an IO delay block and a carry chain towards a small TDC. This TDC constantly measures the timing of the clock edge. At system start-up, the IO delay is used to shift the clock edge to the middle of the TDC range. As a result, the output of the TDC should be constantly indicating half of its range, as the position of the clock edge. 

However, as a result of IR drop and consequent shifts in logic delay, this is not the case. As the IR drop increases, the delay of the elements increases and the edge will shift to a higher TDC output. On the other hand, with lower power consumption, and thus lower IR drop, the internal logic will be faster and the edge will shift to lower TDC values. As the TDC is one of the most precise elements in the FPGA, the changes can be significant even with slight variations in the power consumption, as discussed in the previous section. 

To correct for the changes in power consumption, oscillators in the farm are turned on or off depending on the direction of the shift. With lower measured TDC output, some oscillators are turned on. Higher TDC output will lead to a decrease in the amount of enabled oscillators. 

The TDC is operated with a 400~MHz clock, while decisions are made at 6.25~MHz. To average out noise and jitter in the TDC, the result of 64 consecutive measurements is taken as a measure for the required increase or decrease in the number of enabled oscillators. First, four results are added to cross from the 400~MHz to the 100~MHz clock domain. The remainder is processed using an infinite impulse response (IIR) filter before taking a decision to increment or decrement the number of running oscillators. 

The oscillators are each implemented with 6~LUTs configured as buffers and one NAND gate as the first stage. The number of oscillators required greatly depends on the remainder of the implemented design and the power consumed in the core circuitry. Firstly, the minimum and maximum power consumption of the core circuitry has to be roughly known. Afterwards, the number of implemented oscillators should be capable of bridging the gap between minimum and maximum power consumption with some margin. 

In our design, the core circuit is our analog-to-digital converter \cite{Homulle2016TCAS}, which dissipates in idle mode 250~mW and at maximum conversion rate about 400~mW. To bridge this gap, four farms were implemented, each containing 128~oscillators. Therefore, the total number of active oscillators out of the 512~oscillators can be incremented or decremented  four at a time. 
 \section{Results}\label{sec:results}

\subsection{Voltage regulation}
\begin{figure}[t]
		\centering
		\null\hfill
		\subfloat[]{\label{fig:stab_ICC_a}
			\includegraphics[width=.225\textwidth]{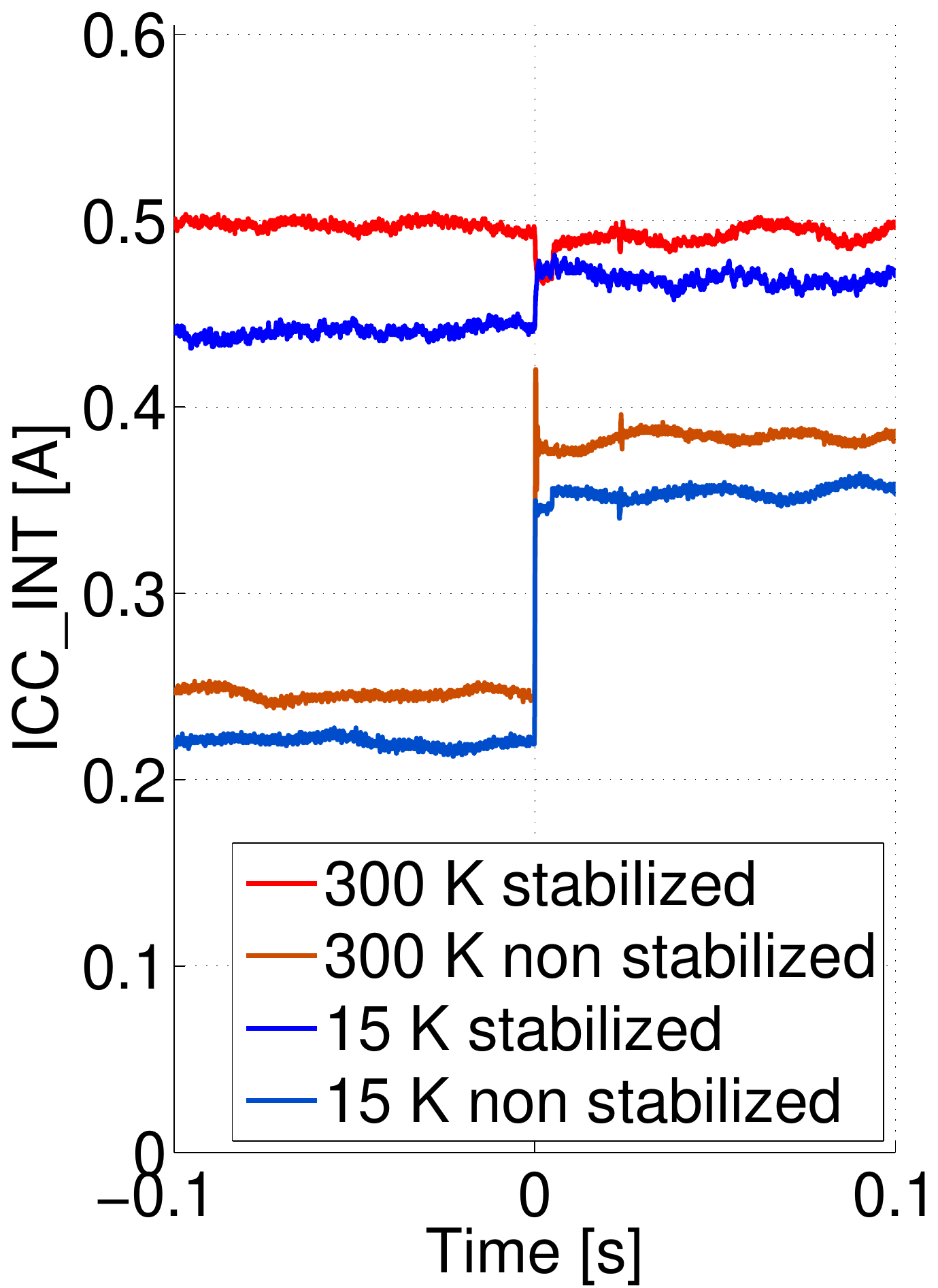}
		}
		\hfill
		\subfloat[]{\label{fig:stab_ICC_b}
			\includegraphics[width=.225\textwidth]{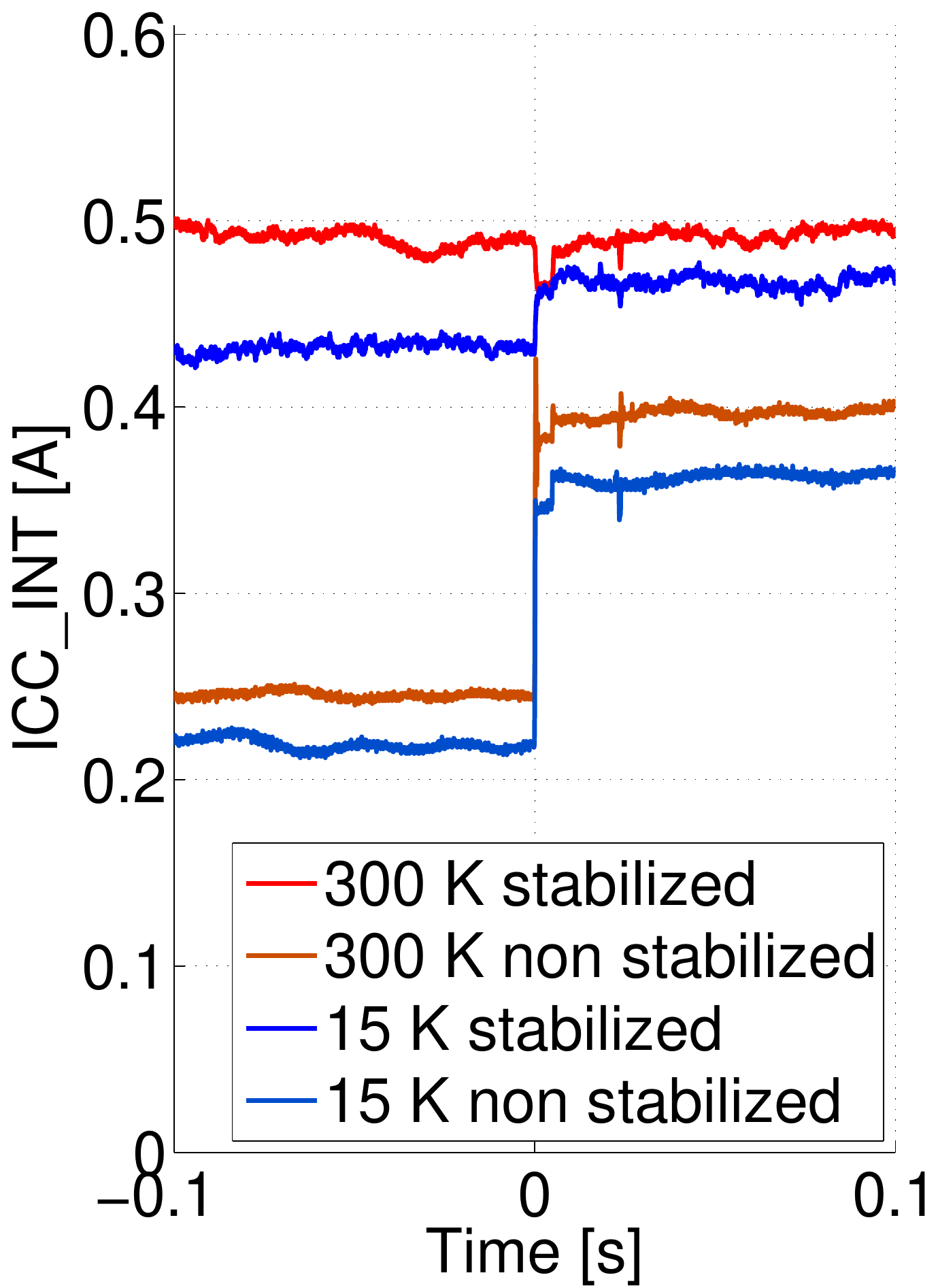}
		}
		\hfill\null
		\caption{Measured FPGA current with stabilizer turned on and off at 300~K and 15~K. At time $t=0$, the ADC starts a measurement with an input sinusoid of 2~MHz in \protect\subref{fig:stab_ICC_a} and 40~MHz in \protect\subref{fig:stab_ICC_a}.  }
		\label{fig:stab_ICC}
	\end{figure}

\noindent The implemented system consists of our 1.2~GSa/s ADC \cite{Homulle2016TCAS} combined with the voltage regulation circuit as discussed in the last section. The ADC is triggered at time $t=0$ to start converting the analog input and store this data in the internal FPGA memory. At time zero, the power consumption is significantly increased for the non-stabilized system, as shown in \autoref{fig:stab_ICC}. In \subref{fig:stab_ICC_a}, the ADC starts converting a sinusoidal signal with a 2~MHz frequency, while in \subref{fig:stab_ICC_b}, the conversion is done with a 40~MHz input signal. Although the difference in power consumption between these two cases is small, about 2.5\%, it will still affect the signal-to-noise ratio (SNR). The main problem for our ADC though is the difference between calibration and final conversion. While the calibration is done with signals in the low kHz frequency range, the difference in the power consumption is more significant. In principle, the final conversion system stability is not the same as the system stability at the time of the calibration, leading to significant differences in signal shapes that cannot be easily accounted for. 

Therefore stabilizing the power consumption in all these cases, can improve the final ADC conversion results. Our stabilization technique indeed stabilizes the current as shown in both figures, however the current is not as flat at 15~K as it is at 300~K. This is mainly caused by increased jitter in some components, inaccurate initial biasing of the power consumption and the larger dependency of carry delay versus supply voltage (\autoref{fig:TDC_resolution}). This makes the control loop not as accurate at 15~K as it is at 300~K.

\subsection{Calibration and conversion}
\noindent Although we can indeed see a significantly more stable current over time, the question remains whether that improves the performance of our ADC. To study the performance of the ADC, tests were done in the four operating conditions, with and without stabilization at 300~K and 15~K. 

	\begin{figure}[tb]
		\centering
		\null\hfill
		\subfloat[Non stabilized @ 300~K.]{\label{fig:sine_time_a}
			\includegraphics[width=.225\textwidth]{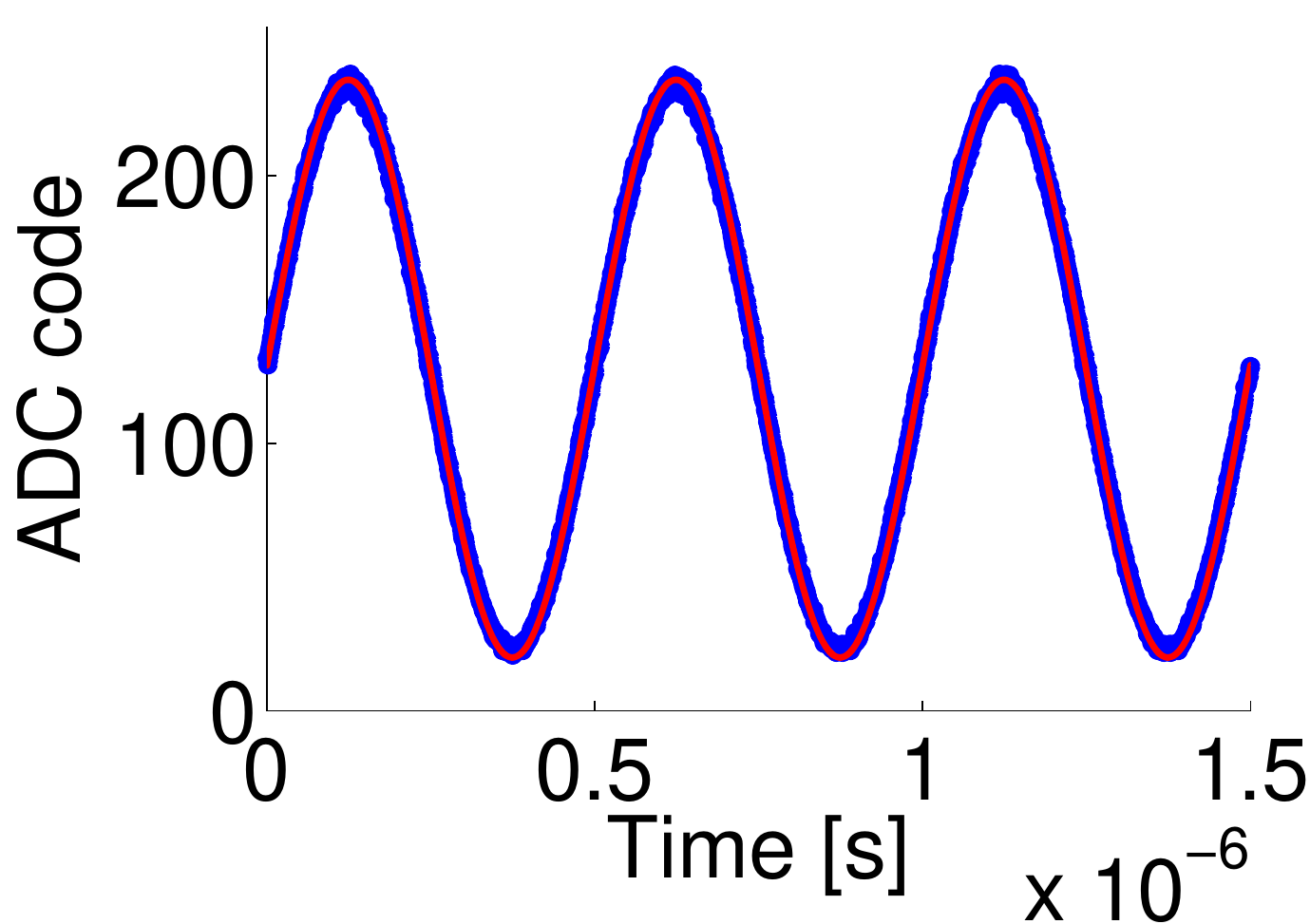}
		}
		\hfill
		\subfloat[Stabilized @ 300~K.]{\label{fig:sine_time_b}
			\includegraphics[width=.225\textwidth]{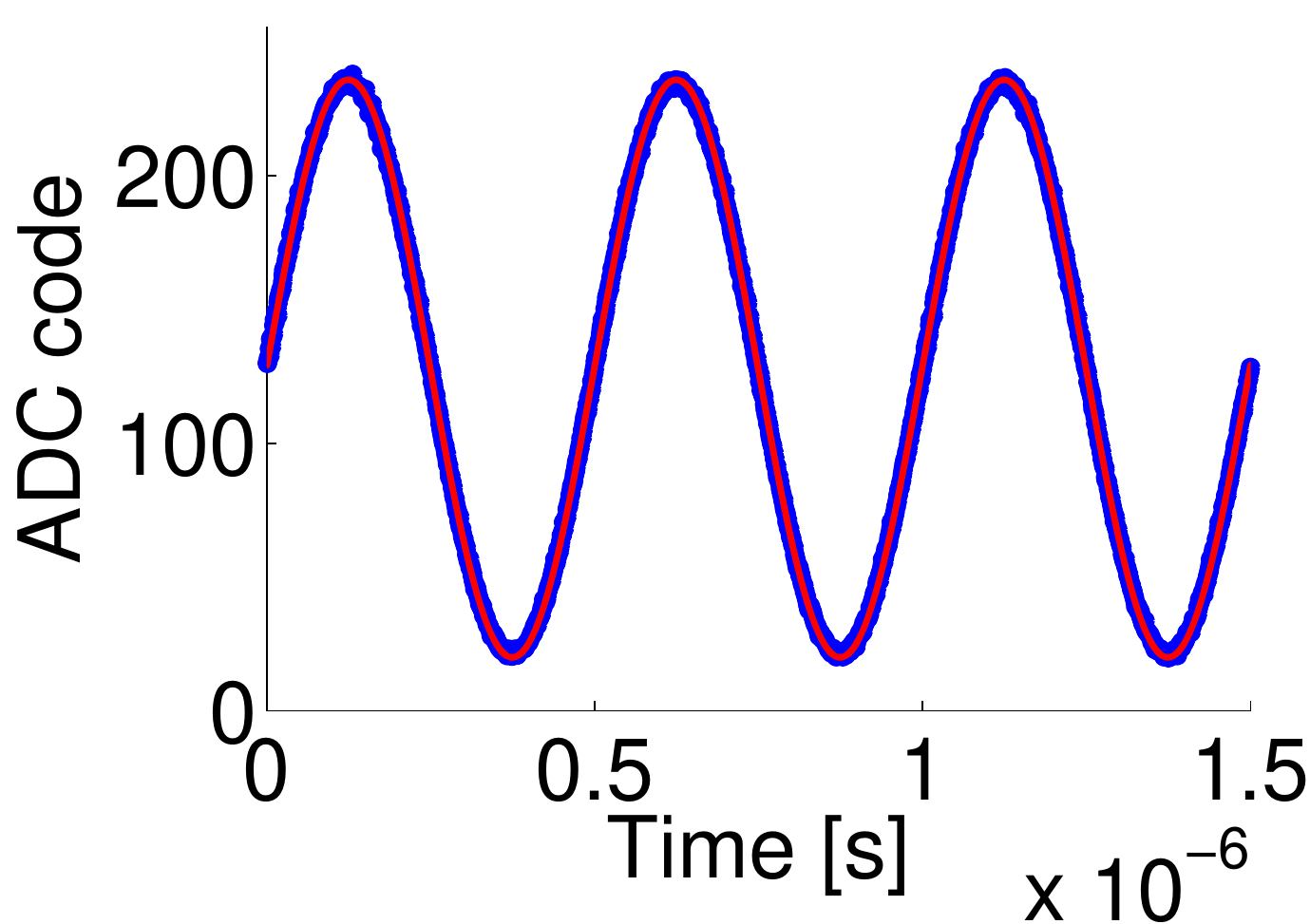}
		}
		\hfill\null
		\\
		\null\hfill
		\subfloat[Non stabilized @ 15~K.]{\label{fig:sine_time_c}
			\includegraphics[width=.225\textwidth]{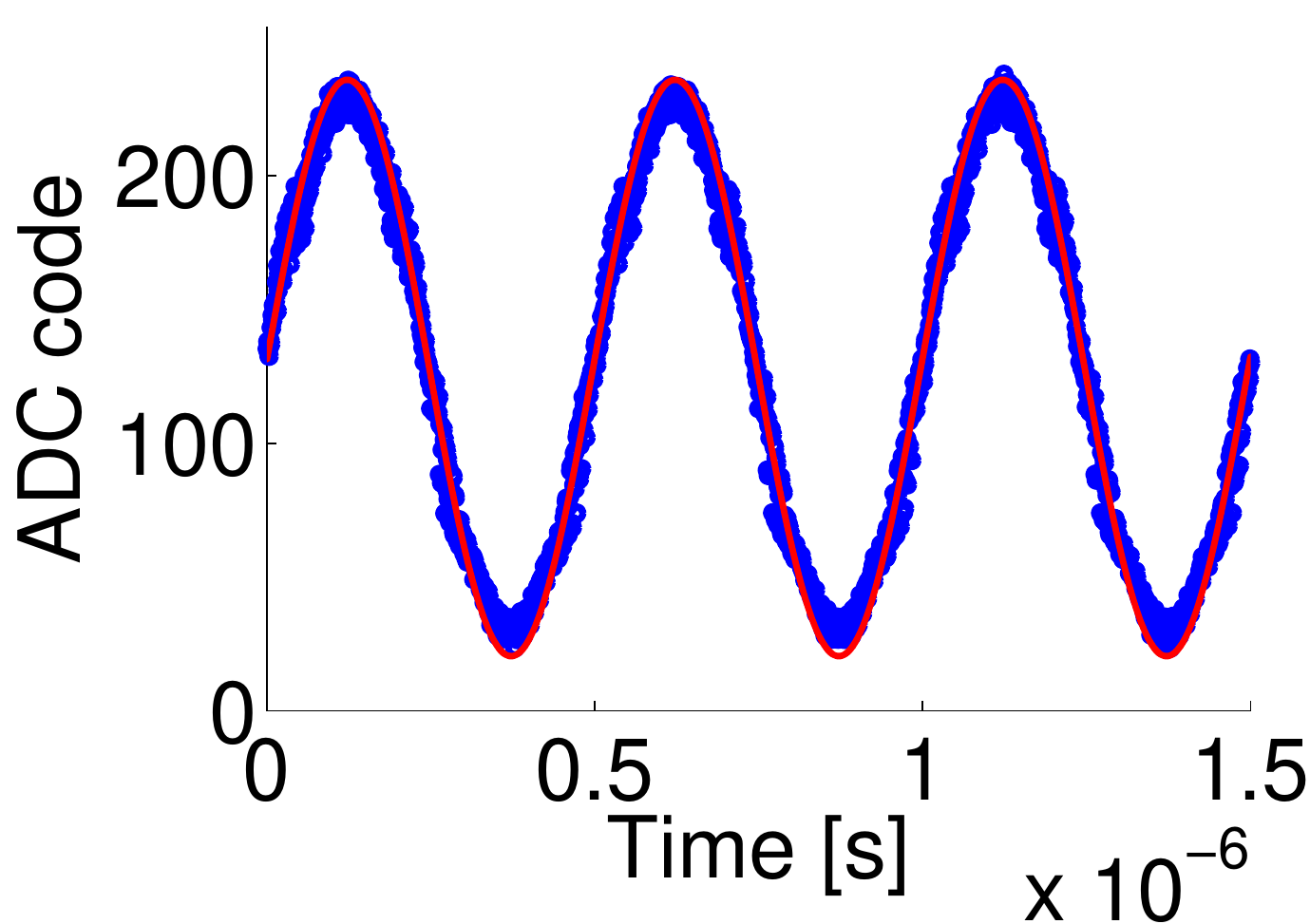}
		}
		\hfill
		\subfloat[Stabilized @ 15~K.]{\label{fig:sine_time_d}
			\includegraphics[width=.225\textwidth]{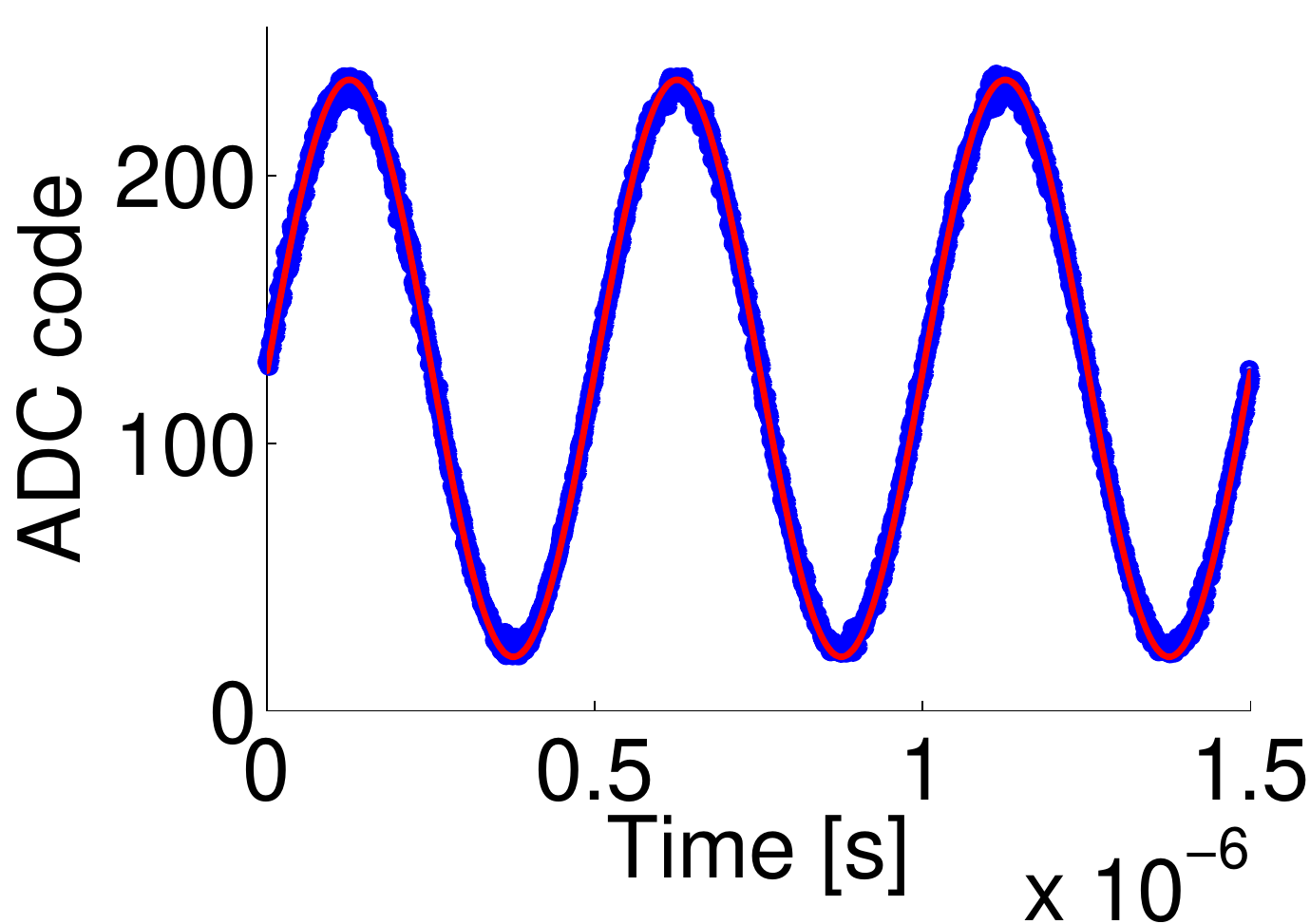}
		}
		\hfill\null
		\caption{Three periods of a 2~MHz sinusoid digitized at 1.2~GSa/s at 300~K and 15~K with the stabilized and non stabilized system. The input sine wave spans 90\% of the ADC input range to prevent the system from clipping etc. The solid line is a best fitted sinusoid for reference.}
		\label{fig:sine_time}
	\end{figure}

First, the ADC is calibrated at both temperatures. Using those calibration sets, measurements were performed with an input sinusoidal signal of 2 and 40~MHz. The results for the 2~MHz conversion are shown in \autoref{fig:sine_time}. As the ADC is built-up from 12 individual sources, i.e. 6 TDC channels and all measuring rising and falling edges, the calibration is essential. In \autoref{fig:sine_time}, the resulting sine wave is shown after merging 12 sources and applying the calibration. The digitized sine waves can be seen to be quasi spur free when the stabilizer is used. However, especially at 15~K, the result is not as clean as at 300~K. This can be partially explained by the non-equalization of the current at 15~K, as shown before in \autoref{fig:stab_ICC}, but is also caused by the decreased decoupling capabilities of our capacitor network and increased jitter in some of the FPGA components. 	

\begin{figure}[tb]
		\centering
		\null\hfill
		\subfloat[Non stabilized @ 300~K.]{\label{fig:sine_fft_a}
			\includegraphics[width=.225\textwidth]{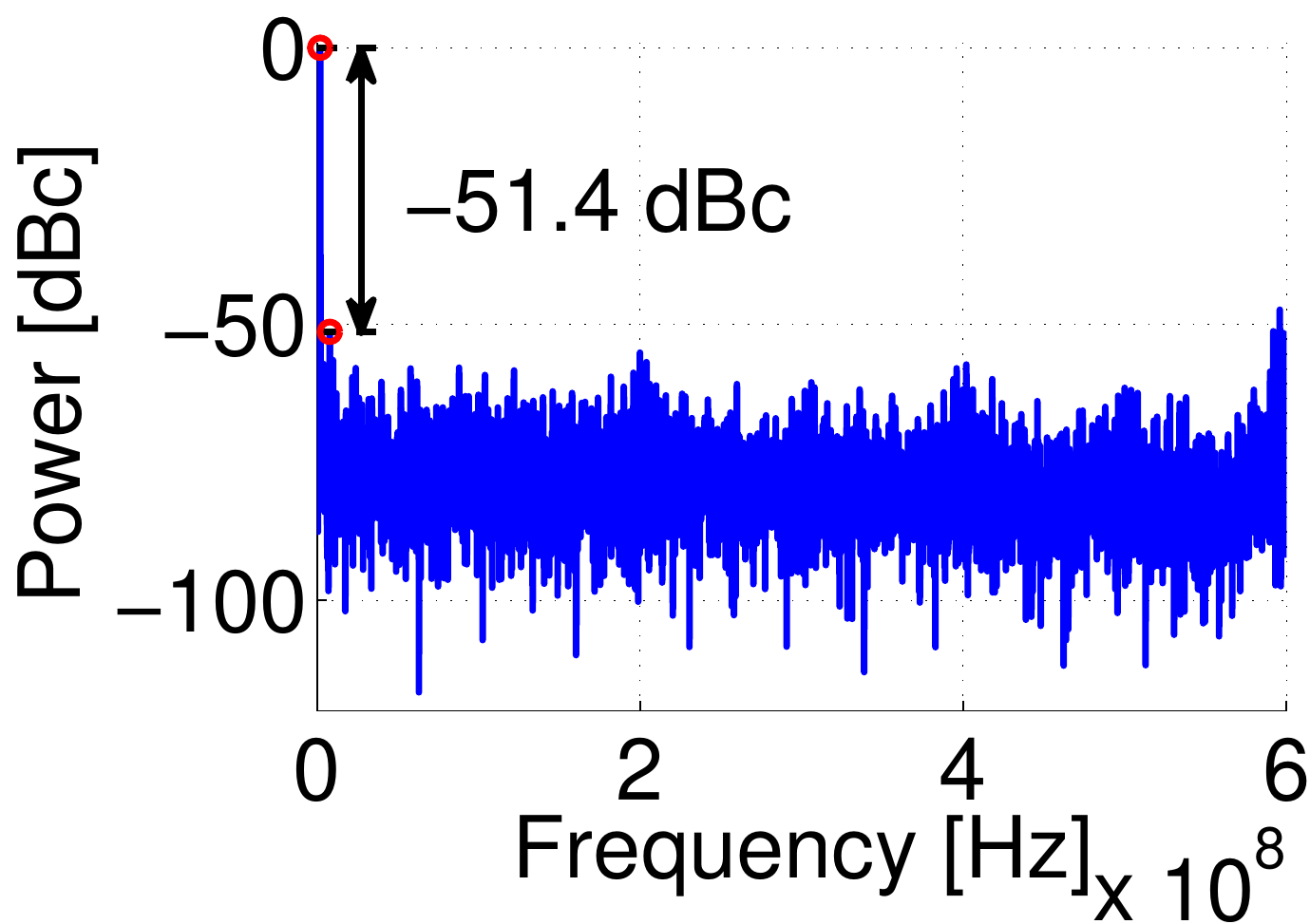}
		}
		\hfill
		\subfloat[Stabilized @ 300~K.]{\label{fig:sine_fft_b}
			\includegraphics[width=.225\textwidth]{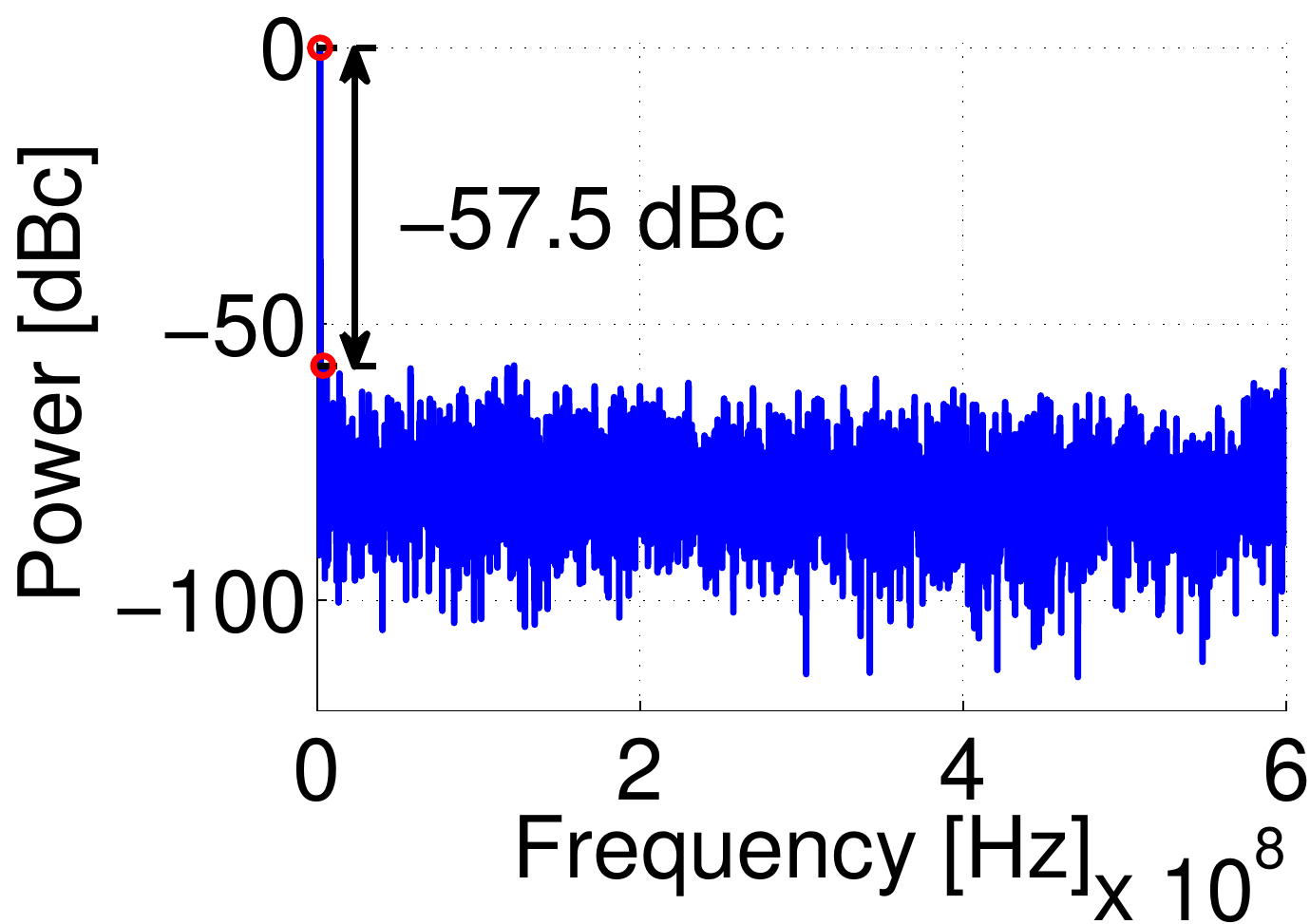}
		}
		\hfill\null
		\\
		\null\hfill
		\subfloat[Non stabilized @ 15~K.]{\label{fig:sine_fft_c}
			\includegraphics[width=.225\textwidth]{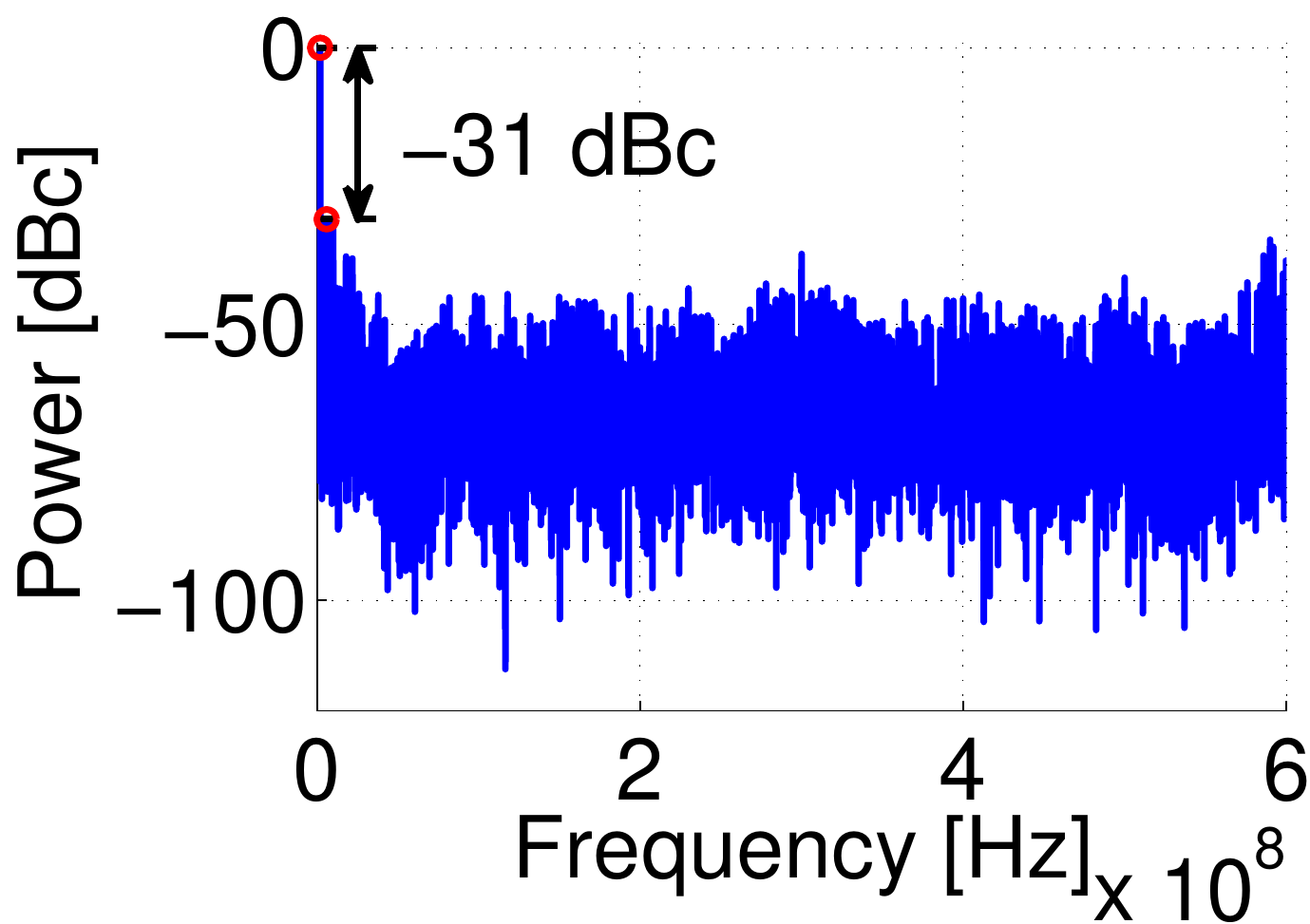}
		}
		\hfill
		\subfloat[Stabilized @ 15~K.]{\label{fig:sine_fft_d}
			\includegraphics[width=.225\textwidth]{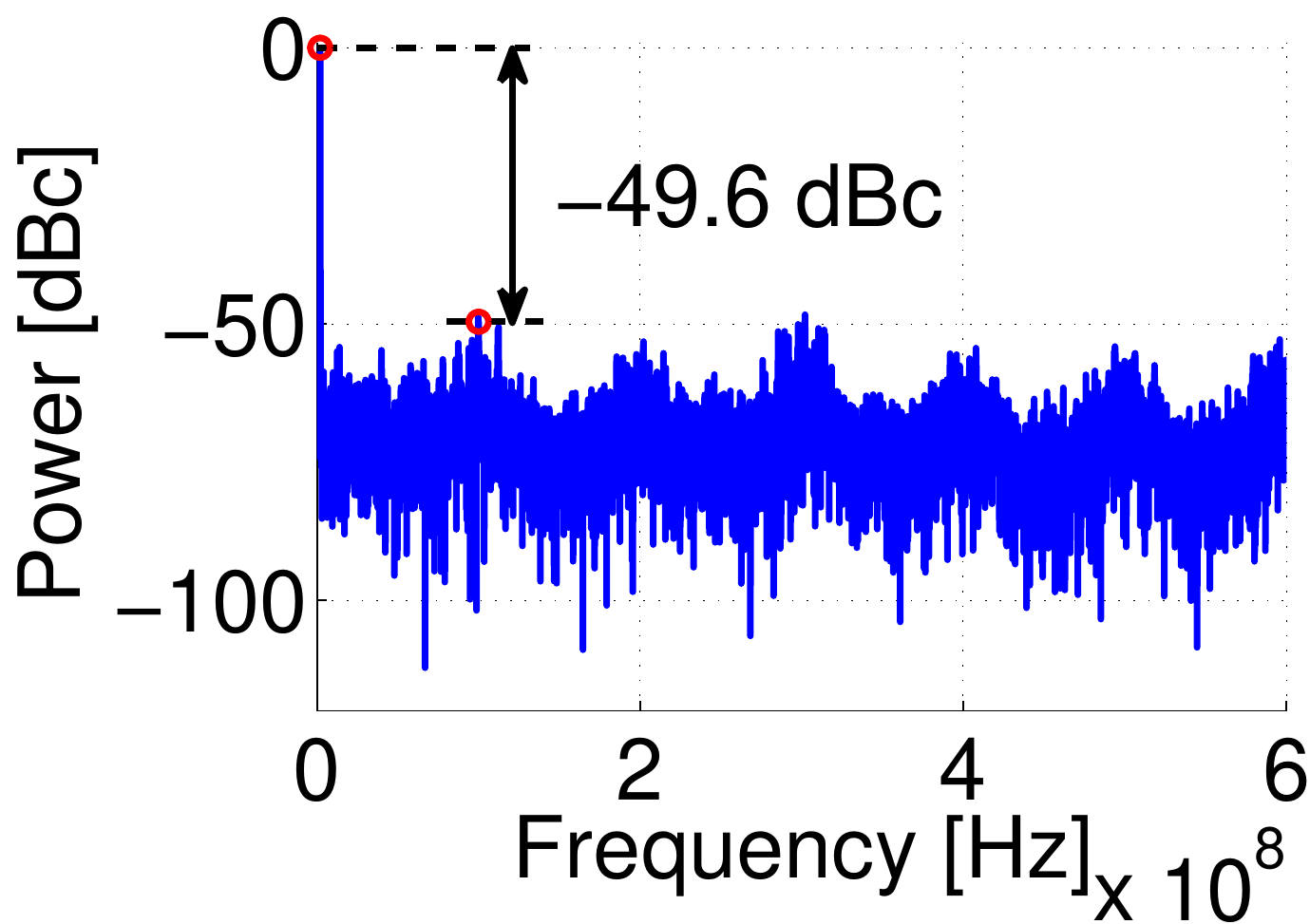}
		}
		\hfill\null
		\caption{Frequency domain representation of the sinusoids depicted in \protect\autoref{fig:sine_time}. Frequency spectra were obtained from 48,000 samples while using a Blackman-Harris window of length $2^{14}$ shifted over the samples. Furthermore, the spurious-free-dynamic-range SFDR is indicated. }
		\label{fig:sine_fft}
	\end{figure}

To quantify the improvement in performance, the sine waves are converted to the frequency domain, as shown in \autoref{fig:sine_fft}. Clearly, the stabilizer brings significant improvement, especially at cryogenic temperatures. 

In summary, the main specifications for our ADC are listed in \autoref{tab:adc_specs} for the use-cases described before. At 300~K, the stabilizer doesn't bring significant improvements. The ENOB is improved by 0.4~bit at low frequencies, but at higher frequencies, there is only a minor improvement of 0.1~bit (which is in the error-margin). At cryogenic temperatures however, the improvement is more significant. For low frequencies, there is a 1.5~bit better ENOB and at high frequencies, the ENOB still improves with 1.2~bit. These results clearly show that the ADC operation at cryogenic temperatures is challenging, especially when not considering power fluctuations and the significant effect of IR drop on the FPGA behaviour. There are still many unknown effects of electronics, materials and the corresponding behaviours while operating at temperatures so far out of the normal industrial temperature range, thus prompting more research in many aspects of cryogenic FPGA design. 

\begin{table}[b]
  \centering
  \scriptsize
  \caption{Summary of the obtained ADC specifications while converting a 2 and 40~MHz sine at 1.2~GSa/s with the stabilizer turned on or off. The stabilizer improves the performance, especially at cryogenic temperatures, with up to 1.5~bit in terms of ENOB or roughly 9~dB SNDR. }
	\begin{tabular}{l@{\hspace{-10pt}}r|rrrr|rrrr}
    \hline
    freq.$_\text{in}$ & [MHz] & \multicolumn{4}{c|}{2}         & \multicolumn{4}{c}{40} \\
	\cline{3-10}
    \multicolumn{2}{l|}{Temperature} & \multicolumn{2}{c}{300~K} & \multicolumn{2}{c|}{15~K} & \multicolumn{2}{c}{300~K} & \multicolumn{2}{c}{15~K} \\
	\cline{3-10}
    Stabilizer &       & \multicolumn{1}{c}{Off} & \multicolumn{1}{c}{On} & \multicolumn{1}{c}{Off} & \multicolumn{1}{c|}{On} & \multicolumn{1}{c}{Off} & \multicolumn{1}{c}{On} & \multicolumn{1}{c}{Off} & \multicolumn{1}{c}{On} \\
	\hline
    SNR   & [dB]  & 38.5  & 14.1  & 24.7  & 32.2  & 30.8  & 31.6  & 22.9  & 28.8 \\
    SNDR  & [dB]  & 36.8  & 38.9  & 21.3  & 30.3  & 28.6  & 29.3  & 19.4  & 26.5 \\
    SFDR  & [dBc] & -51.4 & -57.5 & -31.0   & -49.6 & -36.7 & -37.5 & -27.1 & -35.7 \\
    ENOB  & [bits] & 5.8   & 6.2   & 3.2   & 4.7   & 4.5   & 4.6   & 2.9   & 4.1 \\
    \hline
    \end{tabular}
  \label{tab:adc_specs}
\end{table}
 \section{Conclusion}\label{sec:conclusion}

\noindent We have demonstrated an approach to circumvent the ineffectiveness of voltage regulators at deep-cryogenic temperatures by regulating the FPGA power consumption internally. This has been done by constantly measuring the cell delay and acting upon change by decreasing or increasing the amount of running oscillators in a small oscillator farm. 

The stabilization of the power consumption has been demonstrated and has been quantified by showing significant improvement of the performance of our FPGA ADC system while stabilizing its power consumption and thus its supply voltage. The ENOB of our ADC was at most 1.5~bits higher while using the stabilizer. 

Besides the implementation of the FPGA power regulation, a study was conducted on the behaviour of various components needed in the power distribution network at cryogenic temperatures. The best behaving capacitors are those based on low dielectric constant materials, such as NP0/COG, while those with high dielectric constants tend to drop over 90\% in capacitance and increase the ESR with over 3000\%. Commercial voltage regulators were not found to be working at 4~K, but cease operation around 90~K. 

This study shows the complexity of operating at cryogenic temperatures, especially in terms of power supply over long wires, the unavailability of cryogenic regulators and the degraded performance of the decoupling network. Despite these complications, performance of the FPGA system can be significantly improved while regulating its supply voltage from within itself.  
\bibliographystyle{aipnum4-1}

\appendix
\clearpage
\section{Passives - Detailed results}\label{sec:cap_details}
\begin{figure*}[htbp]
		\centering
		\null\hfill
		\subfloat[Capacitance for 330~$\mu$F]{\label{fig:capacitors_a}
			\includegraphics[width=.45\textwidth]{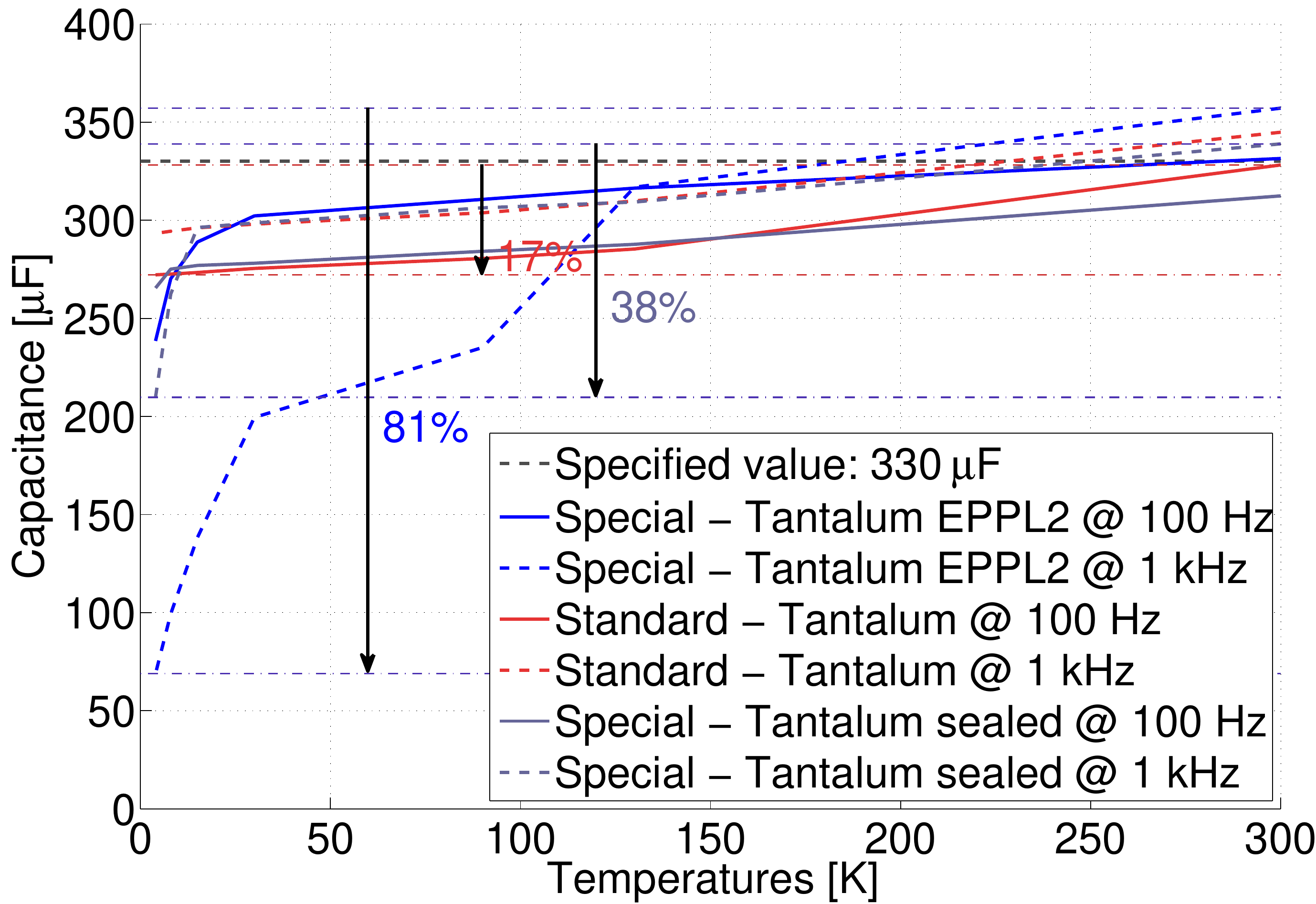}
		}
		\hfill
		\subfloat[ESR for 330~$\mu$F]{\label{fig:capacitors_b}
			\includegraphics[width=.45\textwidth]{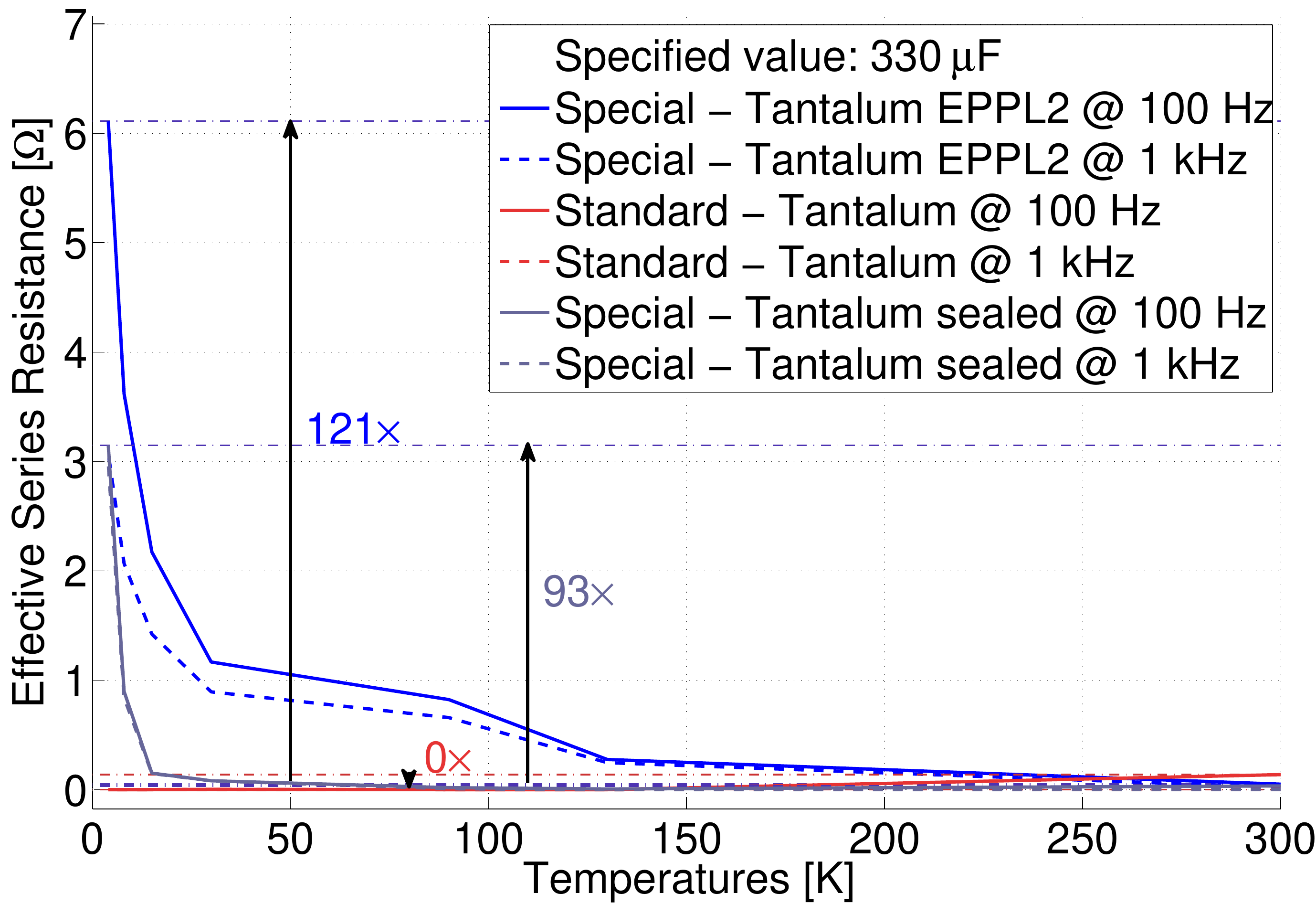}
		}
		\hfill\null
		\\
		\null\hfill
		\subfloat[Capacitance for 100~$\mu$F]{\label{fig:capacitors_c}
			\includegraphics[width=.45\textwidth]{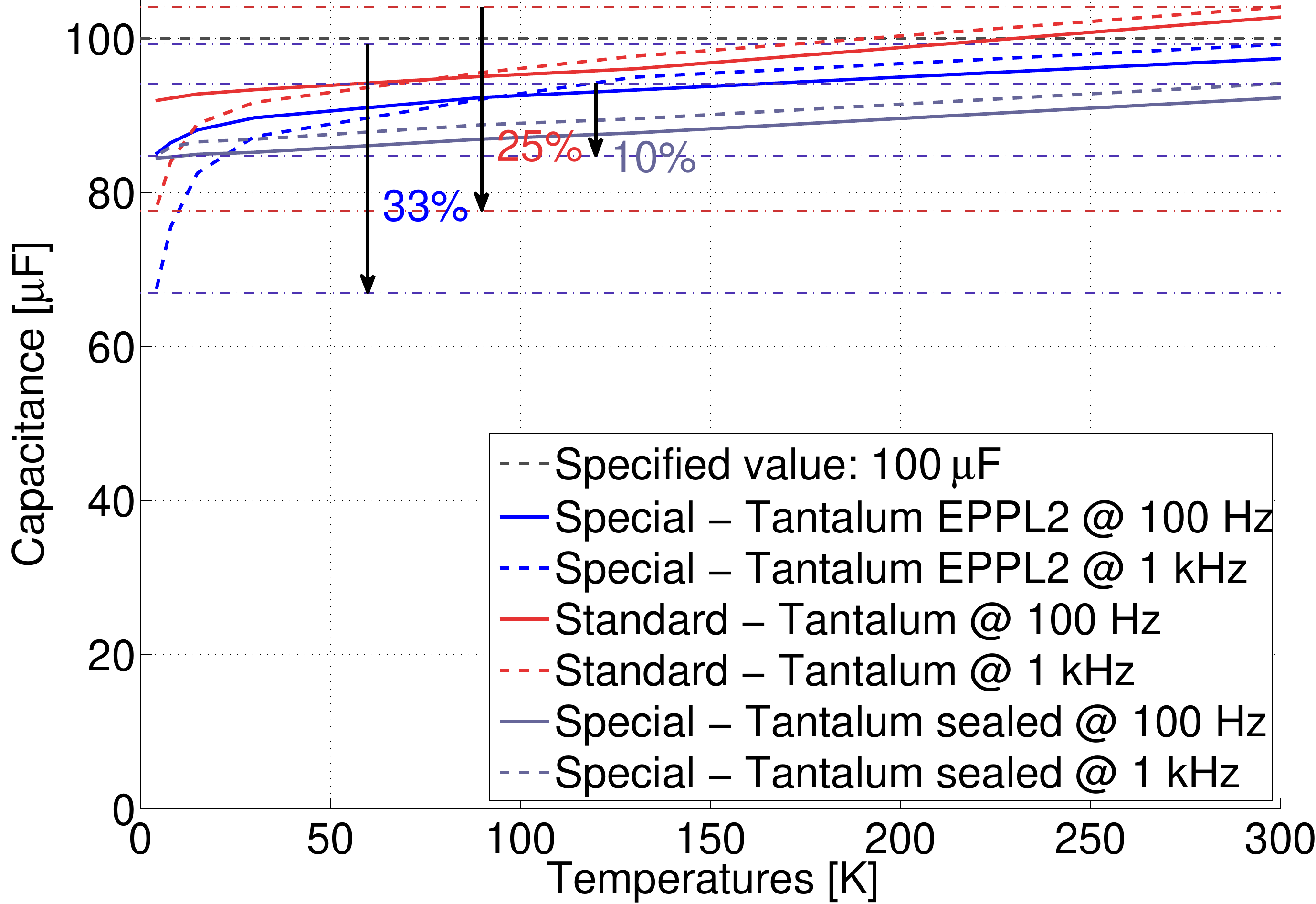}
		}
		\hfill
		\subfloat[ESR for 100~$\mu$F]{\label{fig:capacitors_d}
			\includegraphics[width=.45\textwidth]{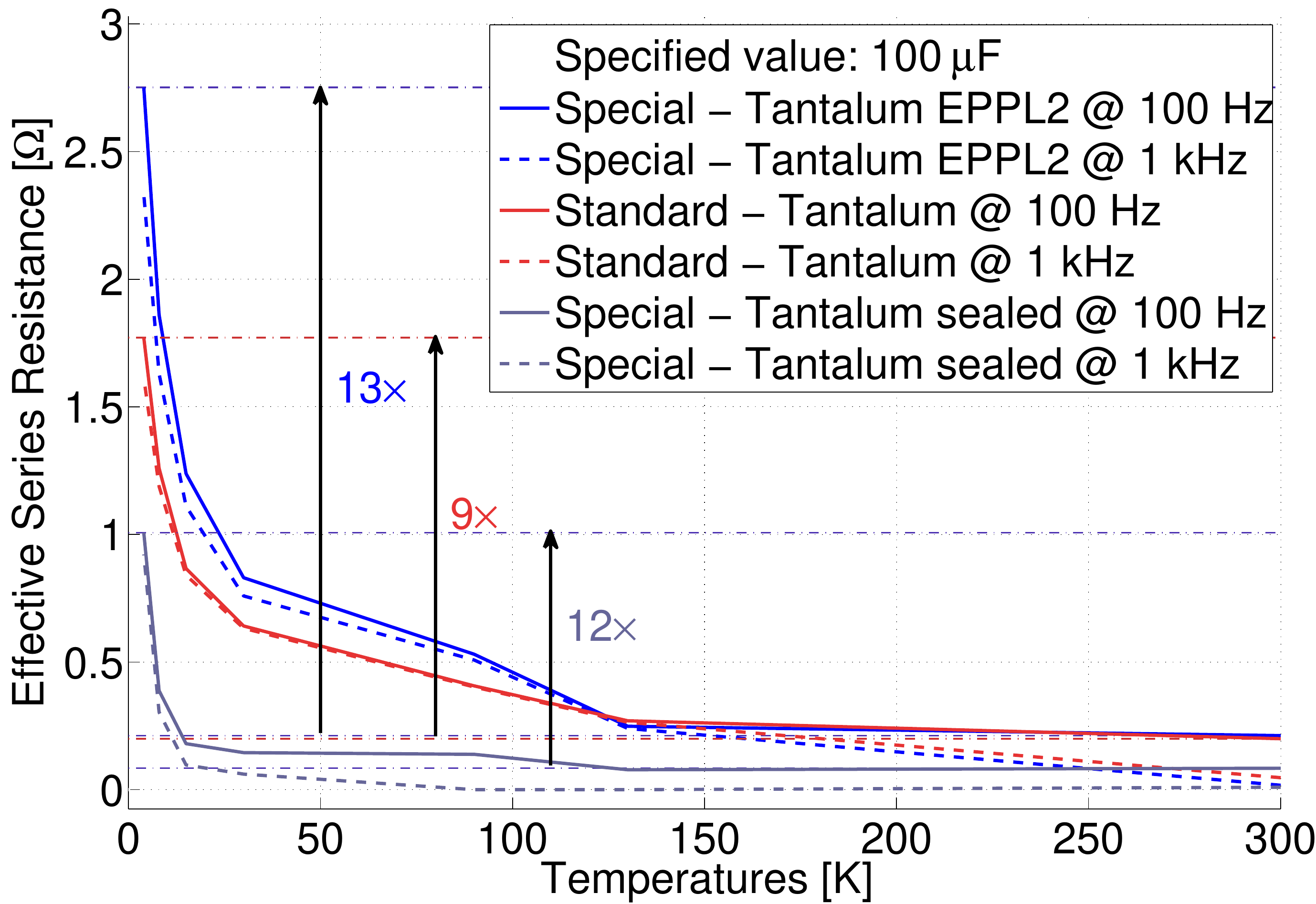}
		}
		\hfill\null
		\\
		\null\hfill
		\subfloat[Capacitance for 47~$\mu$F]{\label{fig:capacitors_e}
			\includegraphics[width=.45\textwidth]{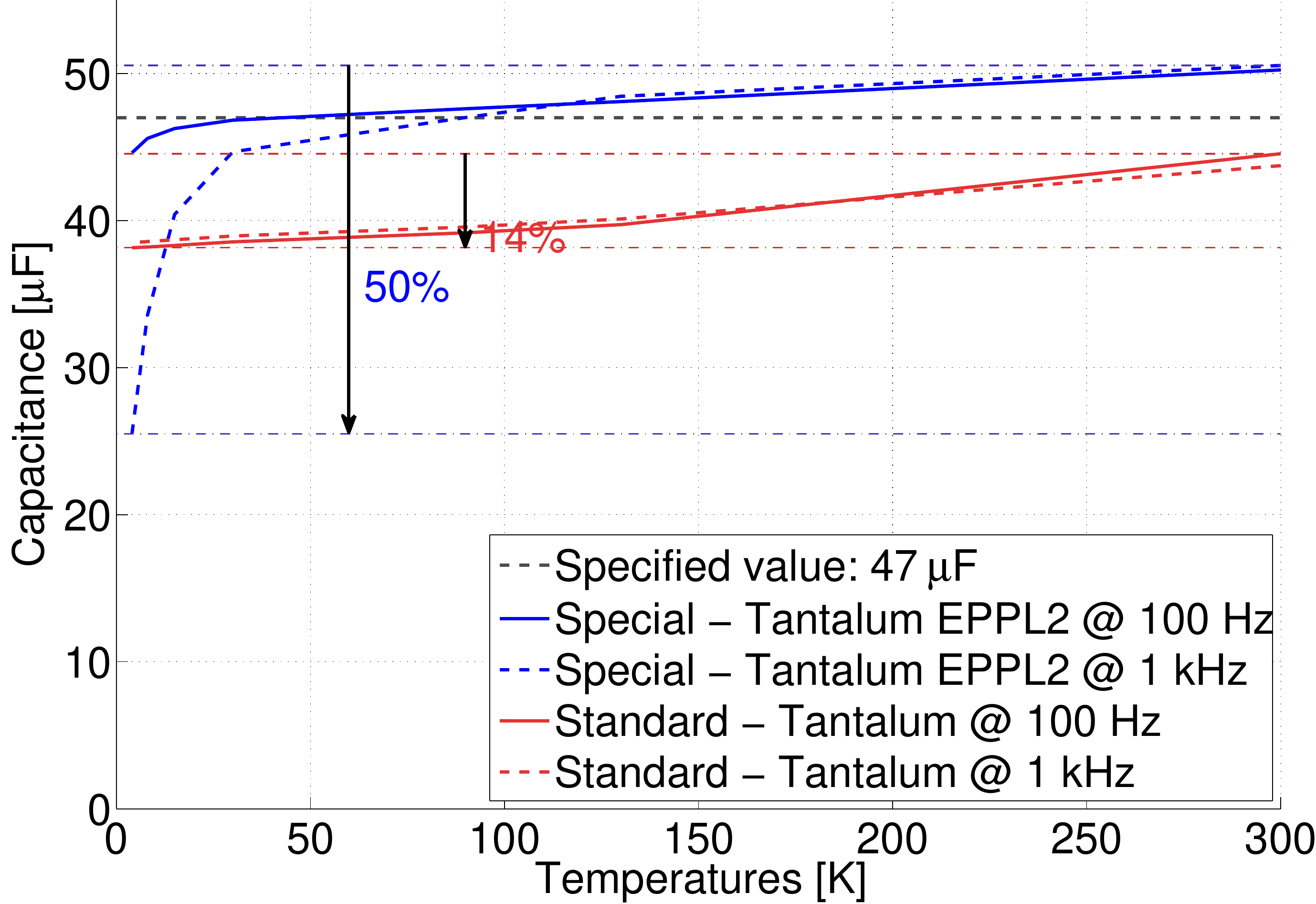}
		}
		\hfill
		\subfloat[ESR for 47~$\mu$F]{\label{fig:capacitors_f}
			\includegraphics[width=.45\textwidth]{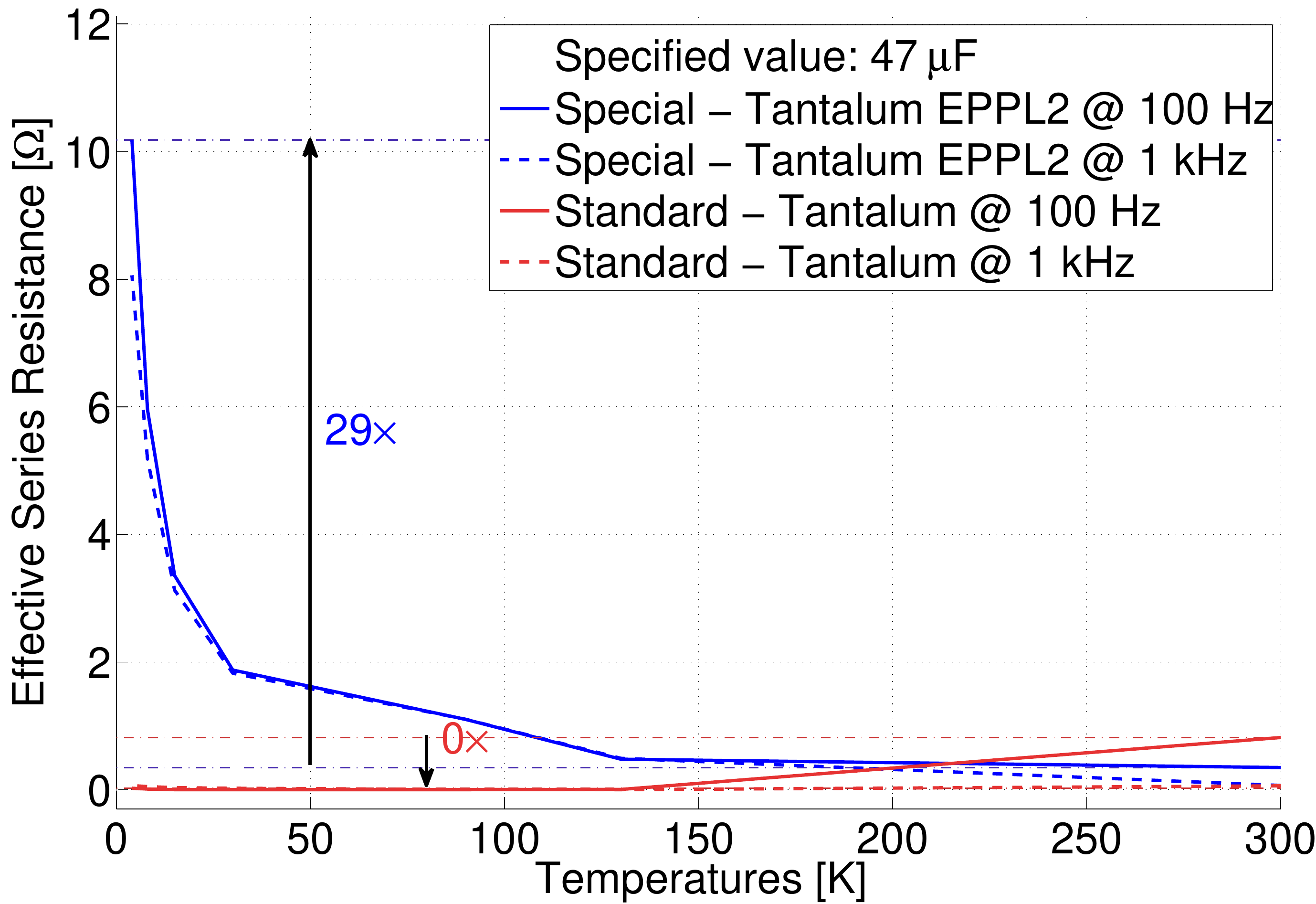}
		}
		\hfill\null
		\caption{Detailed study of various capacitor types and values over a wide temperature range from 4~K to 300~K, showing both its capacitance and ESR.}
		\label{fig:detailedcaps}
	\end{figure*}

\noindent In \autoref{fig:detailedcaps} a detailed overview of the capacitance and ESR of various capacitors is given, the standard capacitors are the ones listed before in \autoref{tab:caps}. While the special capacitors are EPPL2 and sealed caps from AVX for tantalum and silicon type for the 1~$\mu$F (also available in other values). While the special tantalum capacitors perform worse compared to standard tantalum in all cases at 4~K, they (especially the sealed caps) perform indeed better down to 80-100~K temperatures.

	\renewcommand{\thefigure}{\arabic{figure} (Cont.)}
	\addtocounter{figure}{-1}

	\begin{figure*}[htbp]
		\centering
		\null\hfill
		\subfloat[Capacitance for 4.7~$\mu$F]{\label{fig:capacitors_g}
			\includegraphics[width=.45\textwidth]{{{Cap_4.7uF_capacitance_4K_300K}}}
		}
		\hfill
		\subfloat[ESR for 4.7~$\mu$F]{\label{fig:capacitors_h}
			\includegraphics[width=.45\textwidth]{{{Cap_4.7uF_resistance_4K_300K}}}
		}
		\hfill\null
		\\
		\null\hfill
		\subfloat[Capacitance for 1~$\mu$F]{\label{fig:capacitors_i}
			\includegraphics[width=.45\textwidth]{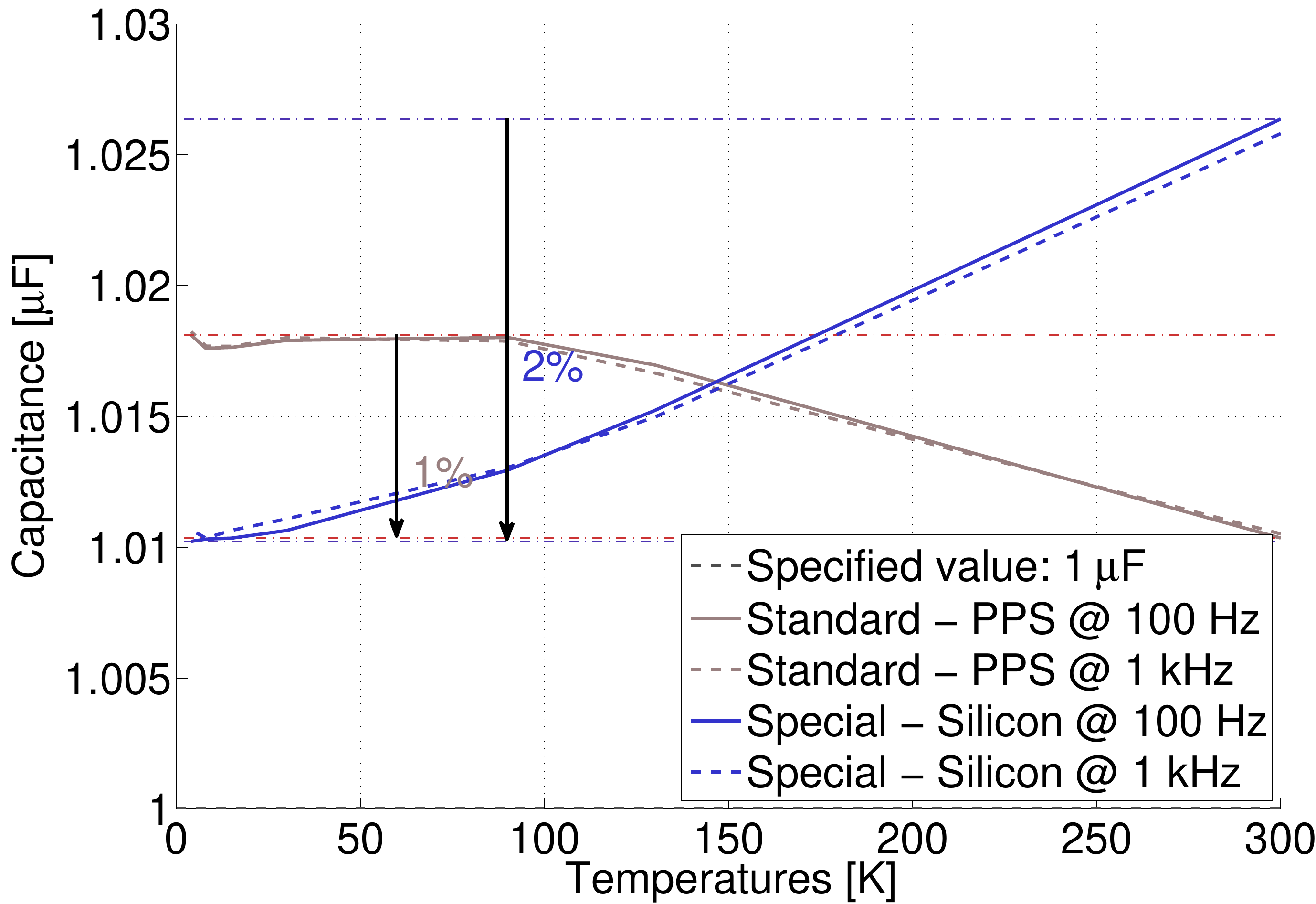}
		}
		\hfill
		\subfloat[ESR for 1~$\mu$F]{\label{fig:capacitors_j}
			\includegraphics[width=.45\textwidth]{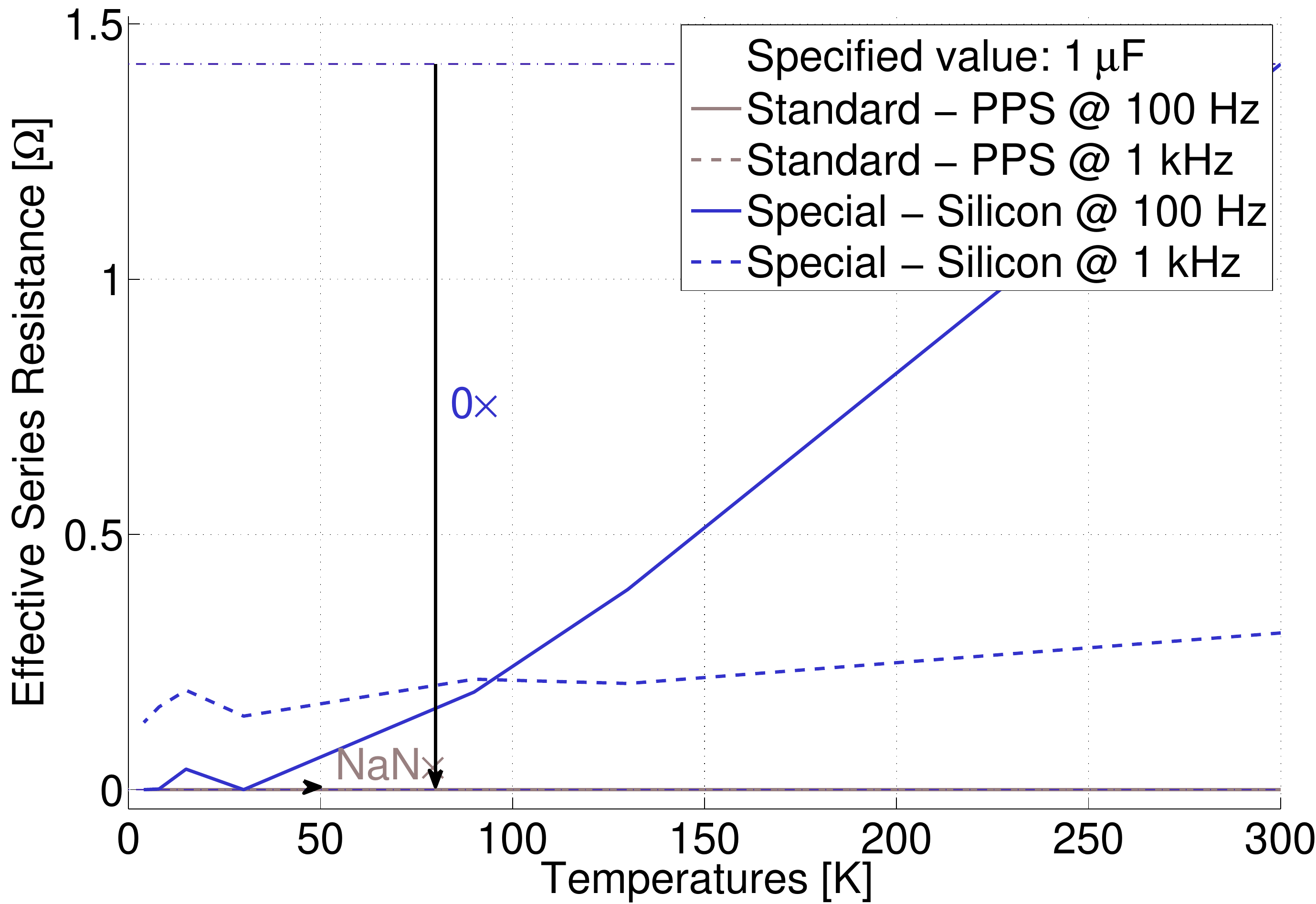}
		}
		\hfill\null
		\\
		\null\hfill
		\subfloat[Capacitance for 0.47~$\mu$F]{\label{fig:capacitors_k}
			\includegraphics[width=.45\textwidth]{{{Cap_0.47uF_capacitance_4K_300K}}}
		}
		\hfill
		\subfloat[ESR for 0.47~$\mu$F]{\label{fig:capacitors_l}
			\includegraphics[width=.45\textwidth]{{{Cap_0.47uF_resistance_4K_300K}}}
		}
		\hfill\null
		\caption{Detailed study of various capacitor types and values over a wide temperature range from 4~K to 300~K, showing both its capacitance and ESR.}
		\label{fig:detailedcapsII}
	\end{figure*}
	\renewcommand{\thefigure}{\arabic{figure}}

\begin{figure*}[tb]
		\centering
		\null\hfill
		\subfloat[]{\label{fig:detailedres_a}
			\includegraphics[width=.3\textwidth]{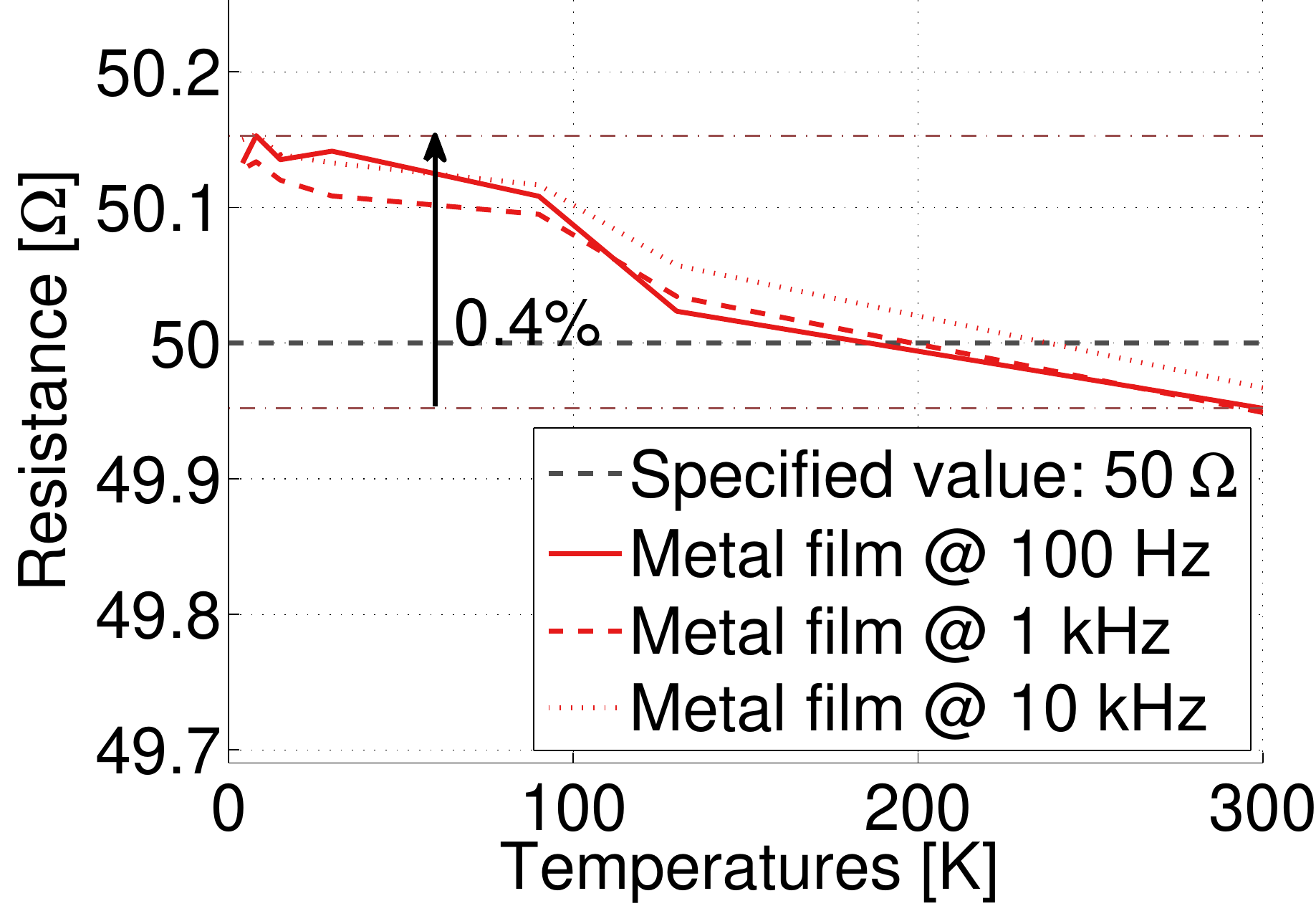}
		}
		\hfill
		\subfloat[]{\label{fig:detailedres_b}
			\includegraphics[width=.3\textwidth]{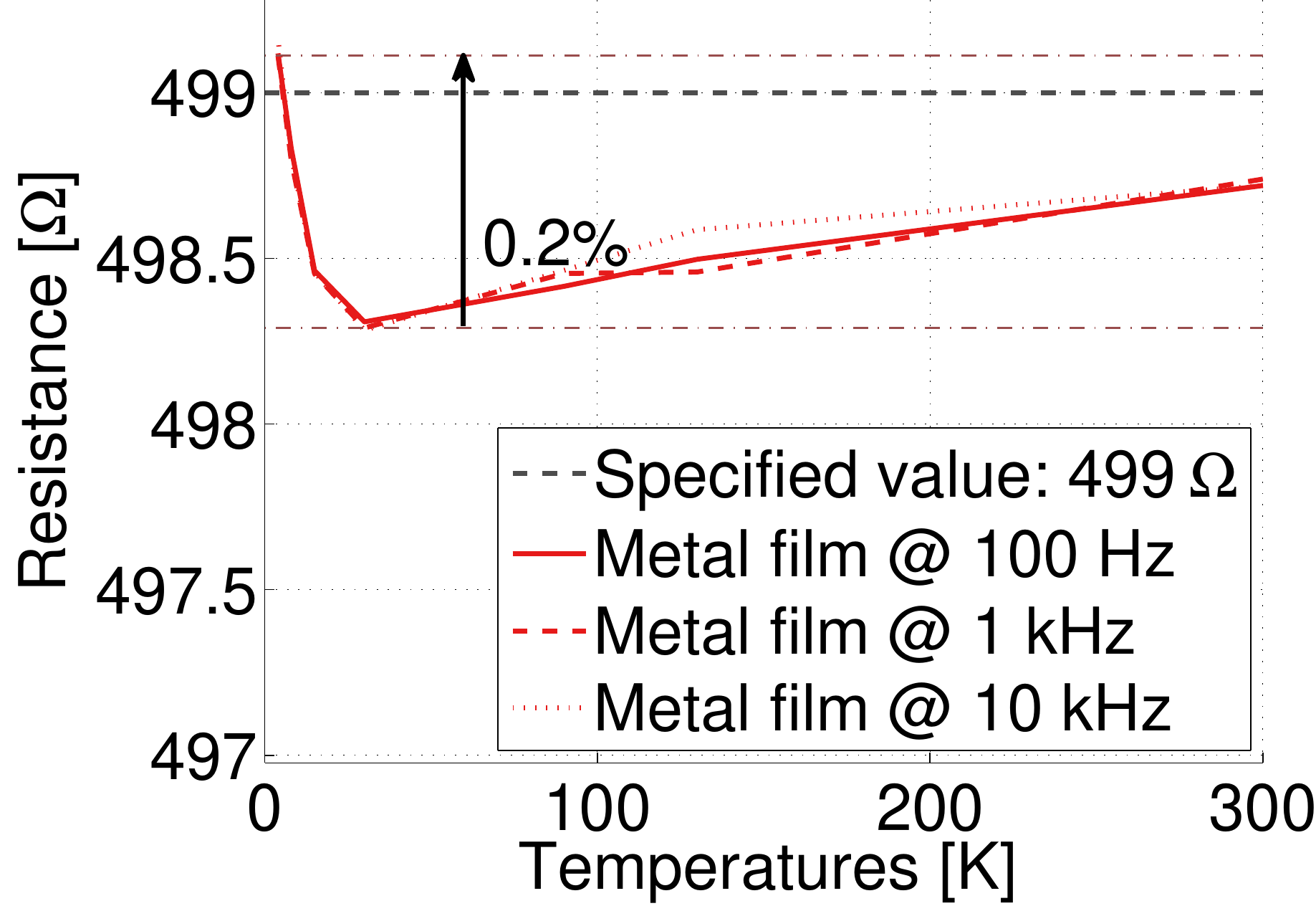}
		}
		\hfill
		\subfloat[]{\label{fig:detailedres_c}
			\includegraphics[width=.3\textwidth]{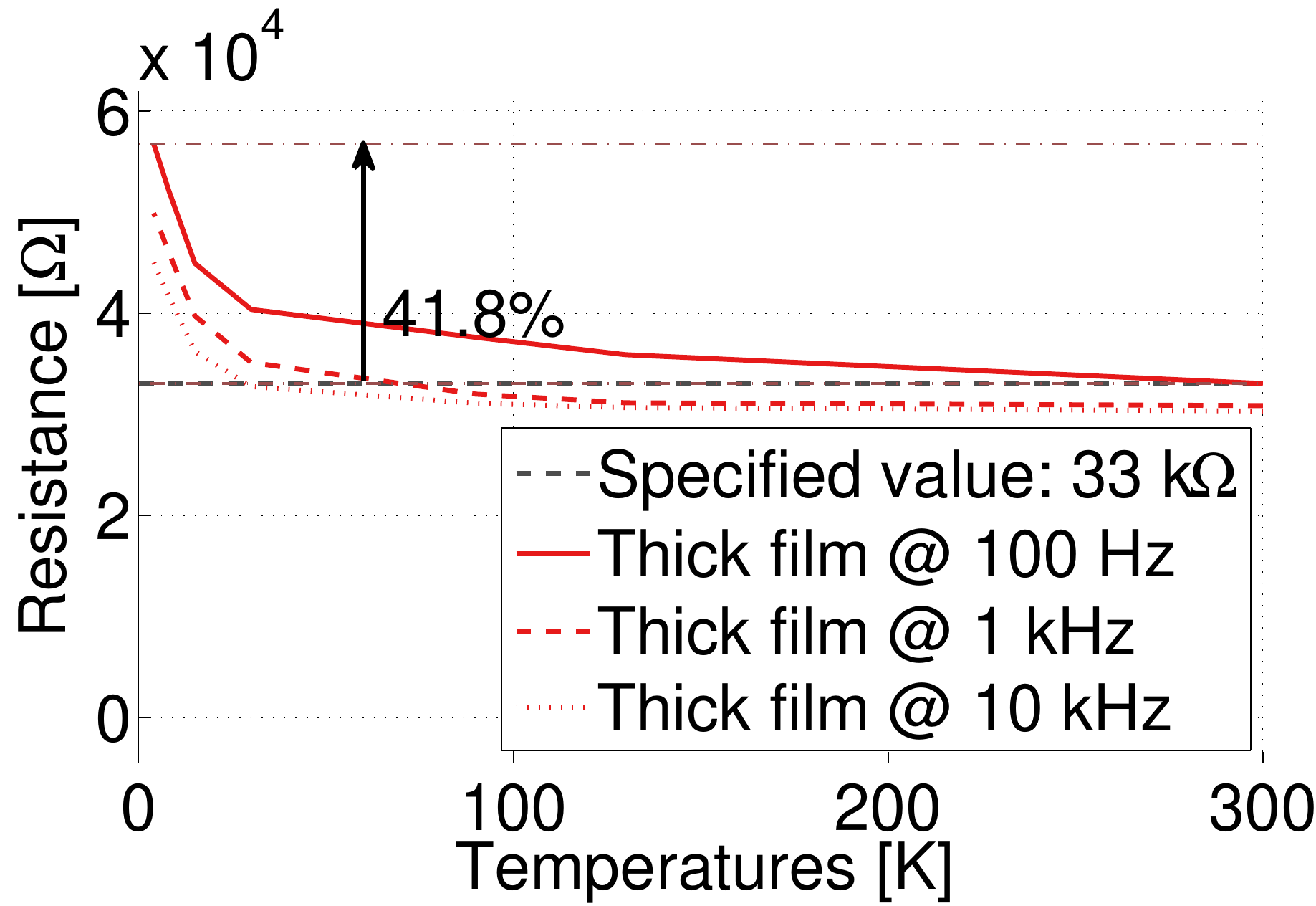}
		}
		\hfill\null
		\caption{Resistance of different resistors, value and resistive material type (metal and thick film) over temperature. }
		\label{fig:detailedres}
	\end{figure*}

In \autoref{fig:detailedres}, the measured resistance for the three resistors shown in \autoref{fig:resistors} are plotted with different testing frequencies. 
	
	 \section{Voltage regulators - Detailed results}\label{sec:reg_details}
\begin{figure*}[htbp]
		\centering
		\subfloat[Buck ISL80030]{\label{fig:regulators_voltage_a}
			\includegraphics[width=.3\textwidth]{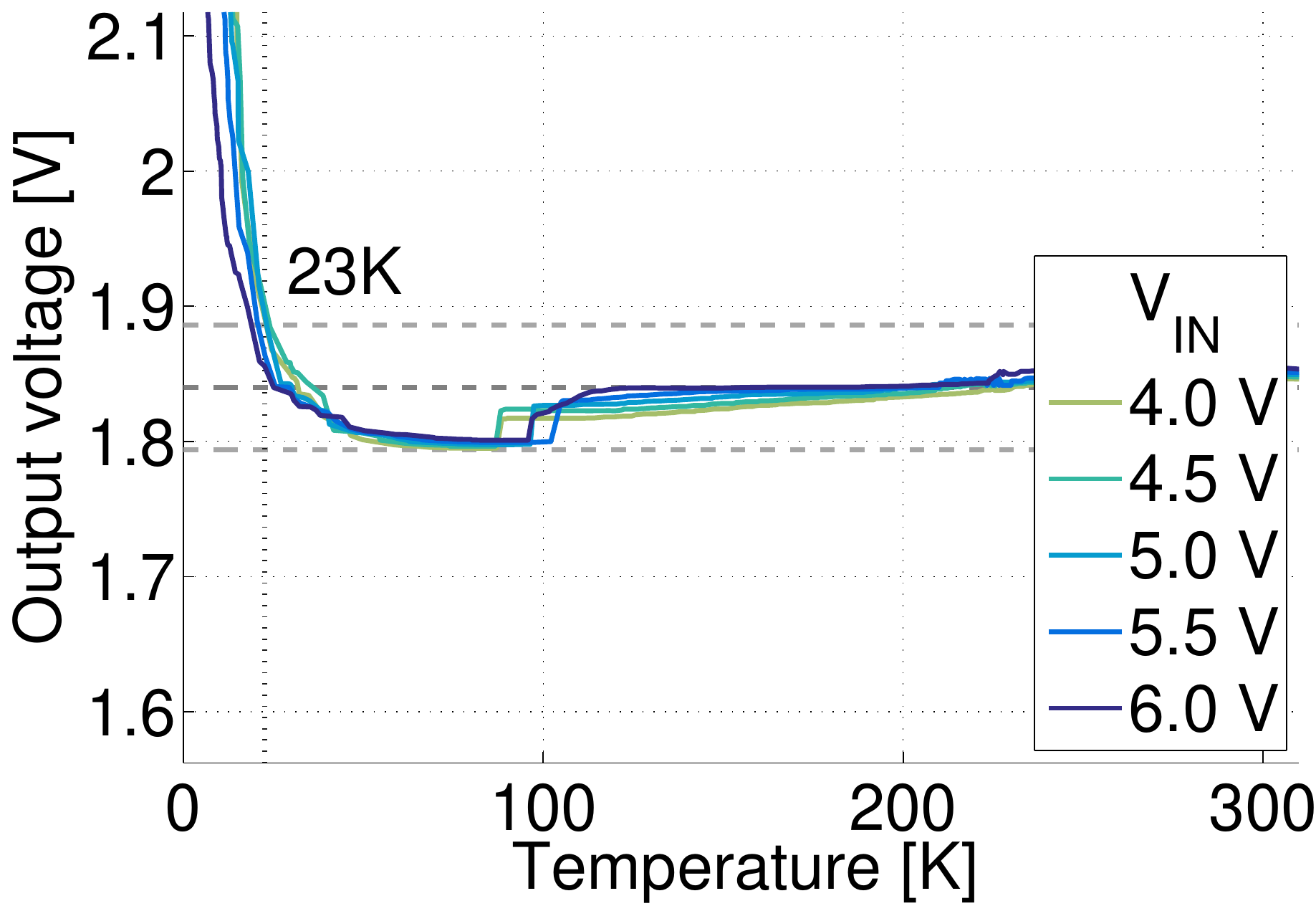} }
		\hfill
		\subfloat[Buck ISL80031]{\label{fig:regulators_voltage_b}
			\includegraphics[width=.3\textwidth]{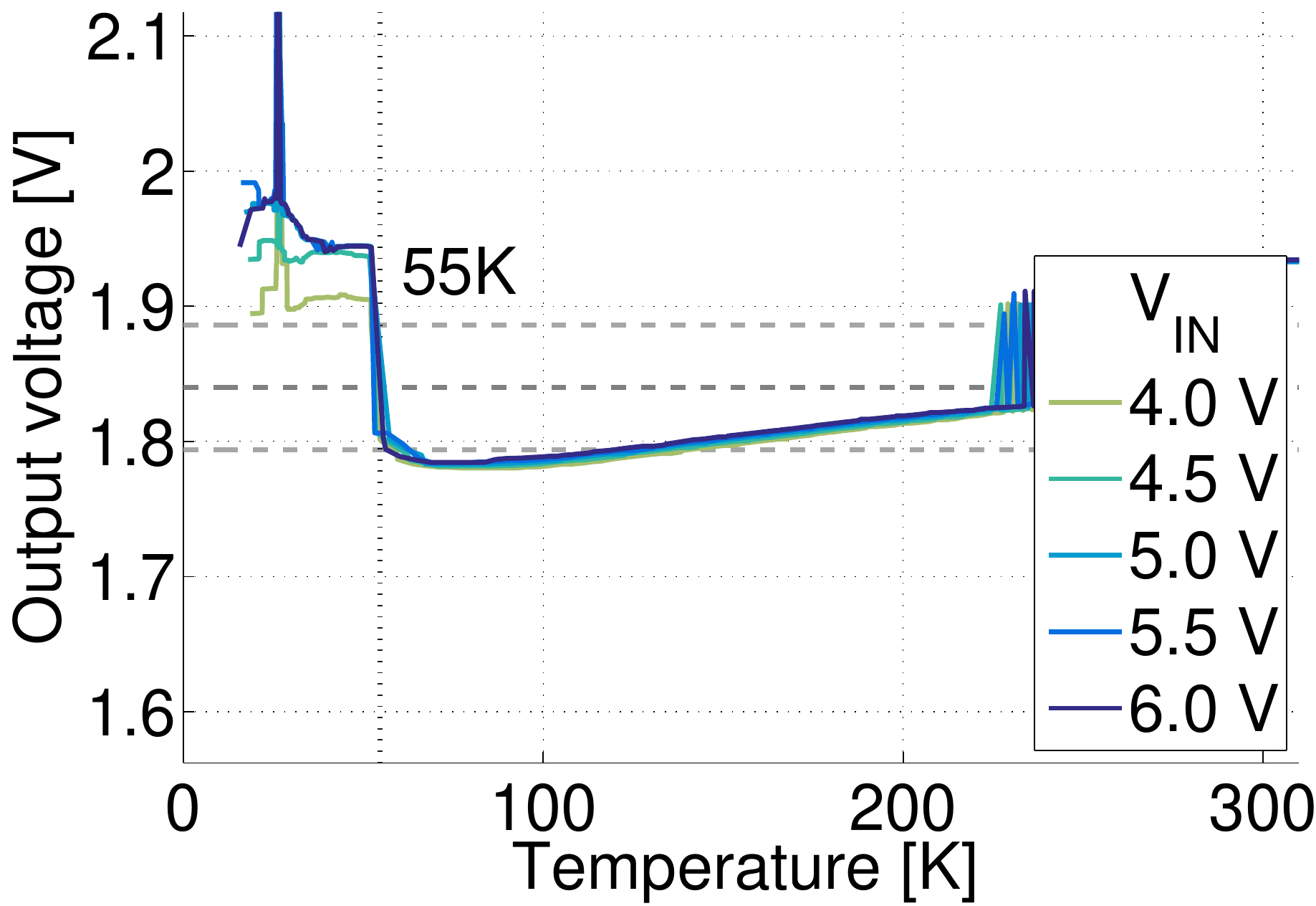} }
		\hfill
		\subfloat[Buck ISL85415]{\label{fig:regulators_voltage_c}
			\includegraphics[width=.3\textwidth]{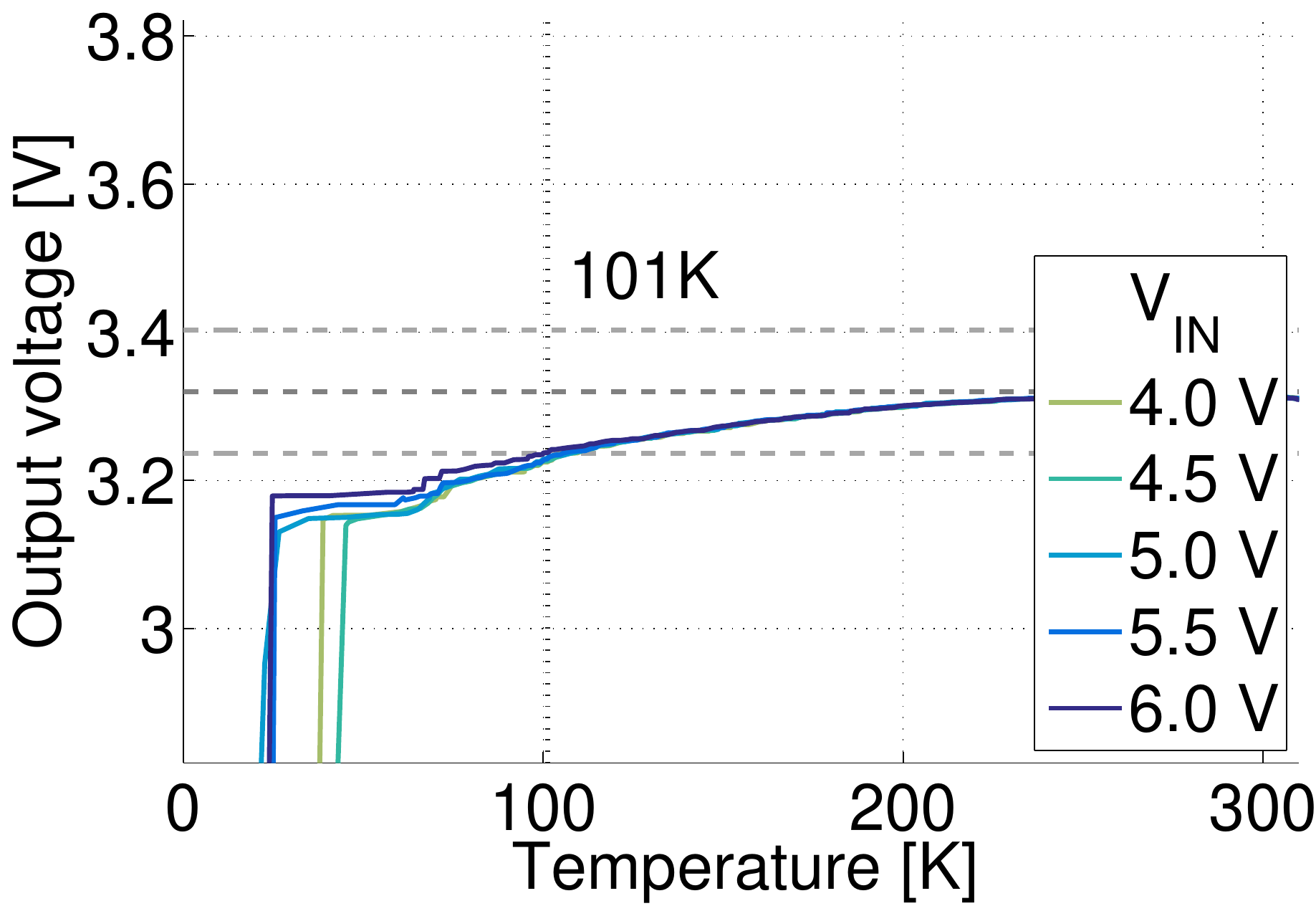} }
		\\
		\subfloat[Buck MIC23150]{\label{fig:regulators_voltage_d}
			\includegraphics[width=.3\textwidth]{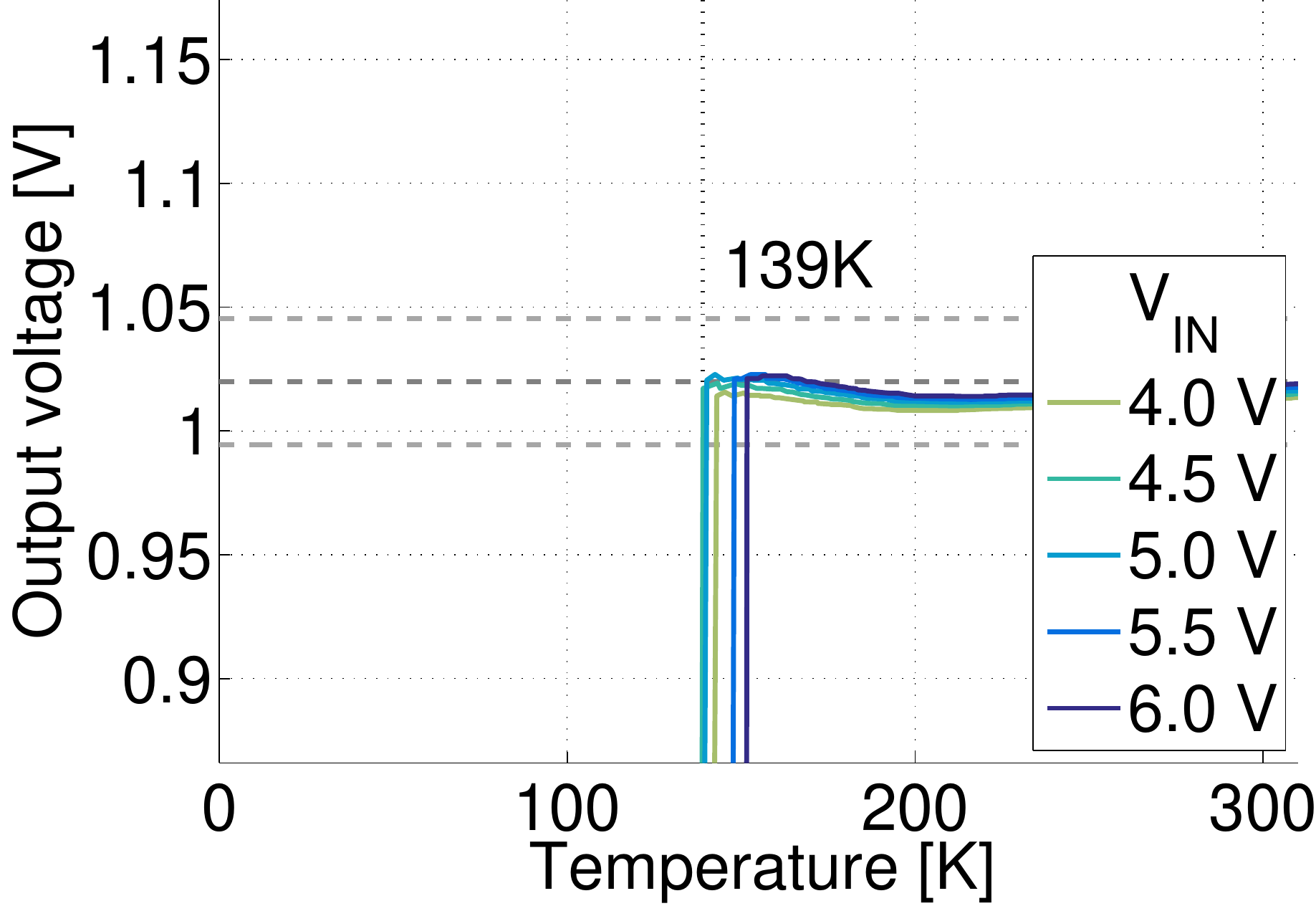} }
		\hfill
		\subfloat[Buck MCP16301]{\label{fig:regulators_voltage_e}
			\includegraphics[width=.3\textwidth]{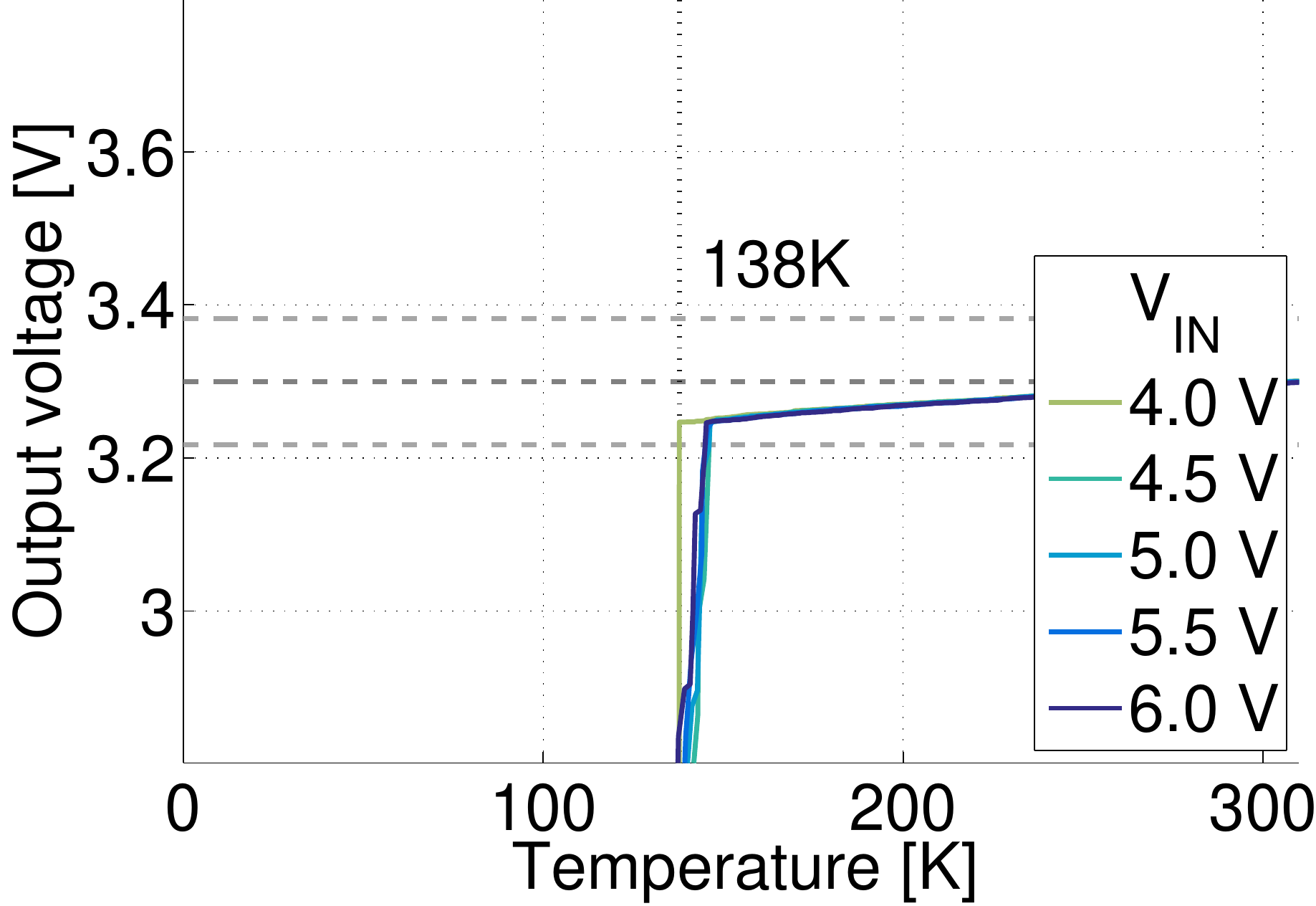} }
		\hfill
		\subfloat[Buck AST1S31HF]{\label{fig:regulators_voltage_f}
			\includegraphics[width=.3\textwidth]{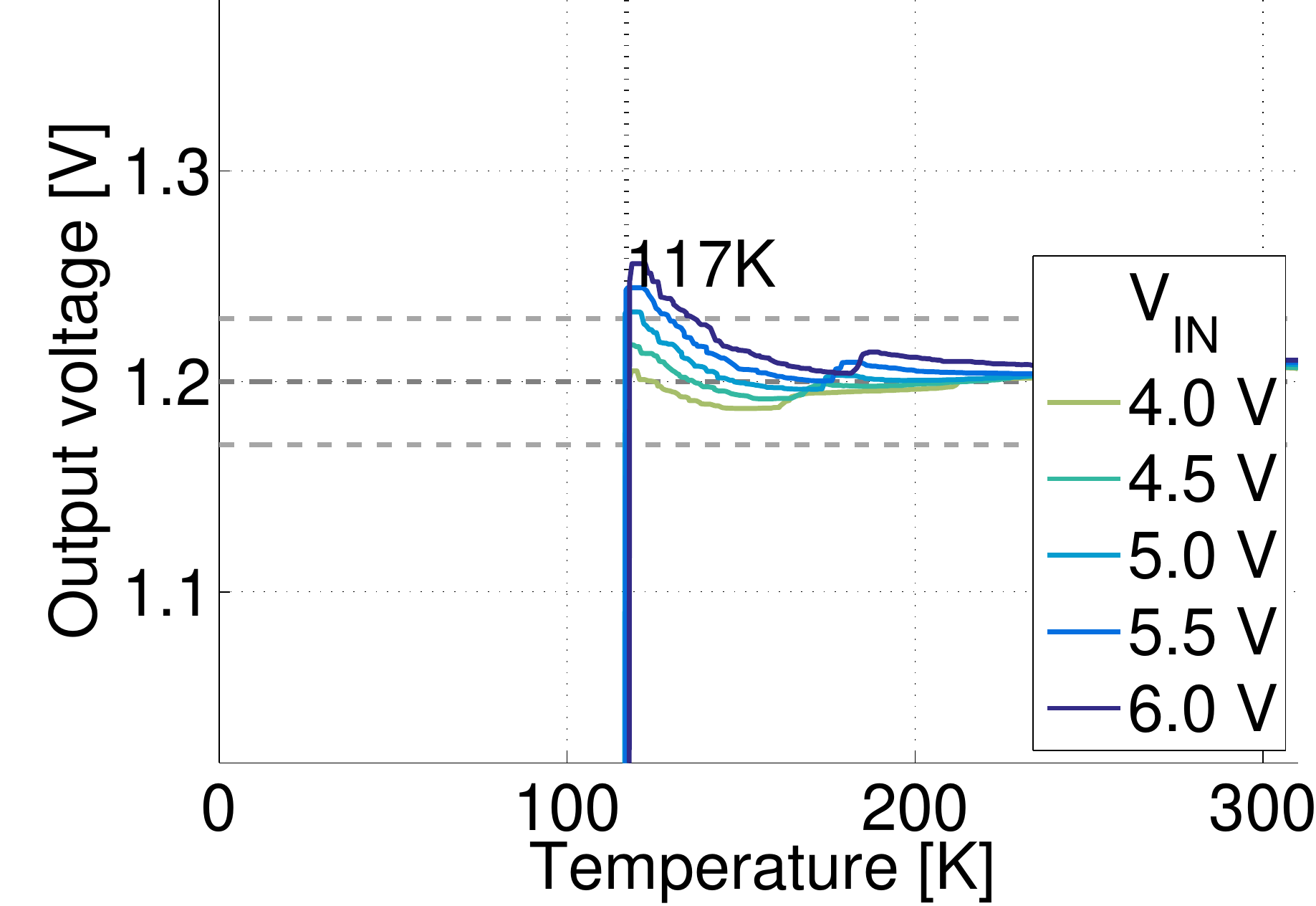} }
		\\
		\subfloat[Buck TPS54614]{\label{fig:regulators_voltage_g}
			\includegraphics[width=.3\textwidth]{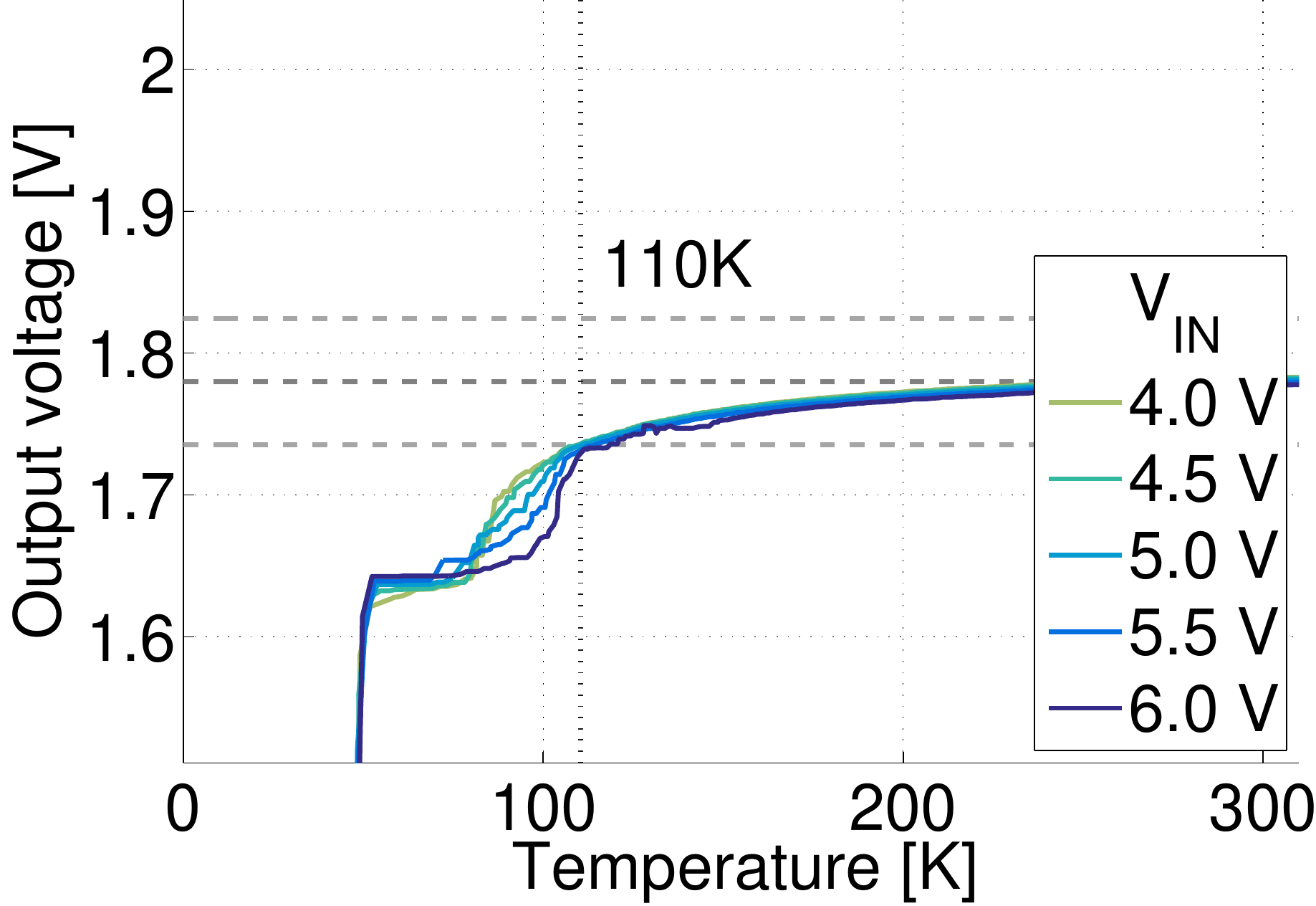} }
		\hfill
		\caption{Detailed stability study of switching (buck) regulators over a wide temperature range from 4~K to 300~K.}
		\label{fig:detailedregulators_buck_voltage}
	\end{figure*}

\begin{figure*}[htbp]
		\centering
		\subfloat[LDO ADP165CB]{\label{fig:regulators_voltage_h}
			\includegraphics[width=.3\textwidth]{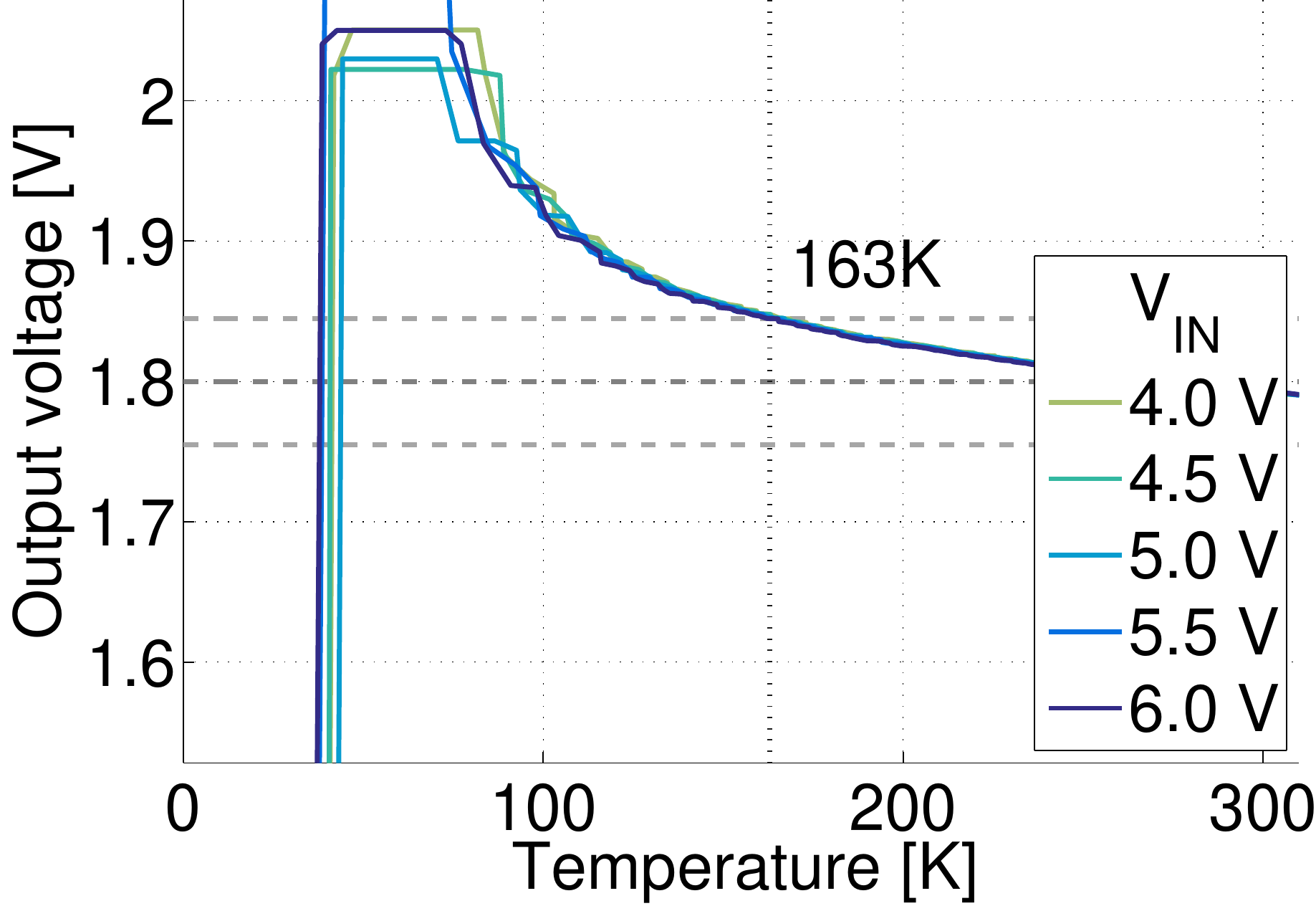} }
		\hfill
		\subfloat[LDO ADP165CP]{\label{fig:regulators_voltage_i}
			\includegraphics[width=.3\textwidth]{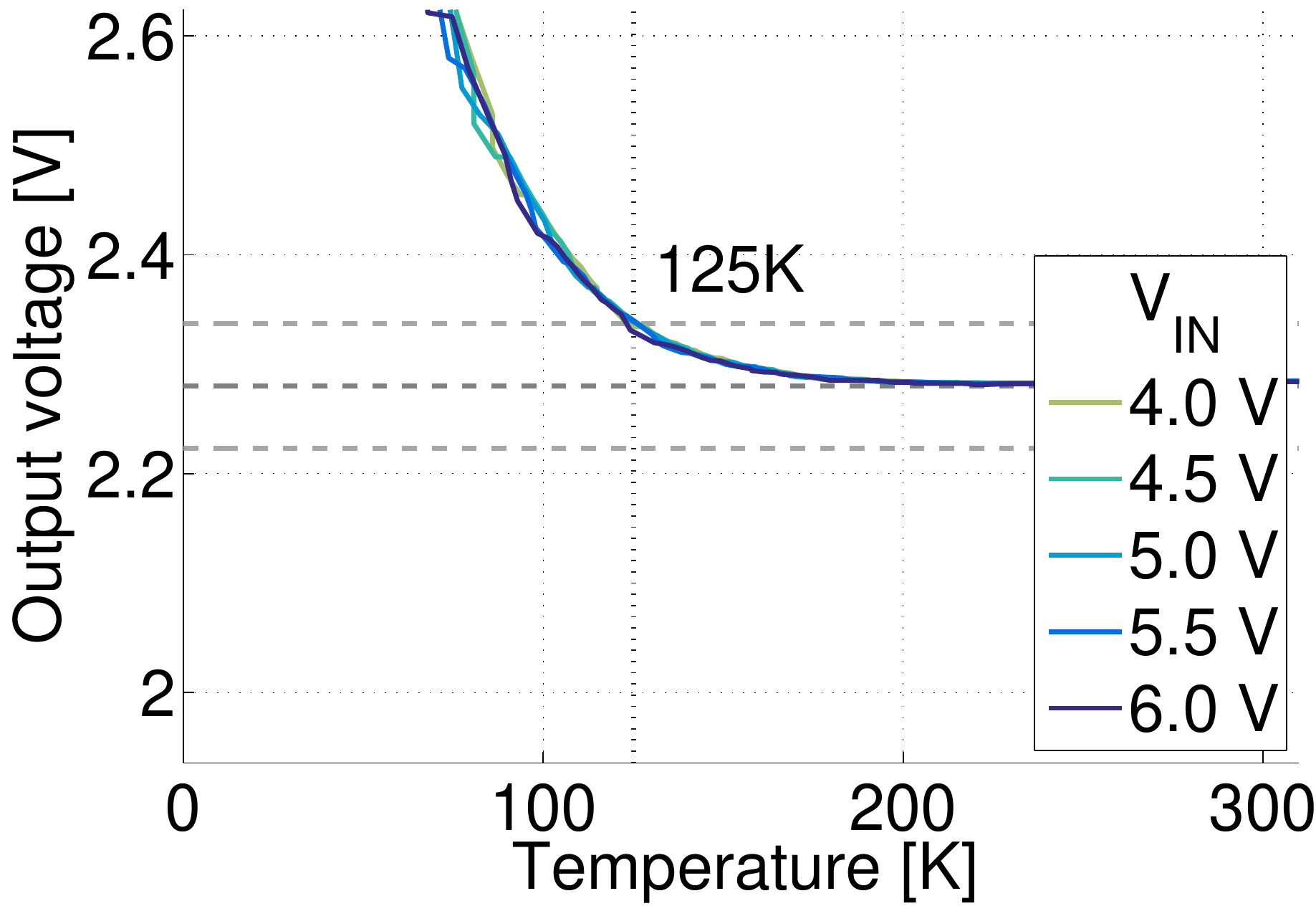} }
		\hfill
		\subfloat[LDO ADP165UJ]{\label{fig:regulators_voltage_j}
			\includegraphics[width=.3\textwidth]{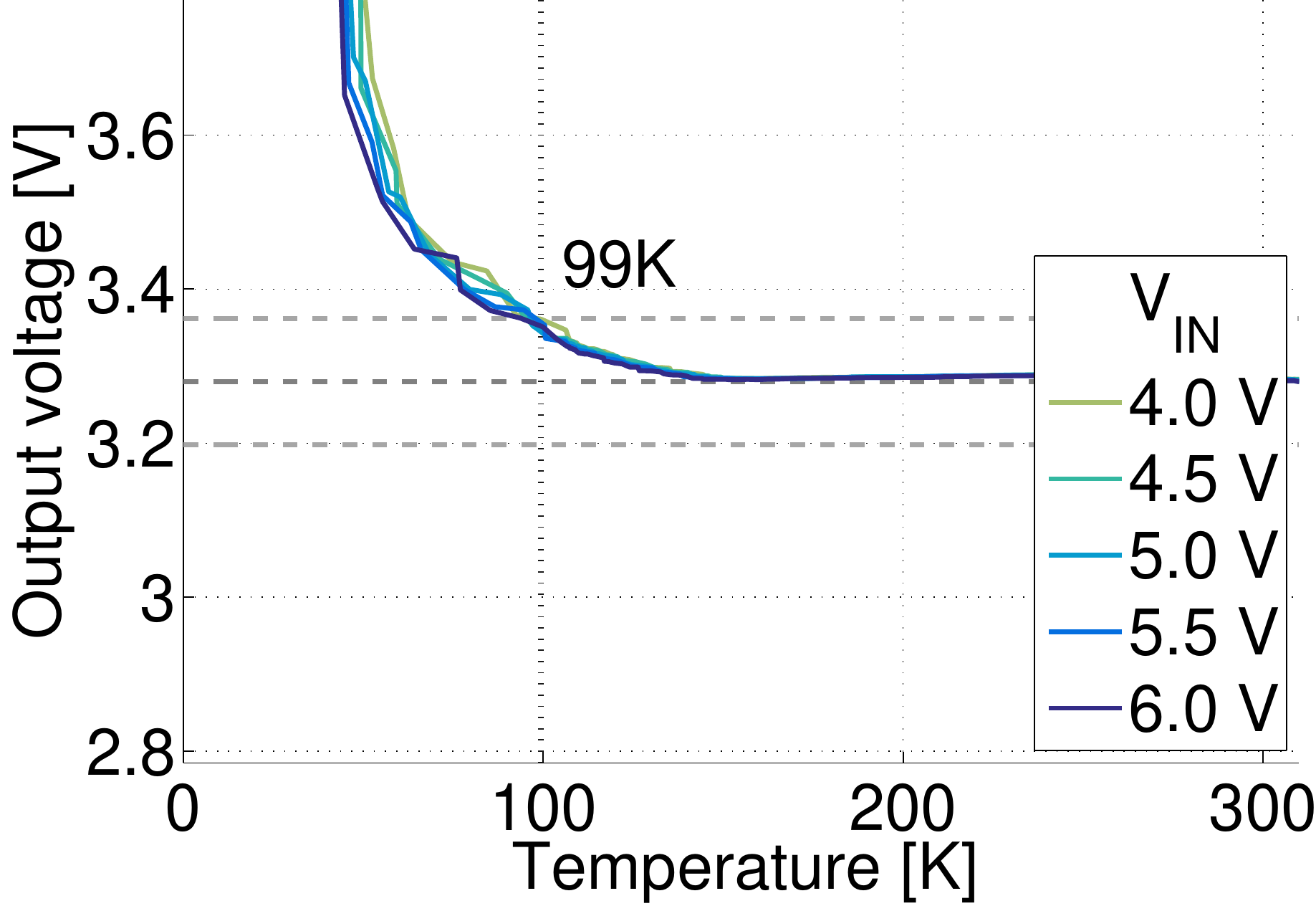} }
		\\
		\subfloat[LDO ISL80510]{\label{fig:regulators_voltage_k}
			\includegraphics[width=.3\textwidth]{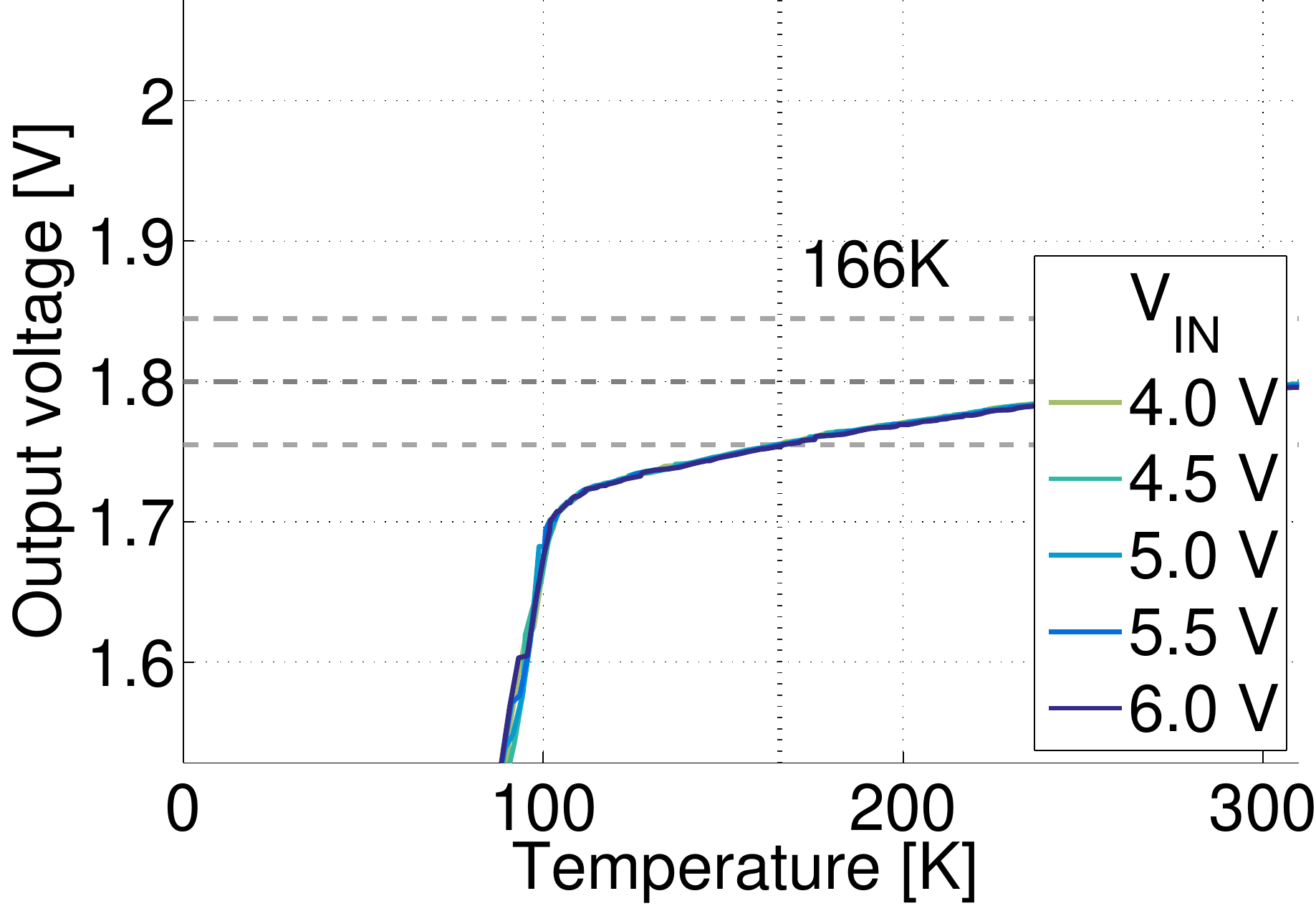} }
		\hfill
		\subfloat[LDO MIC38150]{\label{fig:regulators_voltage_l}
			\includegraphics[width=.3\textwidth]{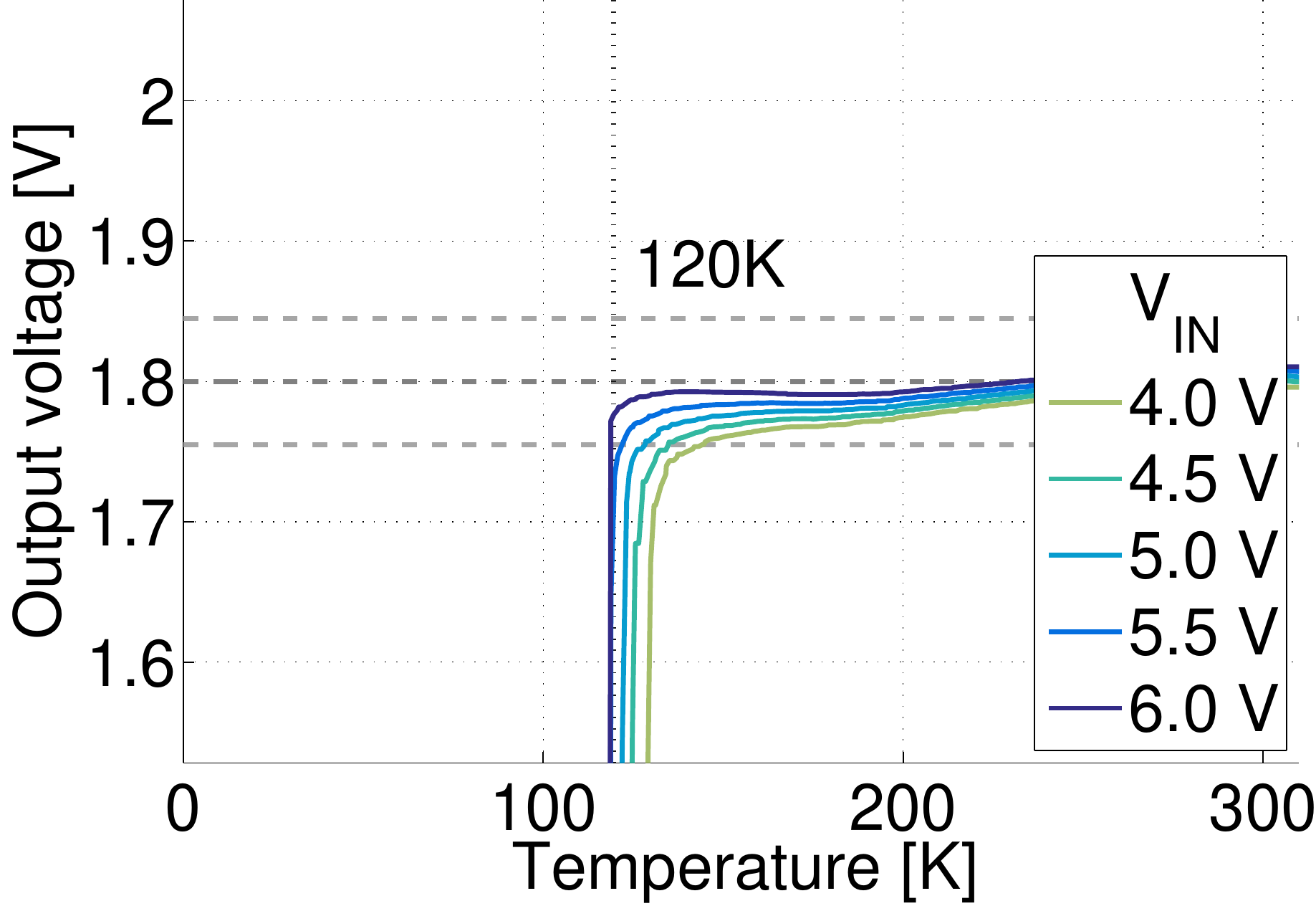} }
		\hfill
		\subfloat[LDO MIC69303]{\label{fig:regulators_voltage_m}
			\includegraphics[width=.3\textwidth]{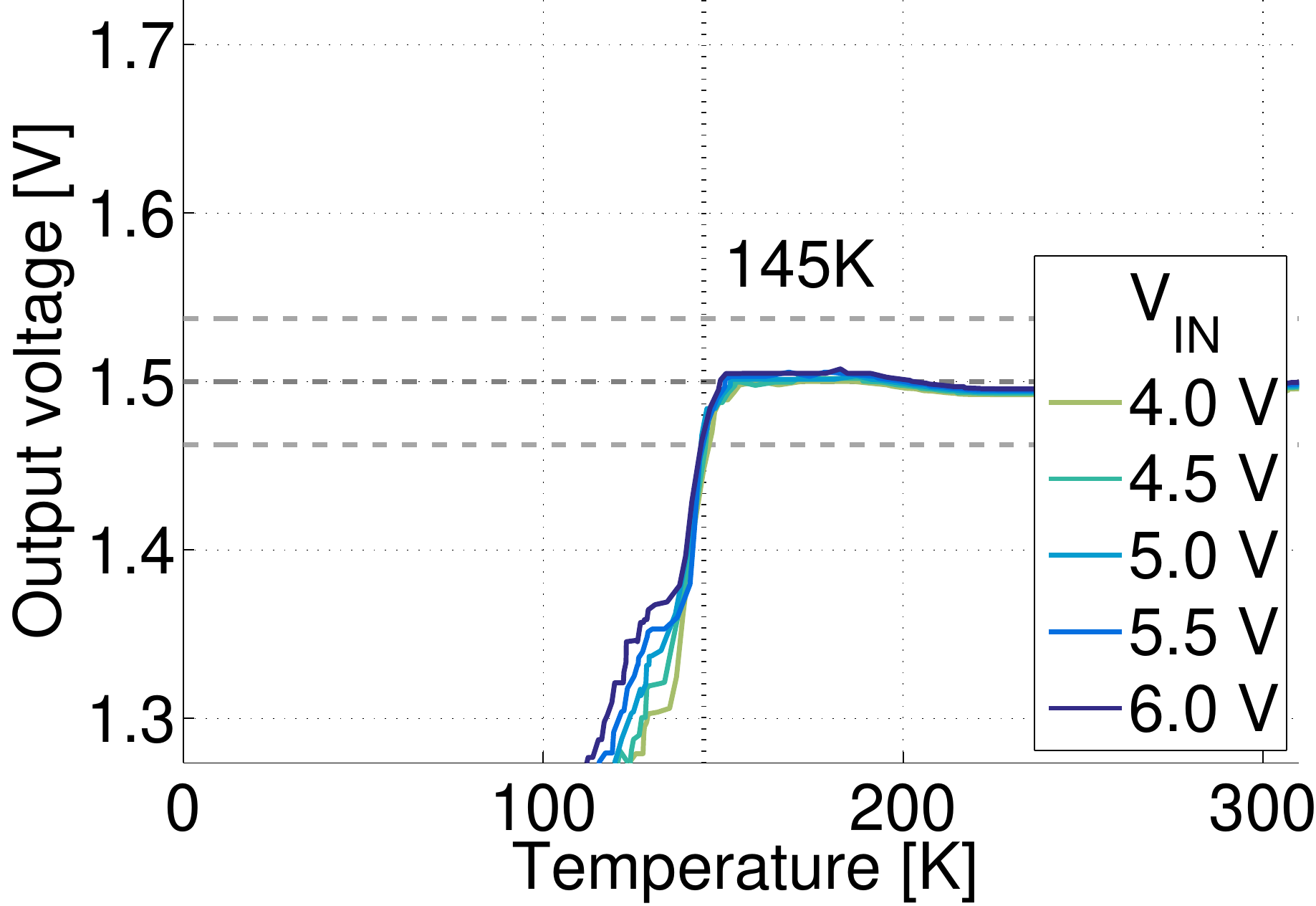} }
		\\
		\subfloat[LDO TPS7A4700]{\label{fig:regulators_voltage_n}
			\includegraphics[width=.3\textwidth]{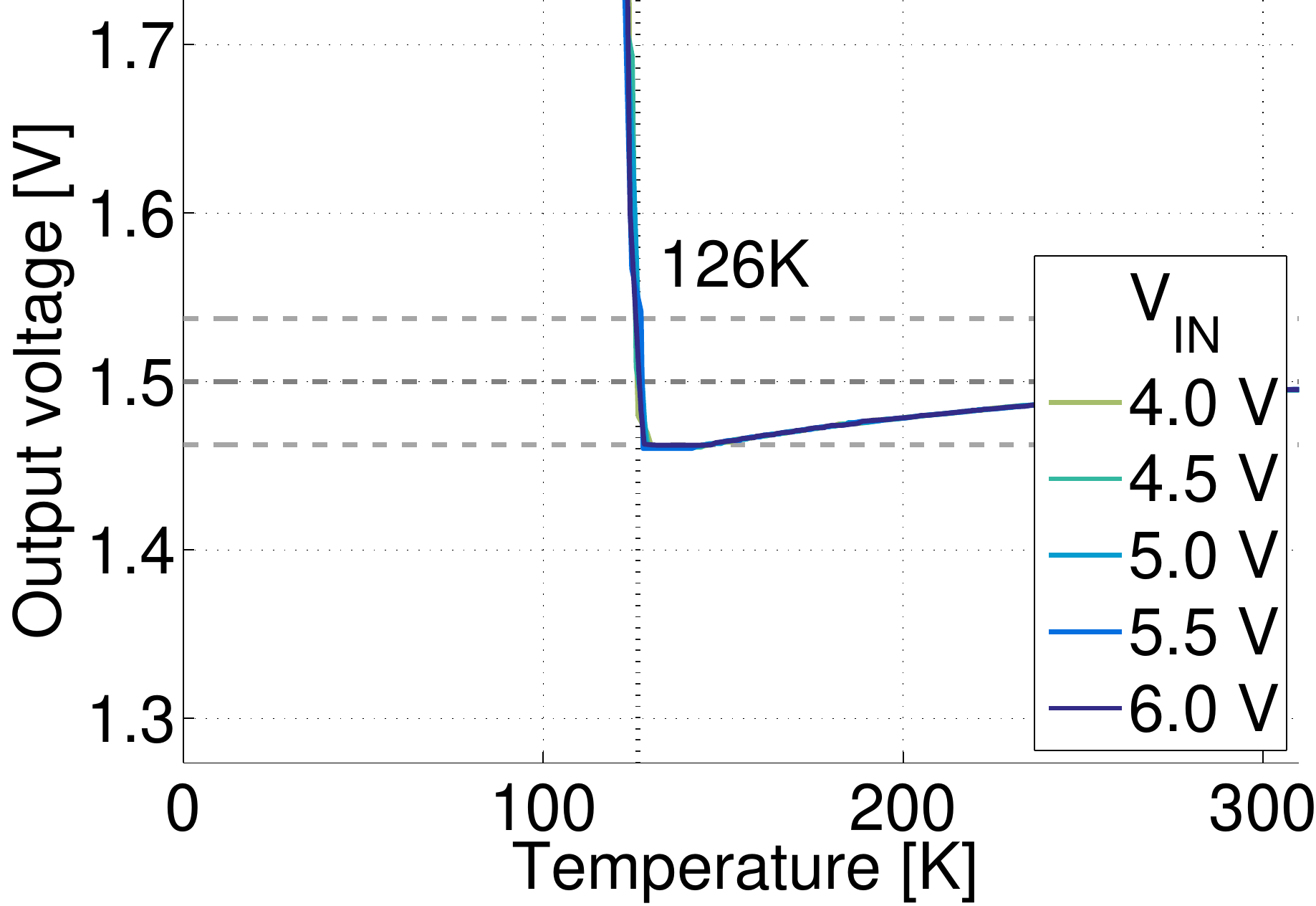} }
		\hfill
		\subfloat[LDO TPS7A7002]{\label{fig:regulators_voltage_o}
			\includegraphics[width=.3\textwidth]{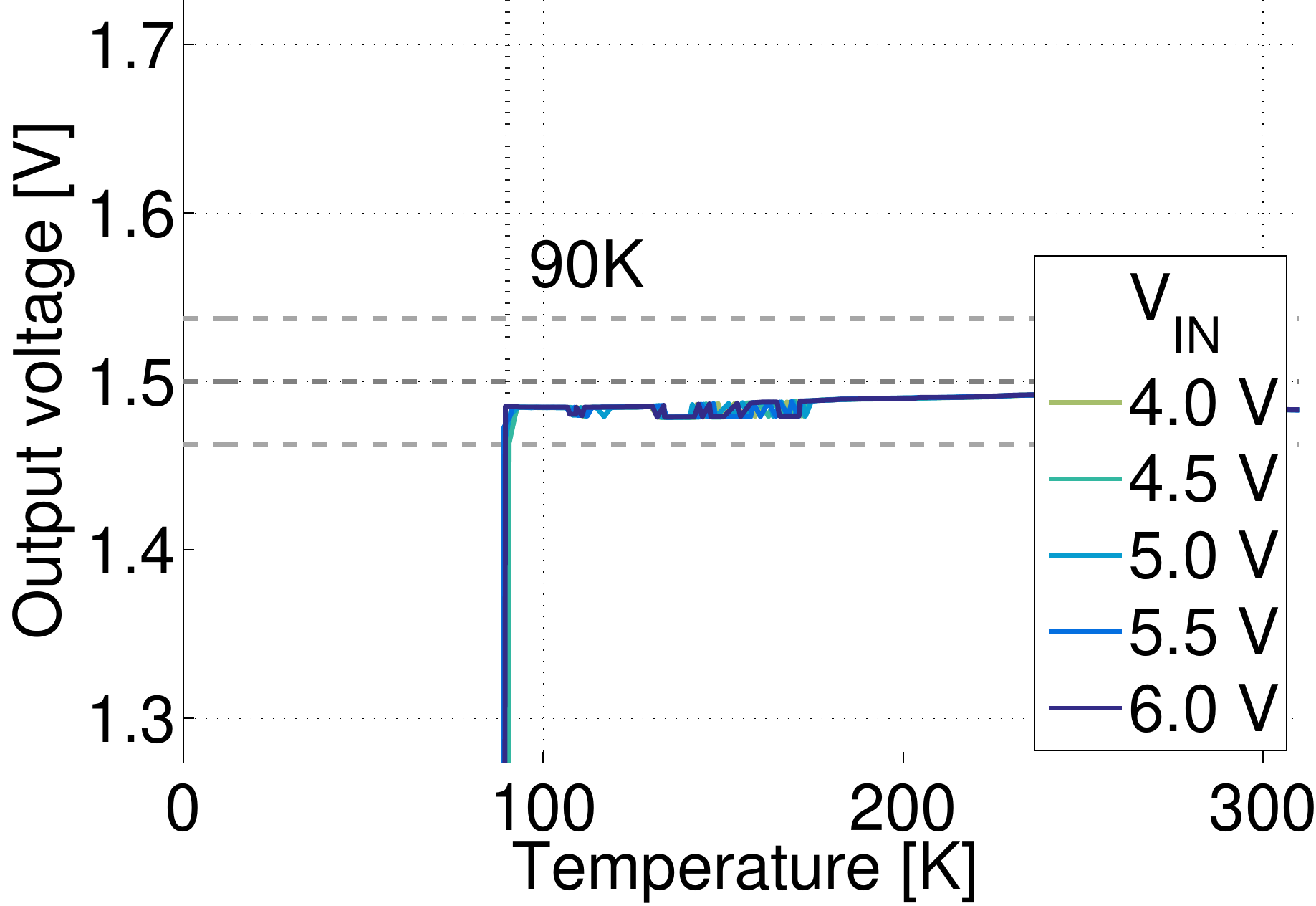} }
		\hfill
		\subfloat[LDO TPS7A8101]{\label{fig:regulators_voltage_p}
			\includegraphics[width=.3\textwidth]{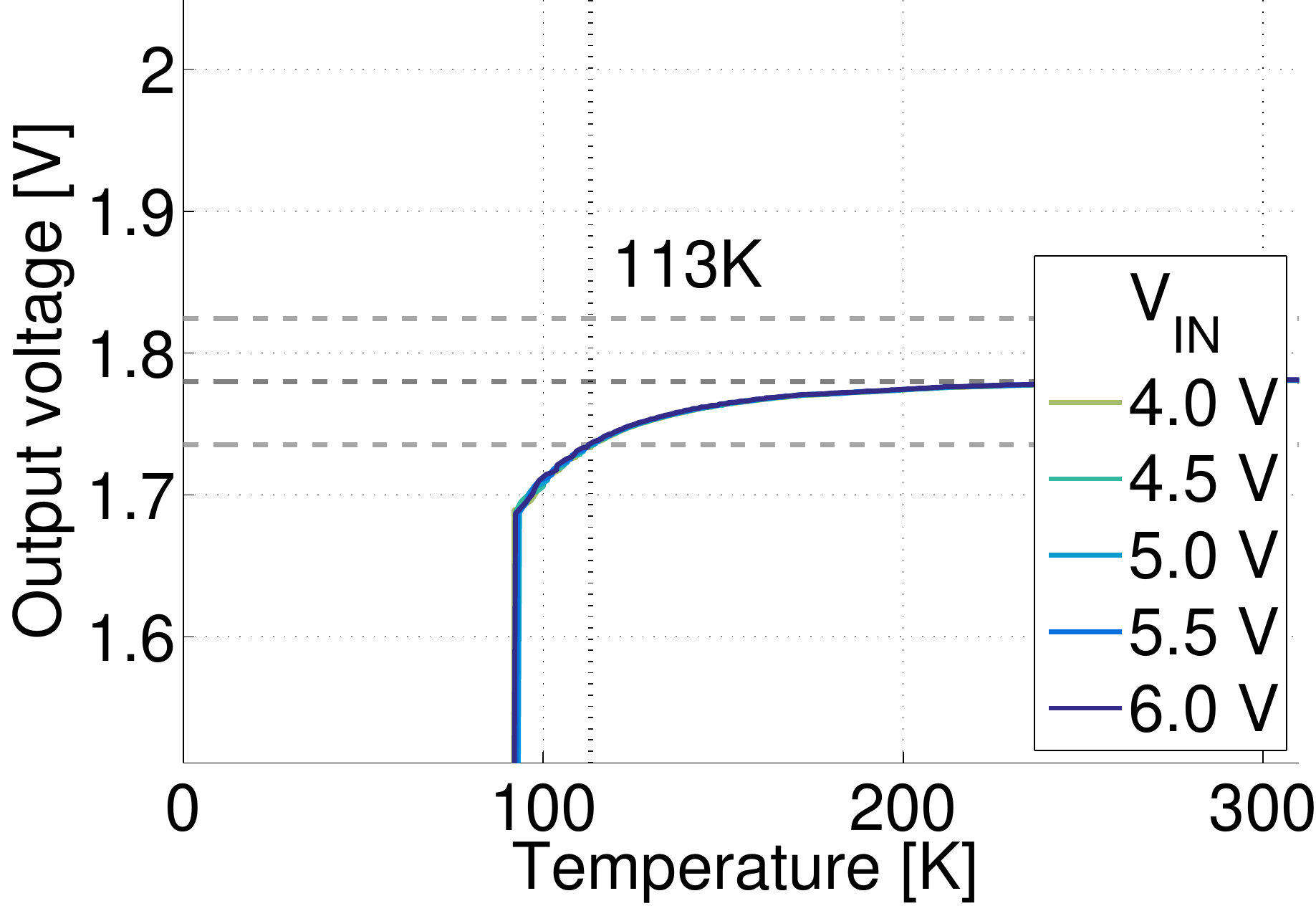} }
		\\
		\caption{Detailed stability study of LDOs over a wide temperature range from 4~K to 300~K.}
		\label{fig:detailedregulators_LDO_voltage}
	\end{figure*}

\noindent Below all results of voltage regulators over temperature can be found. In \autoref{fig:detailedregulators_buck_voltage} and \autoref{fig:detailedregulators_LDO_voltage}, the output voltage of switching (buck) regulators, respectively LDOs are plotted versus temperature. For reference, a margin of 2.5\% from the average output voltage at room temperature is used to indicate the regulator being to far away from what voltage should be seen over the load resistance (R$_\text{load}=100~\Omega$). This boundary is indicated with the best temperature at which the device is still in this margin. 

\begin{figure*}[htbp]
		\centering
		\subfloat[Buck ISL80030]{\label{fig:regulators_power_a}
			\includegraphics[width=.3\textwidth]{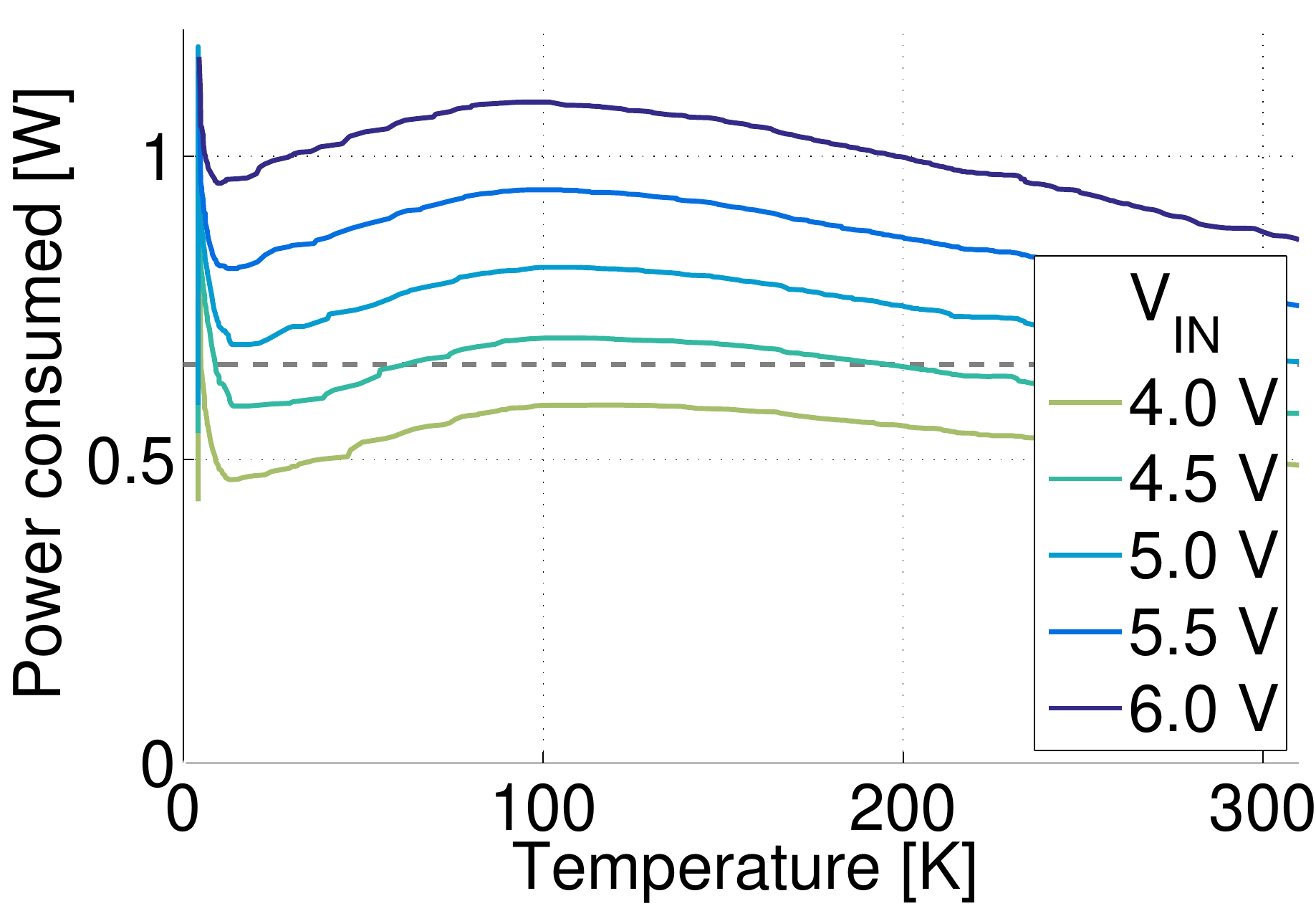} }
		\hfill
		\subfloat[Buck ISL80031]{\label{fig:regulators_power_b}
			\includegraphics[width=.3\textwidth]{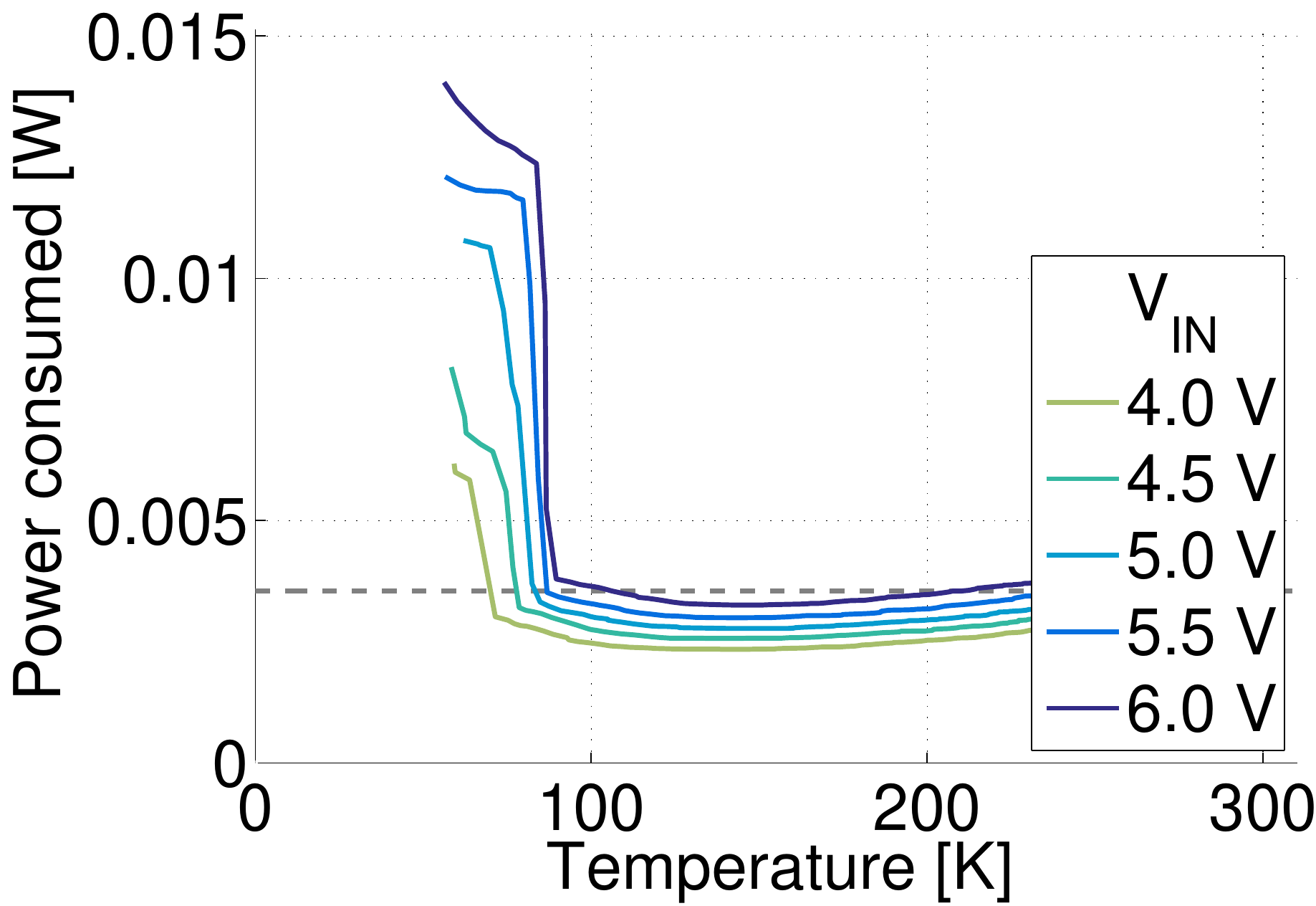} }
		\hfill
		\subfloat[Buck ISL85415]{\label{fig:regulators_power_c}
			\includegraphics[width=.3\textwidth]{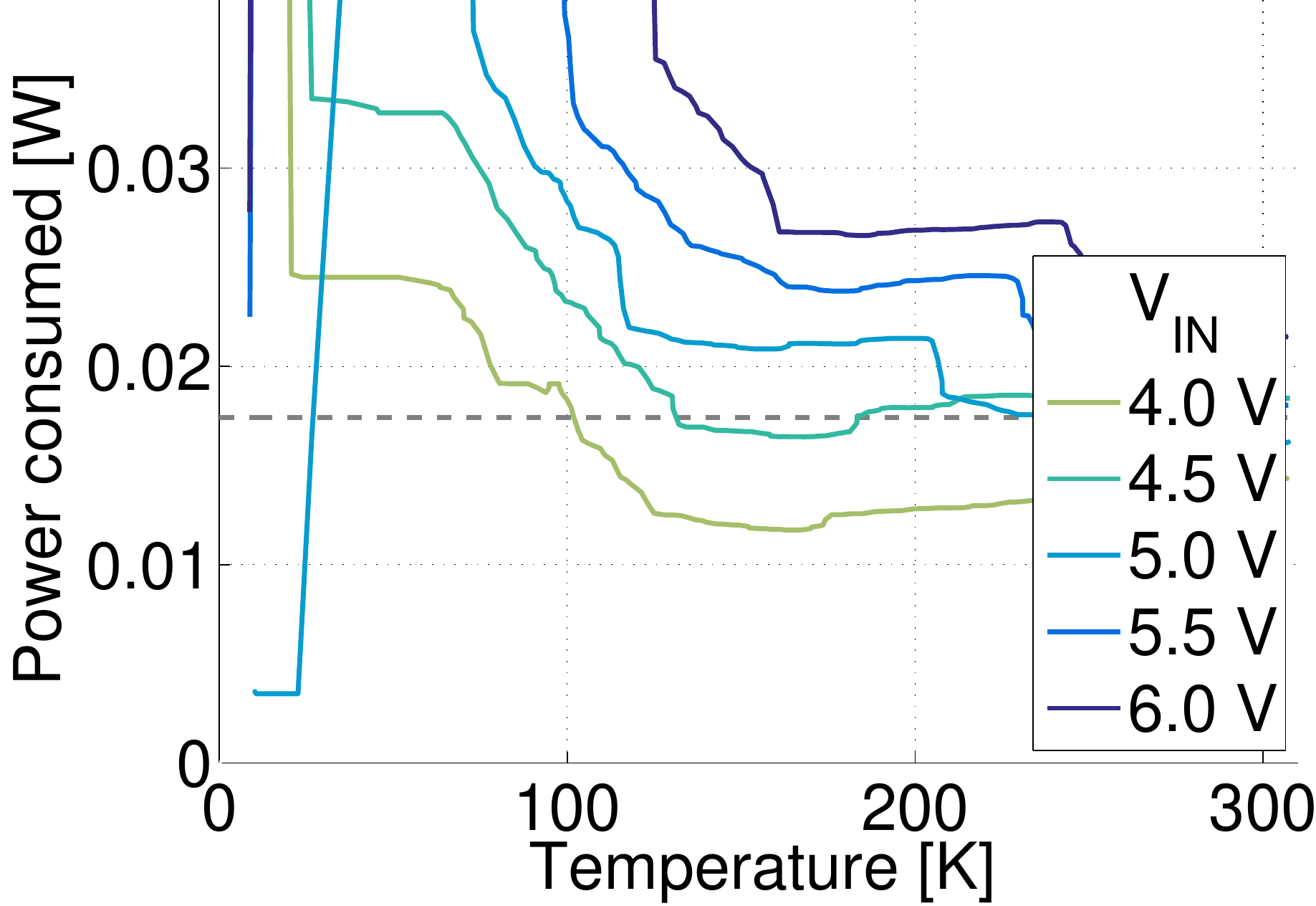} }
		\\
		\subfloat[Buck MIC23150]{\label{fig:regulators_power_d}
			\includegraphics[width=.3\textwidth]{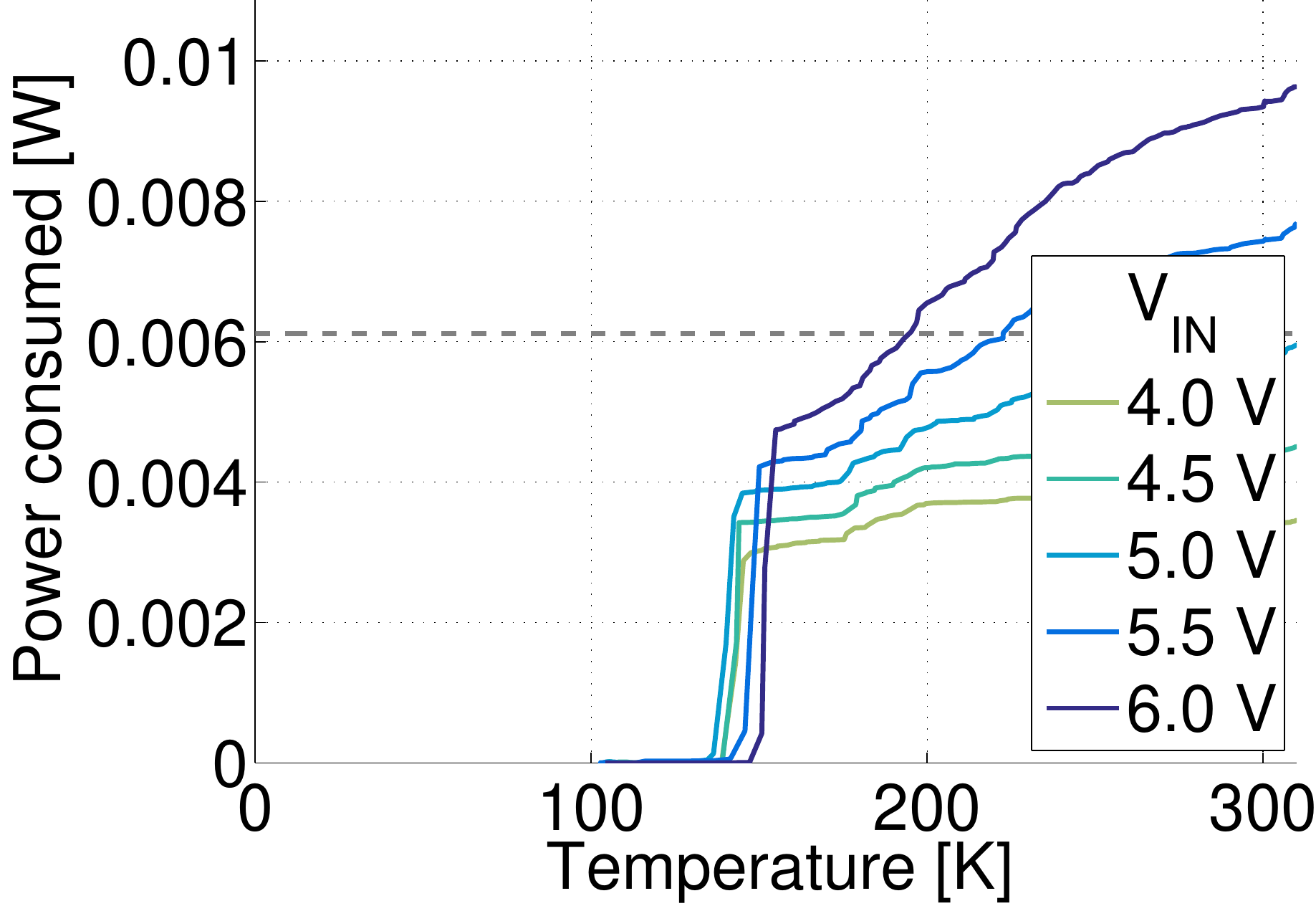} }
		\hfill
		\subfloat[Buck MCP16301]{\label{fig:regulators_power_e}
			\includegraphics[width=.3\textwidth]{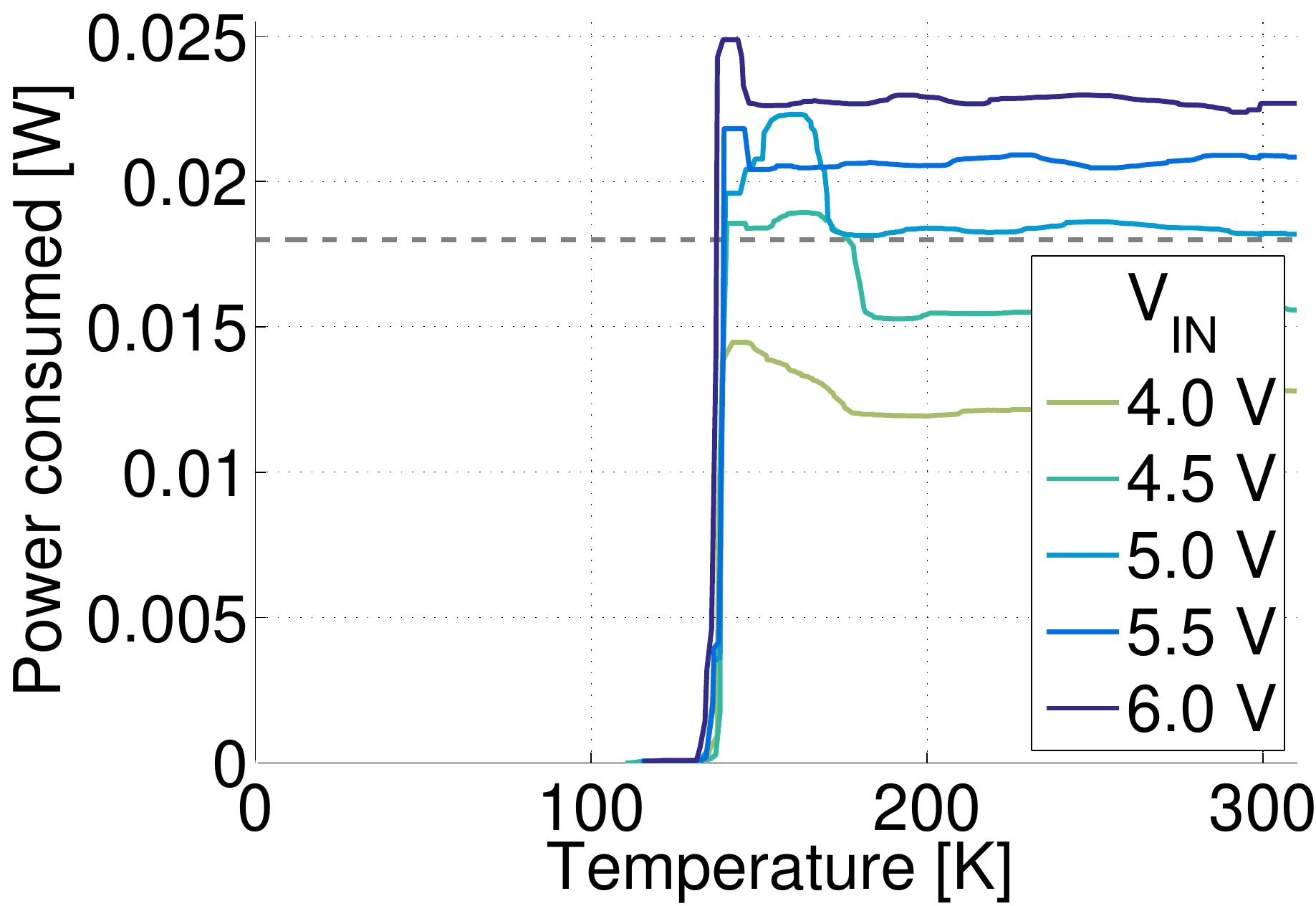} }
		\hfill
		\subfloat[Buck AST1S31HF]{\label{fig:regulators_power_f}
			\includegraphics[width=.3\textwidth]{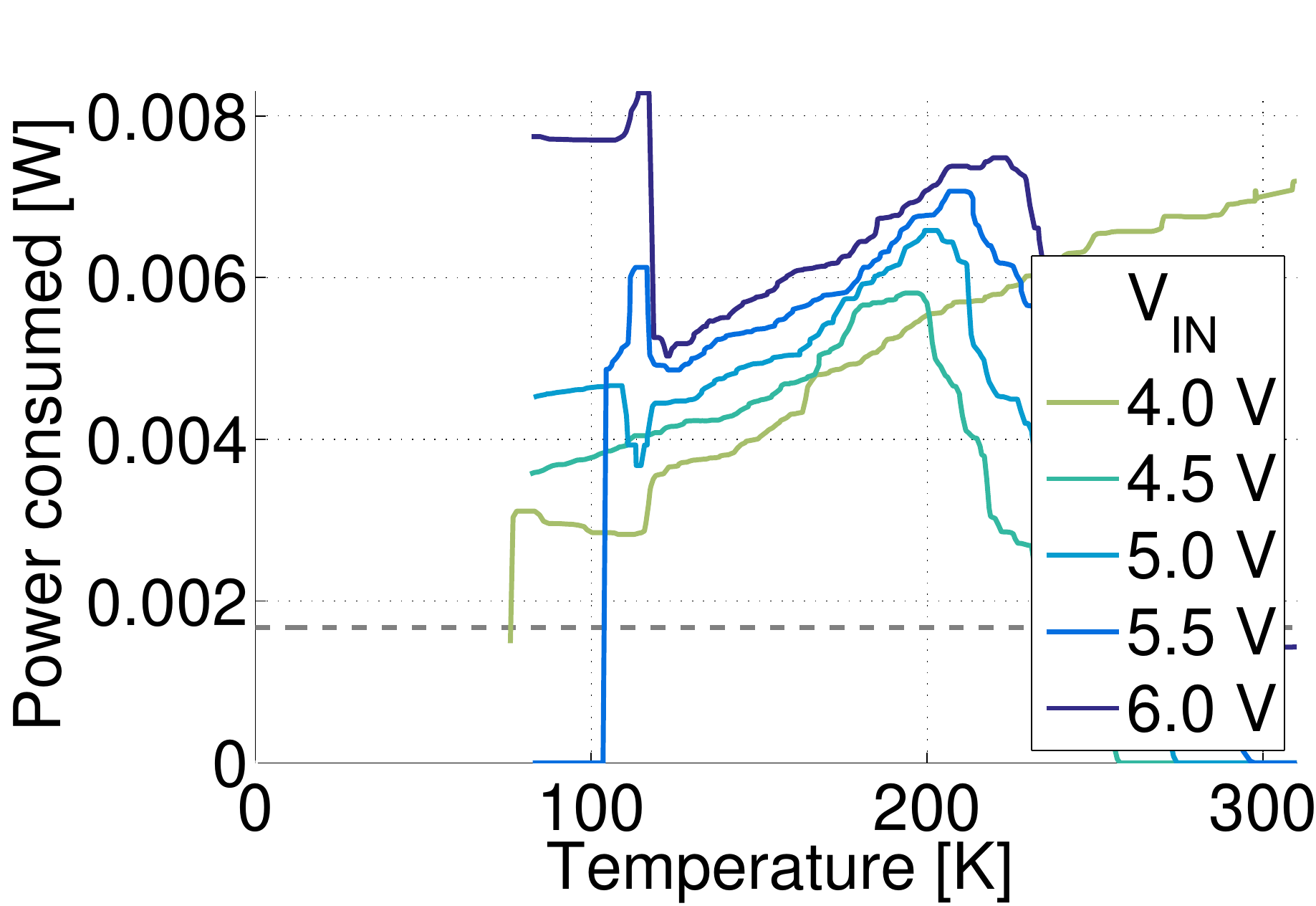} }
		\\
		\subfloat[Buck TPS54614]{\label{fig:regulators_power_g}
			\includegraphics[width=.3\textwidth]{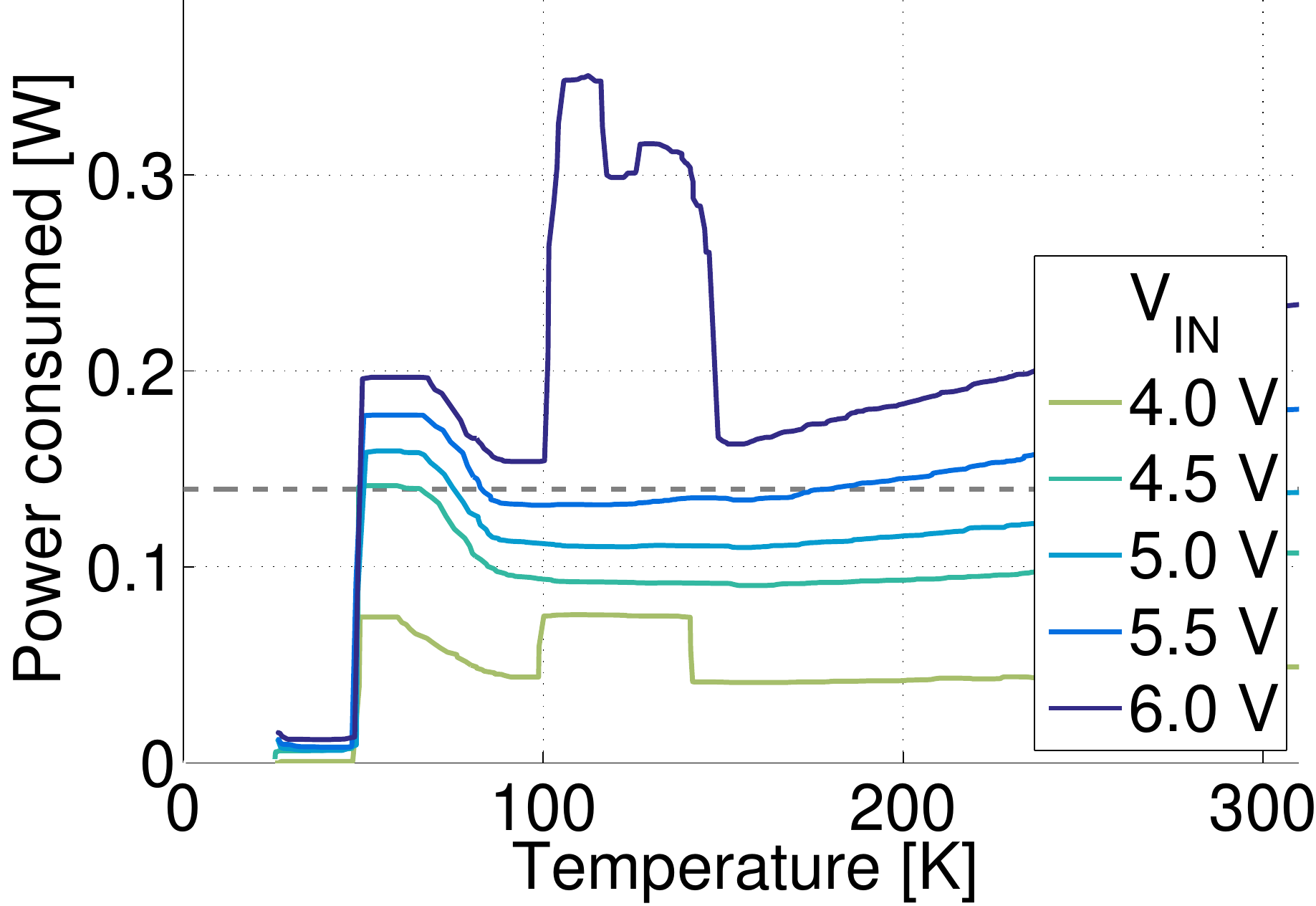} }
		\hfill
		\caption{Power consumption of the voltage regulators shown in \autoref{fig:detailedregulators_buck_voltage}.}
		\label{fig:detailedregulators_buck_power}
	\end{figure*}

\begin{figure*}[htbp]
		\centering
		\subfloat[LDO ADP165CB]{\label{fig:regulators_power_h}
			\includegraphics[width=.3\textwidth]{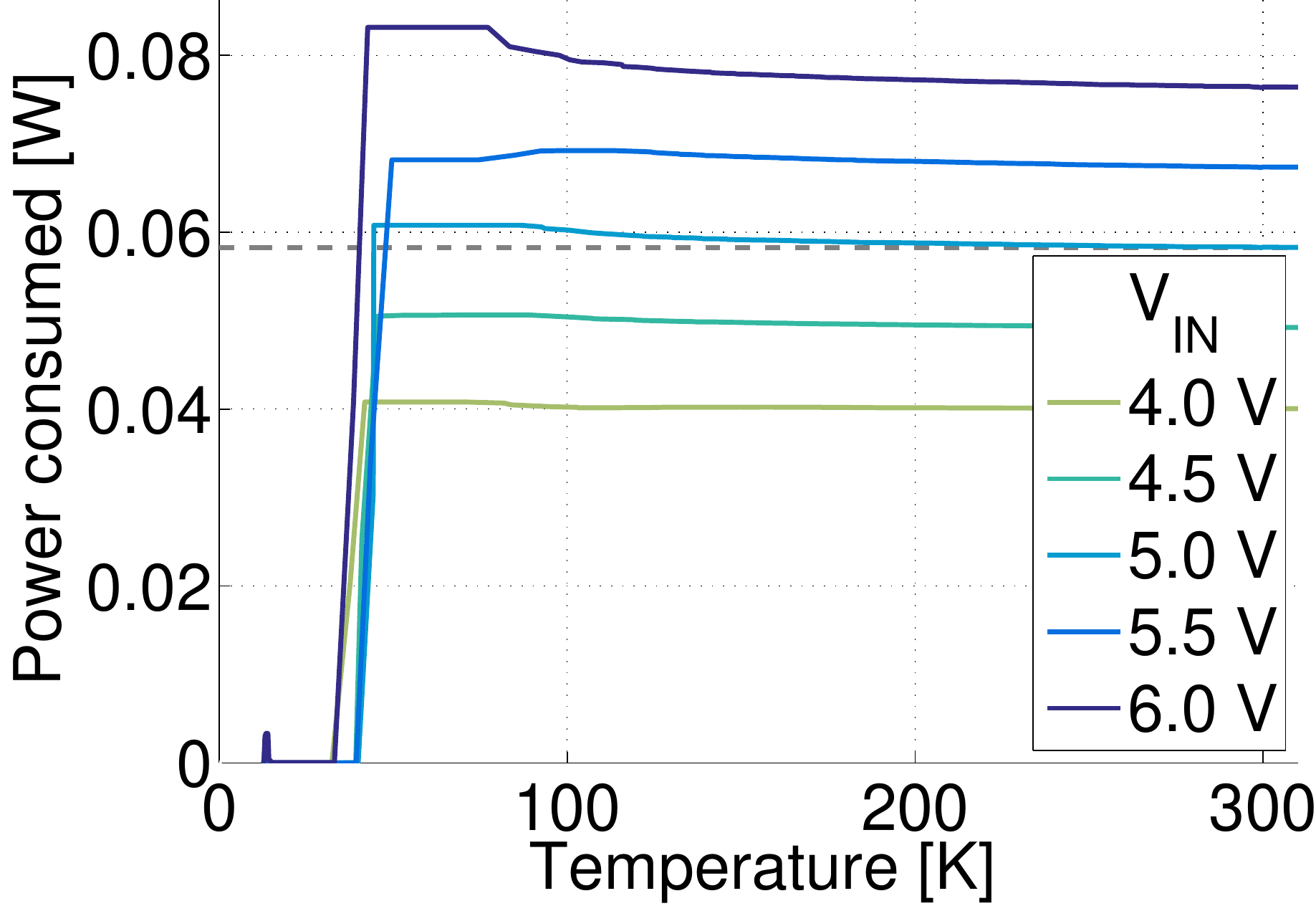} }
		\hfill
		\subfloat[LDO ADP165CP]{\label{fig:regulators_power_i}
			\includegraphics[width=.3\textwidth]{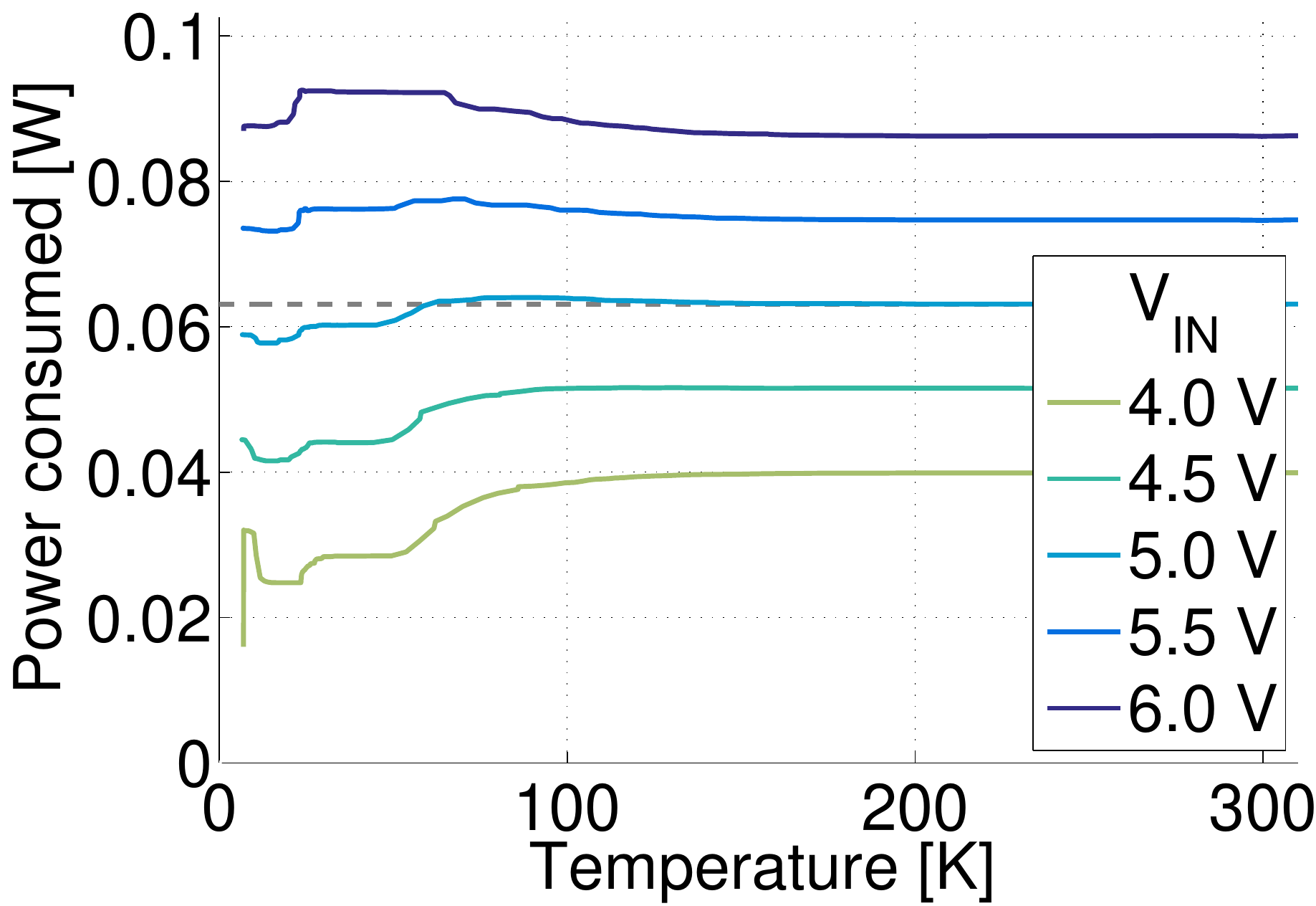} }
		\hfill
		\subfloat[LDO ADP165UJ]{\label{fig:regulators_power_j}
			\includegraphics[width=.3\textwidth]{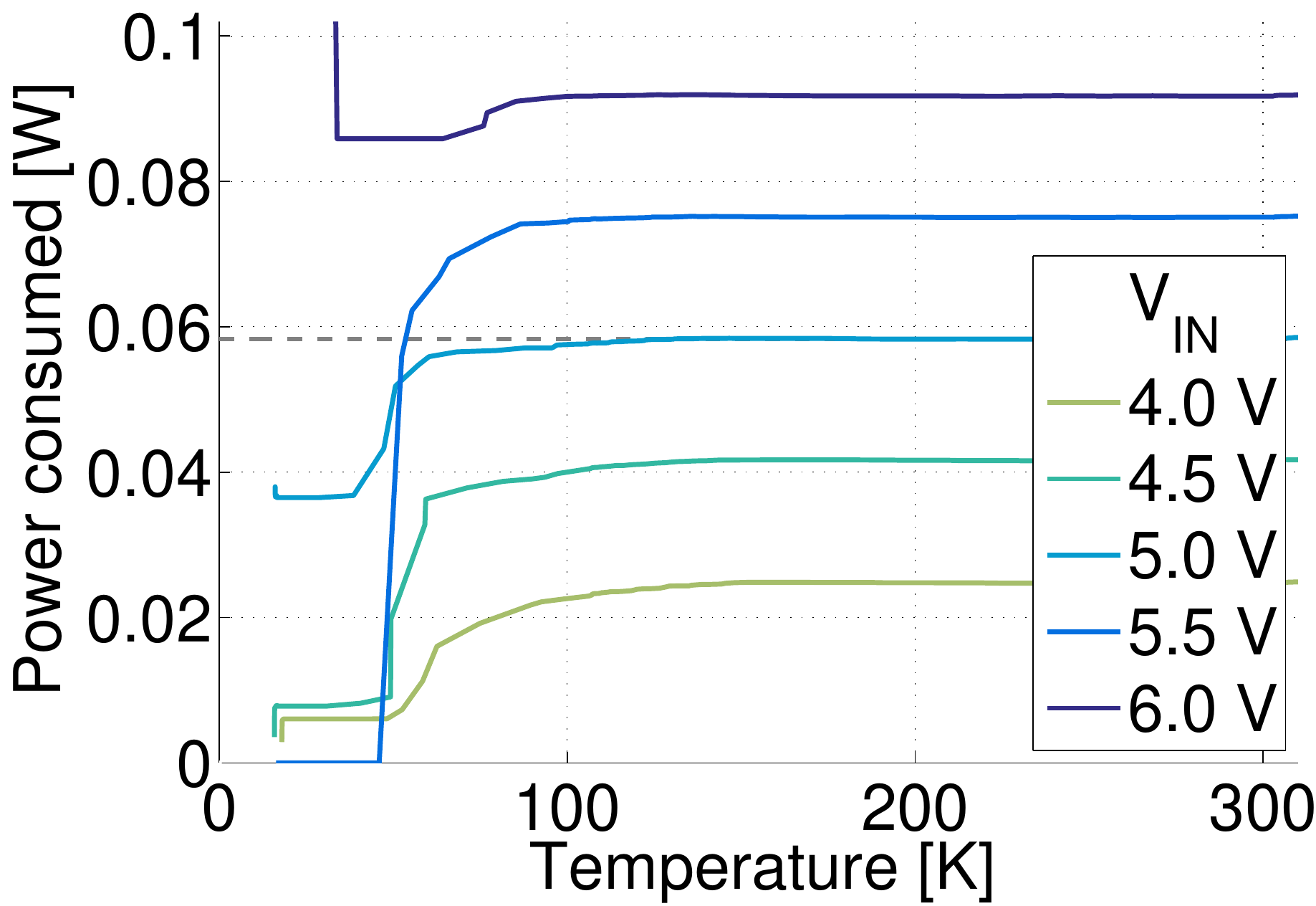} }
		\\
		\subfloat[LDO ISL80510]{\label{fig:regulators_power_k}
			\includegraphics[width=.3\textwidth]{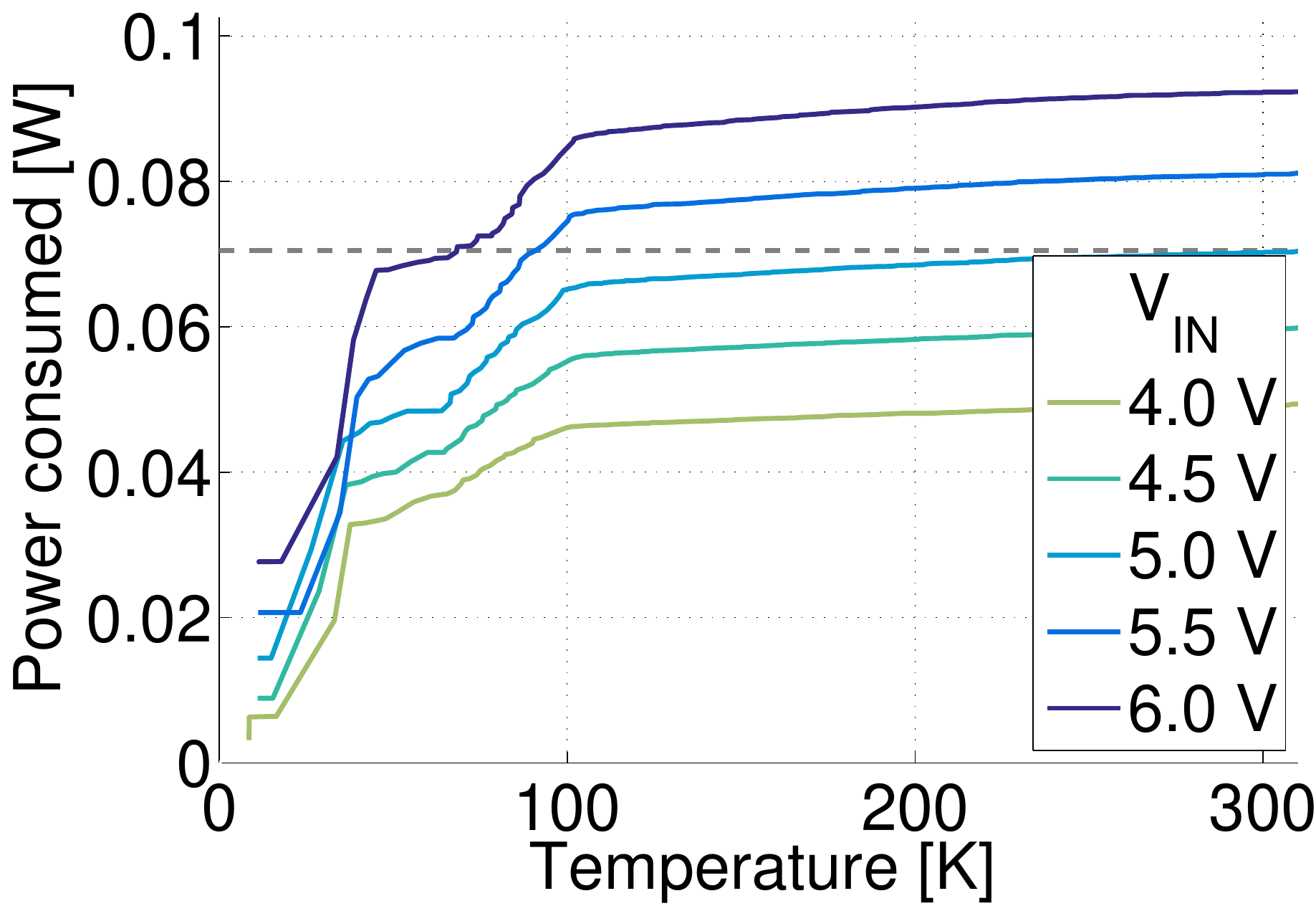} }
		\hfill
		\subfloat[LDO MIC38150]{\label{fig:regulators_power_l}
			\includegraphics[width=.3\textwidth]{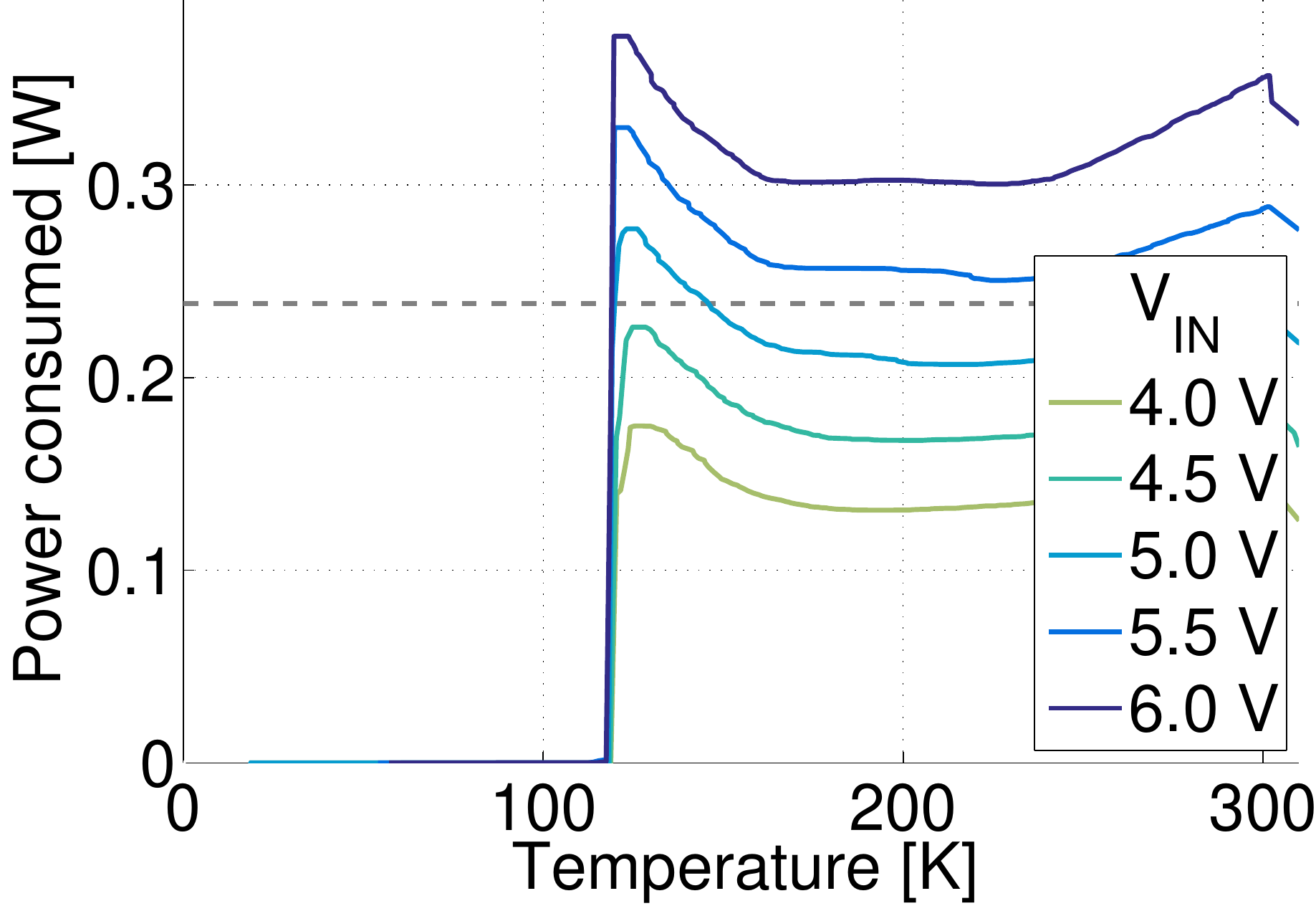} }
		\hfill
		\subfloat[LDO MIC69303]{\label{fig:regulators_power_m}
			\includegraphics[width=.3\textwidth]{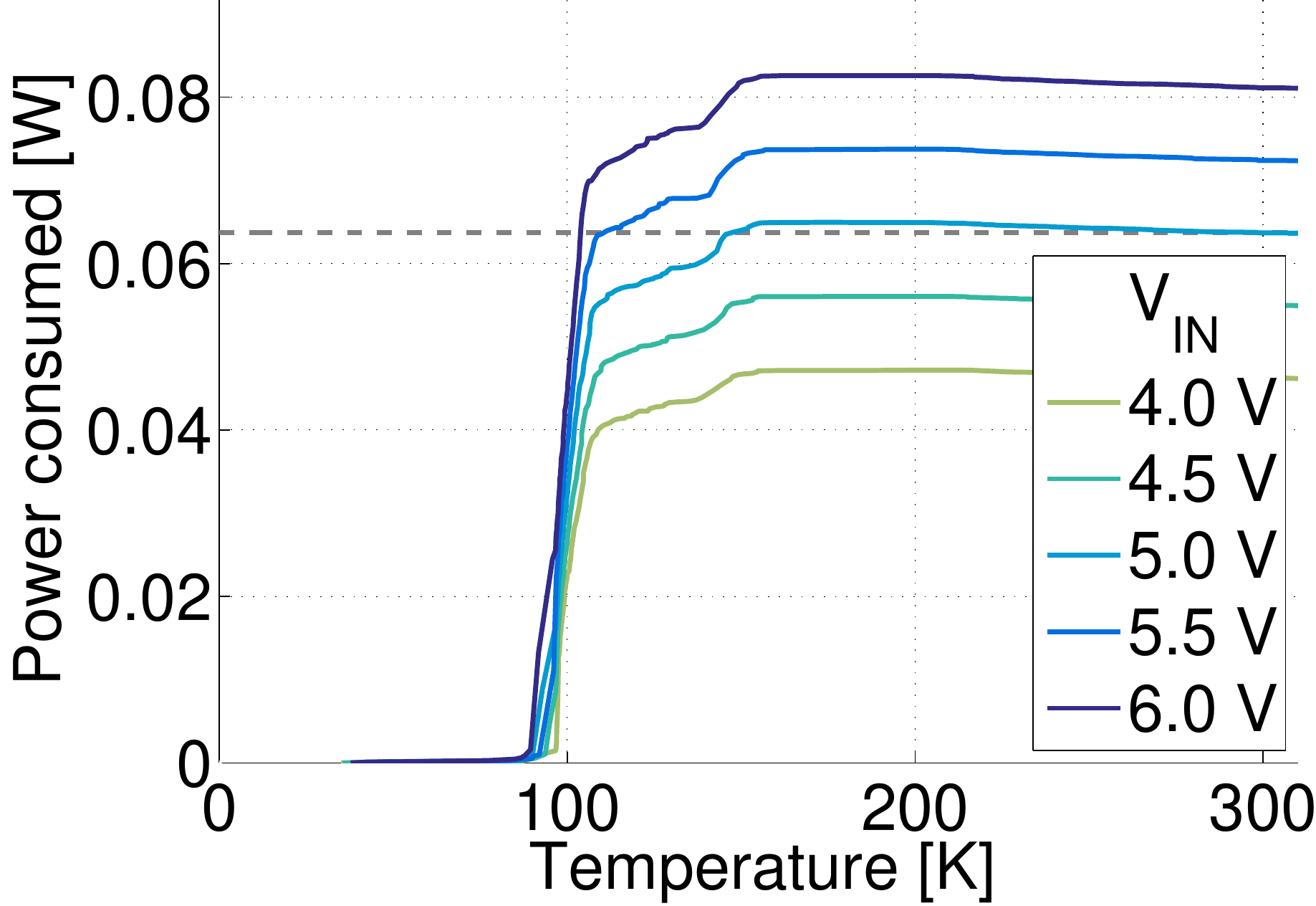} }
		\\
		\subfloat[LDO TPS7A4700]{\label{fig:regulators_power_n}
			\includegraphics[width=.3\textwidth]{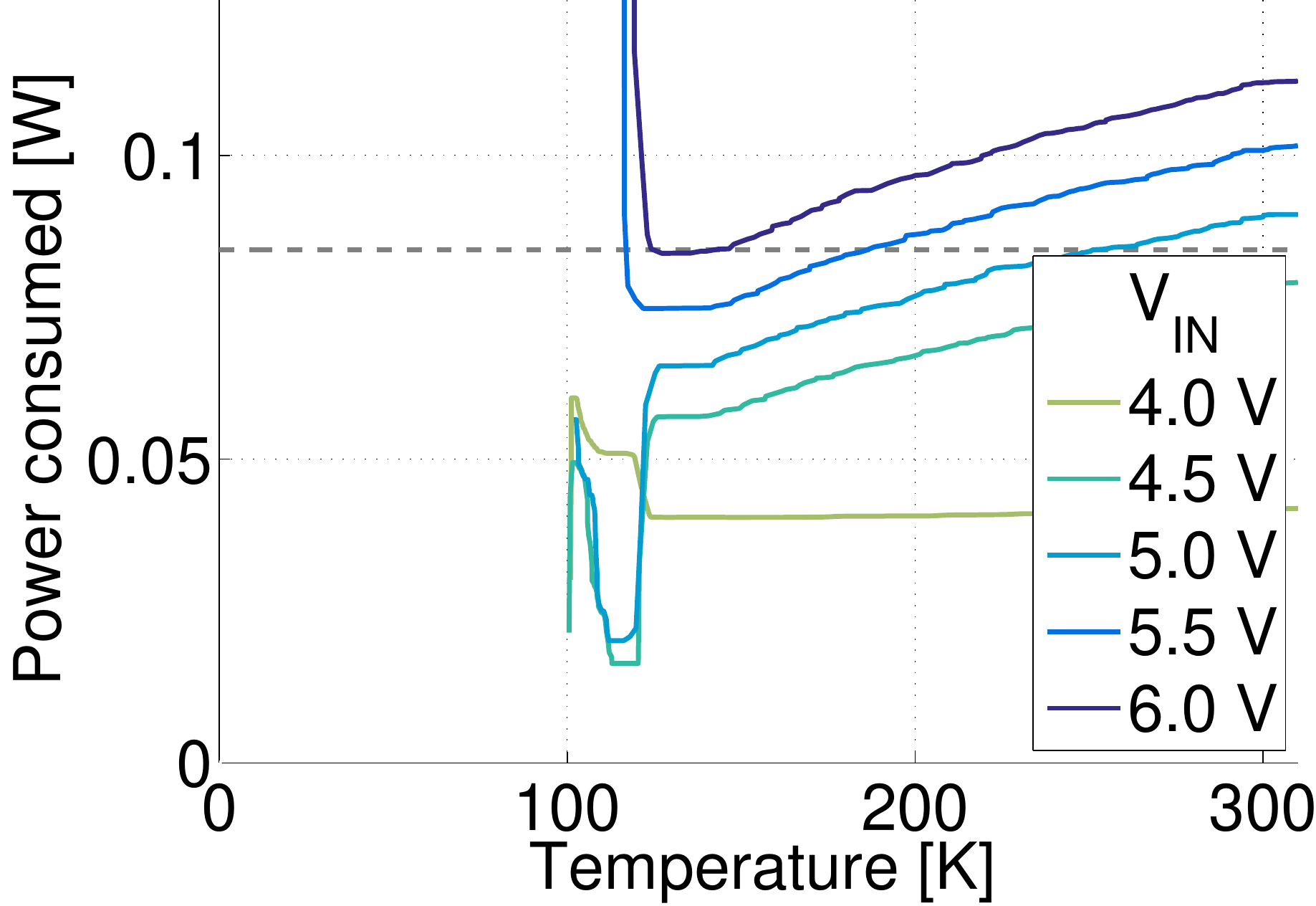} }
		\hfill
		\subfloat[LDO TPS7A7002]{\label{fig:regulators_power_o}
			\includegraphics[width=.3\textwidth]{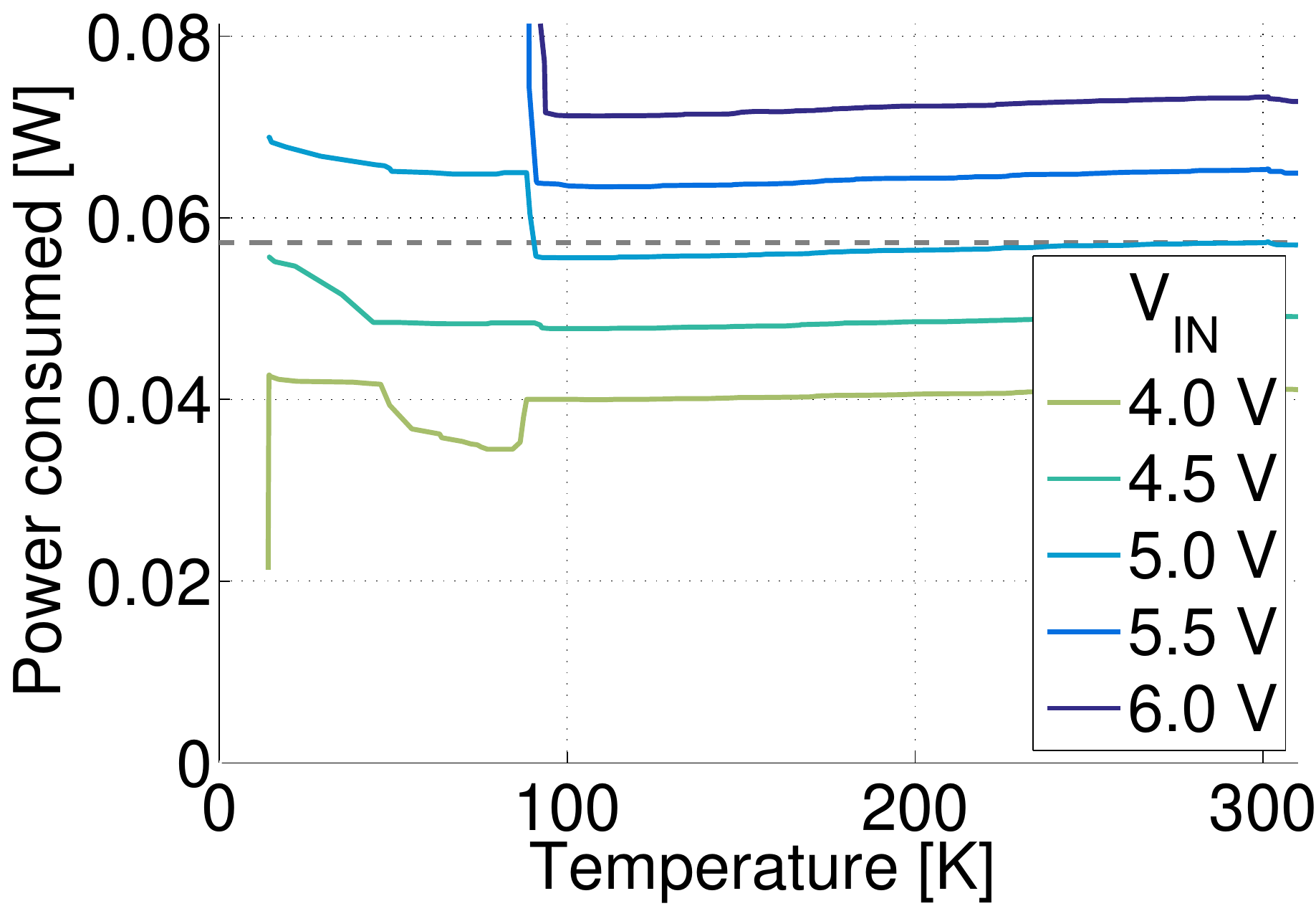} }
		\hfill
		\subfloat[LDO TPS7A8101]{\label{fig:regulators_power_p}
			\includegraphics[width=.3\textwidth]{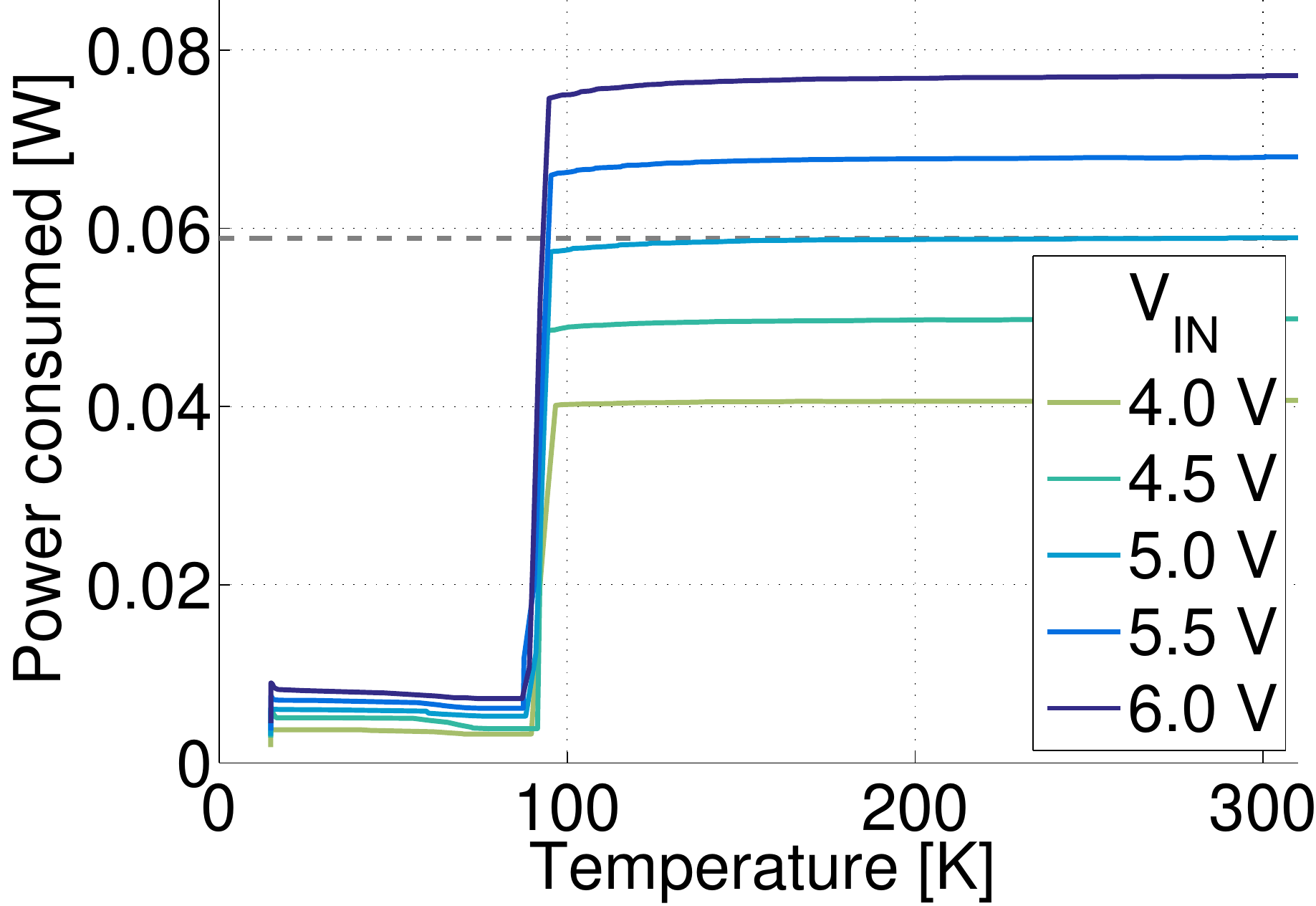} }
		\\
		\caption{Power consumption of the voltage regulators shown in \autoref{fig:detailedregulators_LDO_voltage}.}
		\label{fig:detailedregulators_LDO_power}
	\end{figure*}

\noindent In \autoref{fig:detailedregulators_buck_power} and \autoref{fig:detailedregulators_LDO_power}, the power consumption of the same set of regulators is shown versus temperature. Clearly, in general the switching supplies consume less power as expected compared to the LDOs. A major exception is the switching regulator ISL80030 (\autoref{fig:detailedregulators_buck_power}\subref{fig:regulators_power_a}) from InsterSil which consumes about 1~W, which is extremely high. 

 
\end{document}